\documentclass[twocolumn,8pt]{aastex631}
\usepackage[utf8]{inputenc}
\usepackage{CJK}
\usepackage[flushleft]{threeparttable}
\usepackage{longtable}
\usepackage{gensymb}

\usepackage{multirow}
\usepackage{array}
\usepackage{amsmath}
\usepackage{mathtools}

\shorttitle{Statistics of SESNE nebular spectra}
\shortauthors{Fang et al.}

\begin{document}
\begin{CJK*}{UTF8}{gbsn}
\title{Statistical properties of the nebular spectra of 103 stripped-envelope core-collapse supernovae}
\author[0000-0002-1161-9592]{Qiliang Fang (方其亮)}\affiliation{Department of Astronomy, Kyoto University, Kitashirakawa-Oiwake-cho, Sakyo-ku, Kyoto 606-8502, Japan}
\author[0000-0003-2611-7269]{Keiichi Maeda}\affiliation{Department of Astronomy, Kyoto University, Kitashirakawa-Oiwake-cho, Sakyo-ku, Kyoto 606-8502, Japan}
\author[0000-0002-1132-1366]{Hanindyo Kuncarayakti}\affiliation{Tuorla Observatory, Department of Physics and Astronomy, FI-20014 University of Turku, Finland} \affiliation{Finnish Centre for Astronomy with ESO (FINCA), FI-20014 University of Turku, Finland}
\author[0000-0001-8253-6850]{Masaomi Tanaka}\affiliation{Astronomical Institute, Tohoku University, Sendai 980-8578, Japan}
\author[0000-0001-6099-9539]{Koji S. Kawabata}\affiliation{Department of Physical Science, Hirosima University, Kagamiyama, Higashi-Hiroshima, Hiroshima 739-8526, Japan}
\author{Takashi Hattori}\affiliation{Subaru Telescope, National Astronomical Observatory of Japan, 650 North A'ohoku Place, Hilo, HI 96720, USA}
\author{Kentaro Aoki}\affiliation{Subaru Telescope, National Astronomical Observatory of Japan, 650 North A'ohoku Place, Hilo, HI 96720, USA}
\author[0000-0003-1169-1954]{Takashi J. Moriya}
\affiliation{National Astronomical Observatory of Japan, National Institutes of Natural Sciences, 2-21-1 Osawa, Mitaka, Tokyo 181-8588, Japan}
\affiliation{School of Physics and Astronomy, Faculty of Science, Monash University, Clayton, Victoria 3800, Australia}
\author{Masayuki Yamanaka}\affiliation{Okayama Observatory, Kyoto University, 3037-5 Honjo, Kamogata-cho,
Asakuchi, Okayama 719-0232, Japan}

\begin{abstract}
We present an analysis of the nebular spectra of 103 stripped-envelope (SE) supernovae (SNe) collected from the literature and observed with the Subaru Telescope from 2002 to 2012, focusing on [O {\sc i}] $\lambda\lambda$6300, 6363. The line profile and width of [O {\sc i}] are employed to infer the ejecta geometry and the expansion velocity of the inner core; these two measurements are then compared with the SN sub types, and further with the [O {\sc i}]/[Ca {\sc i}{\sc i}] ratio, which is used as an indicator of the progenitor CO core mass. Based on the best-fit results of the [O {\sc i}] profile, the objects are classified into different morphological groups, and we conclude that the deviation from spherical symmetry is a common feature for all types of SESNe. There is a hint (at $\sim 1 \sigma$ level) that the distributions of the line profile fractions are different between canonical SESNe and broad-line SNe Ic. A correlation between [O {\sc i}] width and [O {\sc i}]/[Ca {\sc i}{\sc i}] is discerned, indicating that the oxygen-rich material tends to expand faster for objects with a more massive CO core. Such a correlation can be utilized to constrain the relation between the progenitor mass and the kinetic energy of the explosion. Further, when [O {\sc i}]/[Ca {\sc i}{\sc i}] increases, the fraction of objects with Gaussian [O {\sc i}] profile increases, while those with double-peaked profile decreases. This phenomenon connects ejecta geometry and the progenitor CO core mass.

\end{abstract}

\keywords{
  line: formation --- supernova: core-collpase: nebular --- supernova: general}

\section{INTRODUCTION}
When the central nuclear fuel is exhausted, a massive star ($\gtrapprox$ 8 $M_{\odot}$) will suffer from core collapse, resulting in a core-collapse supernova (CCSN), expelling the material above the core. The explosion energy and the geometry of the ejecta of this catastrophic event, together with their relations with the properties of the progenitor, are important factors for understanding the final evolution of massive stars. 

Before an SN explodes, the massive star progenitor may suffer from certain degree of envelope stripping either by binary evolution or stellar wind, or the combination of both (\citealt{heger03, groh13, smith14, yoon15, fang19}). If the hydrogen envelope is mostly retained before the explosion, the star will explode as a type II supernova (SN II), with strong hydrogen features in its spectra. Otherwise it will explode as a stripped-envelope supernova (SESN). SESNe can be further classified into type IIb SNe (SNe IIb, with strong hydrogen lines in early-phase spectra which are later replaced by helium lines),  type Ib SNe (SNe Ib, with spectra dominated by helium lines, showing no or weak hydrogen signatures) and type Ic SNe (SNe Ic, with spectra lacking both hydrogen and helium lines). Type Ic SNe can be further divided into normal SNe Ic and broad-line type Ic (SNe Ic-BL). The early-phase spectra of the latter type show broad absorption features, indicating fast-expanding ejecta (by a factor of $\sim 2$ faster than normal SNe Ic at maximum brightness) and large kinetic energy ($\gtrapprox$10$^{52}$ erg, compared with $\sim$10$^{51}$ erg for typical SNe). SNe Ic-BL are sometimes associated with gamma-ray bursts (GRBs, see \citealt{galama98, hjorth03}, or \citealt{woosley06} for a review).

The explosion mechanism of CCSNe is an important open problem in modern astronomy. It is not yet clear how the gravitational energy is transformed to the kinetic energy of the outward-moving material. Placing observational constraints on the explosion geometry is one of the keys to answering this problem. The explosion energy may also depend on the progenitor masses (\citealt{ugliano12, muller16, sukhbold16}). Therefore, it is important to explore possible relations between these quantities from observational data; we thus need to have indicators of the kinetic energy, the ejecta geometry, and the progenitor mass independently from observables. For the mass of the progenitor star, the most robust method is to use a high resolution image of the progenitor, although it still depends on the theoretical calculation of stellar evolution therefore introduces some uncertainties (\citealt{smartt09} and \citealt{smartt15}). The direct detection of the progenitors are only feasible in a relatively small volume, where CCSNe are rare events. The direct images of CCSNe are only available for a few number of cases, especially lacking hydrogen-poor SNe ( \citealt{maund04,maund11,dyk14,folatelli15,kilpatrick17,tartaglia17}. For reviews, see \citealt{smartt09} and \citealt{smartt15}). So far only two SNe Ib, iPTF 13bvn and SN 2019yvr, are identified (\citealt{cao13,kilpatrick21}). 

SNe in their early phases are luminous enough so that they can be observed in distant galaxies. The luminosity scale and the shape of the light curve are dependent on  the amount of radioactive elements, and the mass and the kinetic energy of the ejecta. The light curve shape is also affected by how the radioactive power source is mixed in the envelope. Many works have been conducted which allow investigation of a possible relation between the ejecta mass and the explosion energy based on large samples  (\citealt{drout11,dessart16,lyman16,prentice16,taddia18}). However, the early-phase emission is mainly originated from the outermost region of the optical thick ejecta, and is not directly related to the inner core; thus converting the ejecta mass estimated in this way to the progenitor mass may involve a large uncertainty. Further, the early-phase observables are generally not sensitive to the ejecta geometry except for the polarization signal (\citealt{wang01,wang08,nagao21}). Indeed, most of the codes employed to model the early-phase SN light curve assume that the ejecta are spherically symmetric, which is not necessary valid (\citealt{maeda08, tauben09}). 

Observation during the nebular phase naturally meets all the requirements. After the massive star explodes, the density of the ejecta decreases with time following the expansion. At the same time, recombination also reduces the electron density. These effects together reduce the optical depth of the ejecta. When the ejecta becomes transparent to expose the inner region, the SN enters its nebular phase, and the spectrum is dominated by emission lines, most of which are forbidden lines. An SN usually enters its nebular phase several months to about one year after the explosion, depending on the physical conditions of the ejecta. For SNe II that retain most of its hydrogen envelope before the explosion, the nebular phase usually starts later than its envelope-stripped counterparts (SN IIb/Ib/Ic).  

The optically-thin nature of the late-time ejecta allows a non-biased view on the entire ejecta, especially sensitive to the innermost region. One can therefore obtain indications of the geometry, the mass, and the expansion velocity of the innermost core, using the same late-phase data. The width of an emission line, together with its profile, allow one to explore the velocity scale and the geometry of the emitting region. The absolute or relative strength of emission lines is also related to the mass, volume and physical conditions of the emitting regions (\citealt{fransson89,jerk15,jerk17,dessart21a}). The information thus obtained can be utilized to infer the properties of the progenitor and constrain the explosion mechanism.

In this work, we conduct a study on the properties of the emission lines, including the width, profile and strength based on the so-far largest sample of SESN nebular spectra. In the sample, 88 spectra are collected from the literature, and 15 spectra are newly presented from the observations carried out with the Subaru Telescope from 2002 to 2012. In the present work, we focus on the forbidden line [O {\sc i}] $\lambda\lambda$6300,6364, as it is one of the most luminous emission lines in the optical window of SESN nebular spectra. Further, oxygen is one of the most abundant elements in the ejecta of SESNe, and the [O {\sc i}] dominates the emission from the CO core; the [O {\sc i}] $\lambda\lambda$6300,6364 doublet is thus an ideal tool to trace the geometry of the ejecta and the properties of the progenitor.

The paper is organized as follow: in \S 2, the full sample, the data reduction methods and the measurement of the observables are introduced. The latter includes the width and line profile of the [O {\sc i}] $\lambda\lambda$6300,6364, and the line ratio [O {\sc i}] $\lambda\lambda$6300,6364/[Ca {\sc i}{\sc i}] $\lambda\lambda$7291,7323. In \S 3, we perform statistical analysis on the line profile. The statistics of the [O {\sc i}]/[Ca {\sc i}{\sc i}] ratio, along with its correlations with other observables, including the [O {\sc i}] width and the line profile, are presented in \S 4. \S 5 is devoted to the physical implications of the statistical results of \S 3 and \S 4. The validity and the possible affecting factors of the measurements are discussed in \S 6. Finally the conclusions are given in \S 7.

\section{Data set}

\subsection{Sample Description}
The sample in this work includes the late-time spectra of 103 stripped-envelope supernovae (26 SNe IIb, 31 SNe Ib, 32 SNe Ic, 9 SNe Ic-BL and 5 SNe Ib/c), among which 15 objects are not published in the previous literature. The spectra are selected if the signal-to-noise level is acceptable and the wavelength covers 6000 to 8000 $\rm \AA$ so that the measurements in this work ([O {\sc i}] and [Ca {\sc i}{\sc i}]) are possible. The phases of the spectra are restricted to later than 100 days after the explosion or the peak luminosity (if the light curve is available). The objects that are decidedly nebular are also included, even if early-phase observations do not exist and the exact phase is unknown. If multiple nebular spectra are available for a specific object, we pick the one closest to 200 days. However, the quantities of interest in this work ([O {\sc i}]/[Ca {\sc i}{\sc i}] and the [O {\sc i}] width) are not sensitive to the spectral phase within the range used here, so the effect of temporal evolution is generally negligible (see \citealt{maurer10} and \citealt{fang19} for the time evolutions of [O {\sc i}] width and [O {\sc i}]/[Ca {\sc i}{\sc i}] respectively; see also the discussion in \S 6). The previously published spectra are collected from The Open Supernova Catalogue\footnote{https://sne.space/} (\citealt{guillochon17}) and WiseRep\footnote{https://www.wiserep.org/} (\citealt{yaron12}). The full sample of this work is listed in Table~\ref{tab:listIIb} to Table~\ref{tab:listIbc} in the Appendix.

\subsection{Data reduction}
For the new data set presented in this paper, the spectroscopic observations for the 15 SESNe were performed from MJD 52432 (2002 June 7th) to MJD 56222 (2012 October 22th) with the 8.2m Subaru Telescope equipped with the Faint Object Camera and Spectrograph (FOCAS; \citealt{yoshida00, kashikawa02}). The typical instrumental setup is the following: we used the 0″.8 slit and the B300 (with no filter) and R300 (equipped with the O58 filter) grisms, or the 0″.8 offset slit and the B300 grism equipped with the Y47 filter. The spectral resolution is $\sim$500, or $\sim$13$\rm{\AA}$ at 6300$\rm{\AA}$. The log of the observations is listed in Table~\ref{tab:FOCAS log}. 

The spectra are reduced following the standard procedures using IRAF\footnote{IRAF is distributed by the National Optical Astronomy Observatory, which
is operated by the Association of Universities for Research in Astronomy, Inc., under cooperative agreement with the National Science Foundation. PyRAF is a product of the Space Telescope Science Institute, which is operated by AURA for NASA.} (\citealt{tody86,tody93}), including bias subtraction, flat fielding, sky subtraction, 1D spectral extraction, wavelength calibration using ThAr or HeNeAr lamps and skylines, cosmic-ray rejection using LAcosmic (\citealt{dokkum01}). Flux calibration is performed by using standard stars observed in the same night.

\begin{deluxetable}{cccc}
\tablecaption{Log of spectroscopic observations with FOCAS}
\label{tab:FOCAS log}
\tablehead{
\colhead{Object} & \colhead{Date} & \colhead{Instrumental setup}& \colhead{Exposure time}\\
\nocolhead{e} & \colhead{YY/MM/DD} & \colhead{(grism/filter)} & \colhead{(seconds)}
}
\startdata
2005bj&05/08/25&B300off/Y47&3$\times$1200\\
2005aj&05/10/26&B300off/Y47&2$\times$1200\\
2006G&06/06/30&B300off/Y47&1$\times$1200\\
2006ep&06/12/24&B300off/Y47&1$\times$1200\\
2007D&07/09/18&B300off/Y47&2$\times$1500\\
2007ay&07/11/05&B300off/Y47&1$\times$1200\\
2008fo&09/04/05&B300cen, R300cen/O58&1$\times$1200\\
2008fd&09/07/23&B300cen, R300cen/O58&1$\times$1200\\
2008hh&09/08/18&B300cen, R300cen/O58&2$\times$1000\\
2008im&09/08/18&B300cen, R300cen/O58&2$\times$720\\
2009C&09/10/26&B300cen, R300cen/O58&2$\times$900\\
2009K&09/10/26&B300cen, R300cen/O58&2$\times$900\\
2008ie&09/10/27&B300cen, R300cen/O58&4$\times$1200\\
2009jy&10/05/06&B300cen, R300cen/O58&2$\times$1200\\
2009ka&10/05/06&B300cen, R300cen/O58&1$\times$900\\
\enddata
\end{deluxetable}

\begin{figure}[!htb]
\epsscale{1.1}
\plotone{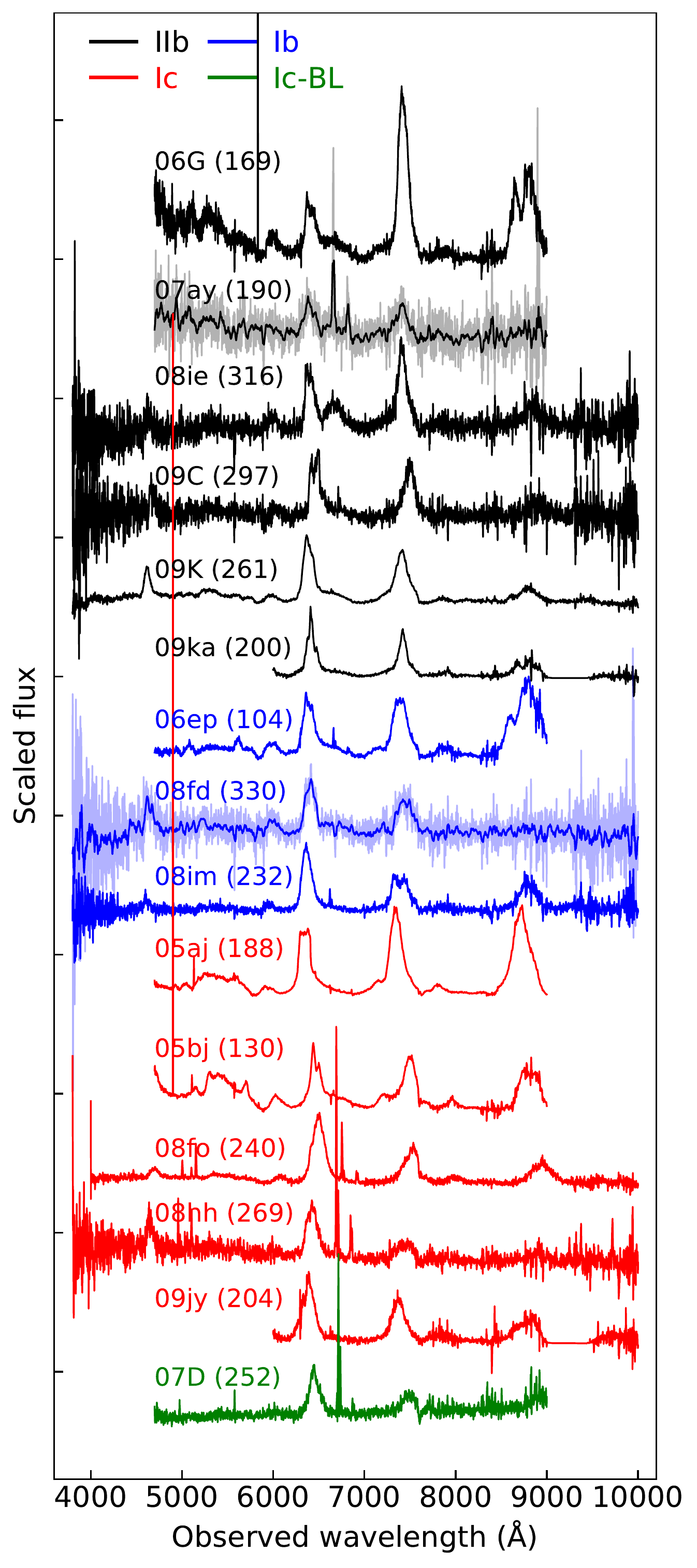}
\centering
\caption{The nebular spectra of the Subaru/FOCAS objects. The fluxes of the spectra are first normalized, and then added by different constants for illustration. The objects are labeled by the last two digits of the discovery year and letter(s). The phase relative to the discovery date or light curve maximum (if available) of each spectrum is listed in the parentheses (unit: days). SNe of different sub types are plotted with different colors. SN2007ay and SN2008fd are smoothed for illustration purpose.}
\label{fig:compare_width}
\vspace{4mm}
\end{figure}

\subsection{Measurement of observables}
The goal of this work is to investigate the physical properties of the ejecta and the progenitors, by using a large data set of nebular spectra of SESNe. In this work, we are not attempting to fit the nebular spectra with full spectral modeling; instead, several observables are employed as the indicators of the physical properties of the ejecta or the progenitor, including the line ratio of [O {\sc i}] $\lambda\lambda$6300,6364 to [Ca {\sc i}{\sc i}] $\lambda\lambda$7291,7323, which is suggested to be related to the CO core mass, and thus the zero-age-main-sequence (ZAMS) mass of the progenitor (see later discussion). Following \citet{tauben09}, the line profile and the width of [O {\sc i}] are utilized to probe the geometry and velocity scale of the ejecta.

The nebular spectra of SESNe are dominated by [O {\sc i}] and [Ca {\sc i}{\sc i}] emissions. 
Before measuring the observables, a nebular spectrum is de-reddened and corrected for redshift at the first step. The color excess $E(B-V)$ of the host galaxy and the Milky way absorption are derived from the previous literature (see the references in Table~\ref{tab:listIIb} to Table~\ref{tab:listIbc}). For SNe without reported $E(B~-~V)$, the extinction is estimated from the equivalent width of Na ID absorption, using the relation derived from \citet{turatto03}, if spectra around the light curve peak are available. Otherwise $E(B - V)$ is set to be 0.36 mag, which is the average value for SN Ib/c by \cite{drout11}. The spectra are then corrected for extinction by applying the Cardelli law (\citealp{cardelli89}), assuming $R_{V}$ = 3.1.

The redshifts for most objects are inferred from the central wavelength of the narrow emissions from their explosion sites (H$\alpha$, [N {\sc i}{\sc i}] etc.). If such narrow lines are absent in the spectrum, the redshift of the host galaxy from HyperLeda\footnote{http://leda.univ-lyon1.fr/} is adopted (\citealp{makarov14}). 

The next step is to remove the underlying continuum emission. Following \citet{fang19}, we first slightly smooth the spectra and find the local minimum at both sides of [O {\sc i}]/H$\alpha$-like-structure (also [Ca {\sc i}{\sc i}]/[Fe {\sc i}{\sc i}]) complex. A line connecting the two minima is defined to be the local continuum emission and is then subtracted. Indeed, the continuum of nebular-phase SNe is not real continuum emission, but made of thousands of weak overlapping lines (\citealt{li96,dessart21a}). Subtracting the straight line defined above may result in some residual, therefore affects the measurement. However, as long as all objects are treated with the same method, the effect of the residual on statistics will be negligible. After these two steps, we can start to measure the line ratios and [O {\sc i}] profiles.

\begin{figure}[!t]
\epsscale{1.1}
\plotone{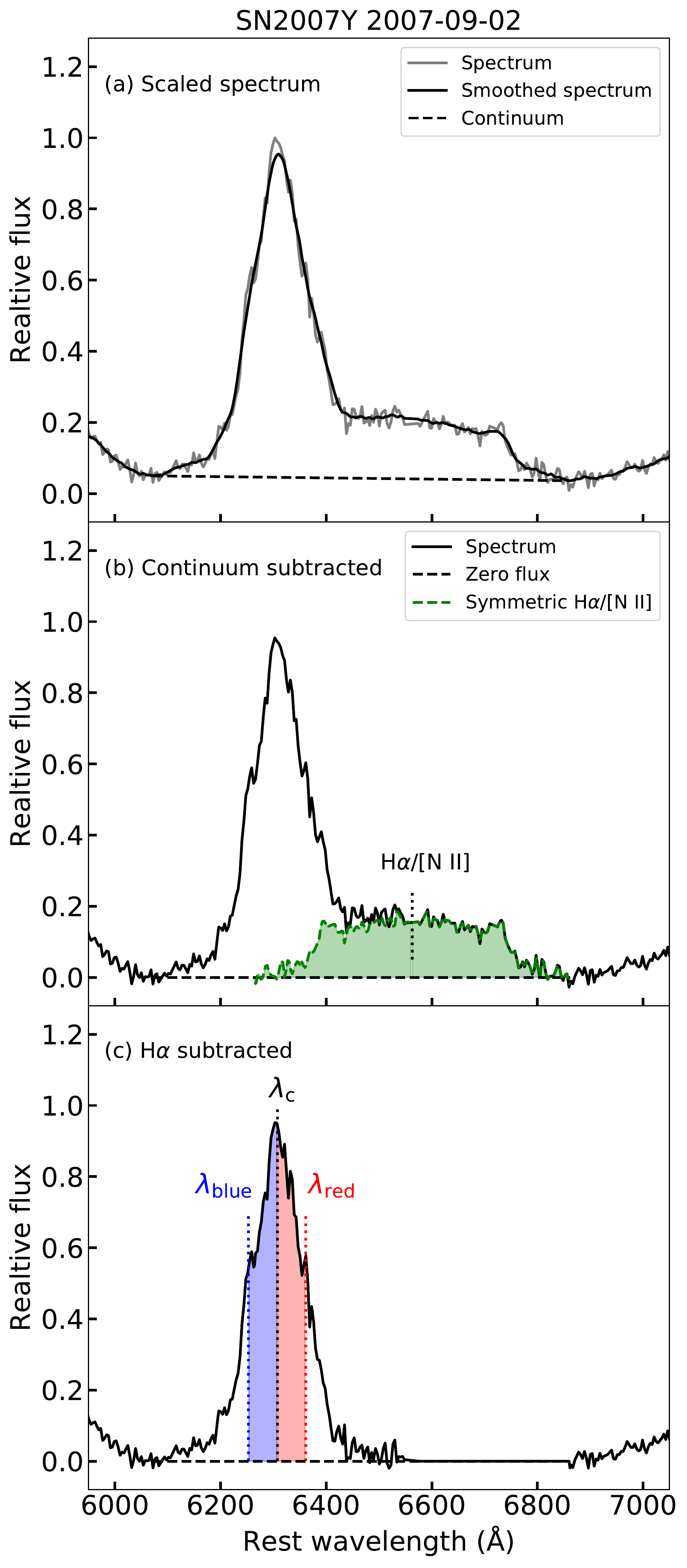}

\caption{A detailed example of observable measurement in \S 2.3. The spectrum of SN2007Y is corrected for extinction and redshift, then multiplied by a constant for illustration purpose. $Upper~panel$: The spectrum is smoothed and the continuum level is determined as illustrated by the dashed line. $Middle~panel$: The continuum is subtracted and the symmetric H$alpha$-like structure is constructed. $Lower~panel$: The H$\alpha$-like structure is subtracted. The line width is determined by $\lambda_{\rm c}$ -$\lambda_{\rm blue}$.}
\label{fig:example_07Y}
\end{figure}

\begin{itemize}
\item {\rm [O ~{\sc I}]}~and~{\rm [Ca ~{\sc I}{\sc I}]} ~~~The relative flux of [Ca {\sc i}{\sc i}] is measured following the same procedure as \citet{fang19}. As for [O {\sc i}], instead of fitting the [O {\sc i}] with a double Gaussian function as illustrated in \citet{fang19}, we assume the H$\alpha$-like structure located at the red side of [O {\sc i}] is symmetric with respect to 6563 ${\rm \AA}$. Its profile is constructed by reflecting the red wing to the blue side with respect to 6363\AA, and the relative flux is then computed. The H$\alpha$-like structure is commonly seen in the nebular spectra of SNe IIb and some SNe Ib, and is identified as H$\alpha$ or [N {\sc i}{\sc i}] (\citealt{patat95, jerk15, fang18}). As will be discussed in \S 6, the measured line width is not sensitive to the assumed symmetric center, therefore the exact identification of this line is not important for the purpose of this work. Given that the symmetric center of the [N {\sc i}{\sc i}] doublets is close to 6563 ${\rm \AA}$, to avoid further complication, we assume the excess emission is symmetric with respect to 6563 ${\rm \AA}$. After the H$\alpha$-like complex is subtracted, the profile and the relative flux of [O {\sc i}] can be determined. 

\item Line~width~of~{\rm [O ~{\sc I}]} ~~~The line width of [O {\sc i}] is measured after the H$\alpha$-like structure is subtracted from the complex. We first define $\lambda_{\rm c}$, such that the integrated fluxes at both sides are equal. We then find $\lambda_{\rm blue}$ and $\lambda_{\rm red}$, where the integrated fluxes between $\lambda_{\rm blue}$...$\lambda_{\rm c}$ and $\lambda_{\rm c}$...$\lambda_{\rm red}$ take 34\% of the total emission. The line width measured in this way defines 1$\sigma$ if the [O {\sc i}] profile is Gaussian. A detailed example of line width measurement is presented in Figure \ref{fig:example_07Y}. Throughout this work, the blue width $\Delta\lambda_{\rm blue}$ ($\equiv$ $\lambda_{\rm c}$ - $\lambda_{\rm blue}$) is employed as the measurement of the line width, instead of using the half width $\Delta\lambda_{\rm half}$ ($\equiv$ $\frac{\lambda_{\rm red} - \lambda_{\rm blue}}{2}$) or the red width $\Delta\lambda_{\rm red}$ ($\equiv$ $\lambda_{\rm red}$ - $\lambda_{\rm c}$). This is because $\Delta\lambda_{\rm blue}$ is less affected by the subtraction process or the profile of the H$\alpha$-like structure. A detailed discussion is left to \S 6. 

\end{itemize}

The emission lines are broadened by the instrument. The measured line width can be corrected to account for the resolution of the instrument as

\begin{equation}
    \Delta\lambda_{\rm intrinsic}=\sqrt{\Delta\lambda_{\rm observed}^2-\Delta\lambda_{\rm narrow}^2}, 
\end{equation}
where $\Delta\lambda_{\rm intrinsic}$, $\Delta\lambda_{\rm observed}$ and $\Delta\lambda_{\rm narrow}$ are the intrinsic line width, observed line width and the width of the narrow emission from the explosion site (H$\rm \alpha$, [N {\sc II}], etc). Here, the width of the narrow emission reflects the instrumental broadening. According to the definition of $\Delta\lambda_{\rm blue}$, the emission within this range takes 34\% of the total flux, which is the same as 1$\sigma$ if the line is Gaussian. Therefore the narrow H$\alpha$ is fitted by a Gaussian function, and the derived variance $\sigma$ is set to be $\Delta\lambda_{\rm narrow}$. The narrow lines from the explosion site are absent for some objects in the sample. For these objects, the instrumental resolution is derived from the source paper, which is usually measured from the FWHM of the sky line, and transformed to the Gaussian $\sigma$ as
\begin{equation}
    \sigma = \frac{\rm FWHM}{2\sqrt{2{\rm ln2}}}.
\end{equation}
The average $\Delta\lambda_{\rm narrow}$ is 4.02 $\rm \AA$, and the variation is {\rm 1.87} $\rm \AA$.

The uncertainties of the measurements are estimated using a Monte Carlo method. A nebular spectrum is slightly smoothed at the first step by convolving with a boxcar filter. The smoothed version of the spectrum is then subtracted from the original one. The standard deviation at the range of 6000...7800 $\rm \AA$ of the residual flux is employed as the noise level of the spectrum. 10$^4$ simulated spectra are generated by adding noise on the smoothed spectrum. We further change the endpoints of the (continuum) background by -25...25 $\rm {\AA}$ range, which is assumed to be distributed uniformly with $\Delta\lambda$ = 1 $\rm \AA$ increments. The symmetry center of the H$\alpha$-like structure, initialized as 6563 $\rm {\AA}$, is also allowed to be shifted by -45...45 $\rm {\AA}$ following the uniform distribution. The above measurements of the observables are then performed on the simulated spectra. Finally the measured line width is corrected for the effect of instrumental broadening. The 84 and 16 percentages of the results of the 10$^4$ measurements are taken as the upper and lower limits of the observables respectively.

In Figure \ref{fig:compare_width}, the measured line widths are compared with those of the previous works with overlap objects. For the comparison work, we take the result of the one-component fit of \cite{tauben09}, and the full-width-half-max (FWHM) from full spectral modeling of \citet{maurer10}. The measurement of the line width in this work agrees well with \citet{tauben09}, while it is systematically smaller than that in \citet{maurer10}; however, a correlation can still be discerned, and the systematic offset may simply be due to different definitions of the line velocity/width. The line width measurement in this work does not assume the geometry or the detailed physical conditions of the [O {\sc i}] emitting region, and thus allows more general discussion on the velocity scale and structure of the ejecta than the previous works.

\subsection{Fitting the {\rm [O {\sc i}]} $\lambda\lambda$ 6300, 6364} 
After the background and the H$\alpha$-like structure are subtracted, the [O {\sc i}] is fitted with multiple Gaussian profiles using the method described in \citet{tauben09}. We define a single `doublet' component as two Gaussian functions with same standard deviation, central wavelengths separated by 3000 km s$^{-1}$, and intensity ratio of 3:1 which is expected if the ejecta are optical thin. A single component has 3 parameters; the center wavelength $\lambda_{\rm peak}$, the width $\sigma$ and the (scaled) intensity. The fitting procedure involves up to two components, and then we have 5 free parameters in total (note that one parameter is reduced since only the relative intensity matters). 

The fitting is started from one component. If the residual exceeds the noise level, an additional component is introduced as follows. We first set four types of initial guess: (1) two components red- and blue-shifted by 2000 km s$^{-1}$, with $\sigma$=1000 km s$^{-1}$ and the same intensity; (2) A broad component centered at $v=0$ km s$^{-1}$ and $\sigma$=2500 km s$^{-1}$, with a narrow component ($\sigma$=500 km s$^{-1}$) centered at $v=0$ km s$^{-1}$. The intensity of the narrow component is initialized to be 30$\%$ of the broad base; (3) Same as above, but the narrow component centered at $v=2000$ km s$^{-1}$; (4) Same as above, but the narrow component centered at $v=-2000$ km s$^{-1}$. We then start the fitting with these initials guesses. For case (1), the two components are forced to blue- and red-shifted by larger than $v=1000$ km s$^{-1}$ (resolution $R\sim$300), and the relative contribution of each component to the flux is forced to be $\gtrapprox$ 0.3, otherwise considered unacceptable. For cases (2) (3) and (4), the center of the broad base is allowed to vary within -1600...600 km s$^{-1}$. Here the broad base is allowed to suffer from bulk blueshift up to 1000 km s$^{-1}$ to account for the effect of residual opacity in the core of the ejecta (see Figure 3 of \citealt{tauben09}). The additional $\pm 600$ km s$^{-1}$ corresponds to the spectroscopic resolution of $\sim$500. The result with the smallest residual is taken to be the final result.

According to the results, the line profiles are classified into four classes; Gaussian, narrow-core, double-peaked and asymmetry (hereafter GS, NC DP and AS, respectively). In the following, the definitions and the physical implications of the line profiles are briefly summarized. The readers may refer to \citet{tauben09} for more details. In \S 5.1, we will further discuss the expected profiles from a specific bipolar-type explosion, given as an example in the list below. Some examples of the line profiles are shown in Figure \ref{fig:morpho_example}.

\begin{itemize}
    \item Gaussian~(GS)~~~~~~The line can be well fitted by one component. The emitter is expected to originate from the Gaussian distribution in the radial direction of a spherically symmetric ejecta. While there is no need to introduce deviation from spherical symmetry to explain the GS profile, it does not reject a possible asphericity; for example, a bipolar-type explosion (with the torus-like distribution of oxygen) viewed from the intermediate angle also results in a similar profile.
    \item Narrow-core~(NC)~~~~~~The line can be fitted by two components: a broad base and a narrow additional one with very close center wavelengths (in this work, it is defined to be offset $\textless$ 1000 km s$^{-1}$)\footnote {In \citet{tauben09}, the narrow-core is defined to have the narrow component with offset $\textless$ 22$\rm \AA$ ($\sim$1000 km s$^{-1}$) with respect to the \textit{rest~wavelength}. However, such offset can also be the result of residual opacity, which will affect both broad and narrow components, rather than pure geometrical effect. We therefore employ offset relative to the center of the \textit{broad~base} as the criterion for the narrow-core.} A straightforward interpretation is the emission from spherically symmetric ejecta with an enhanced core density. The axisymemtric configuration as described above but viewed from the polar direction (perpendicular to the O-rich torus) can also produce a similar profile. Indeed, the profile simply requires that there is a massive O-rich component with a negligible velocity along the line of sight, and thus even a single massive blob moving perpendicular to the line of sight is not rejected. 
    \item Double-peaked~(DP)~~~~~~The line can be well fitted by two components with similar intensities, one blue shifted and the other red shifted by similar amount (case (1) in the above text). If interpreted simply as a geometrical effect, this profile is not reproduced under the assumption of spherical symmetry, and requires two components having the symmetry in the  line-of-sight velocity distribution. A simple configuration leading to this profile is the axisymmetric explosion mentioned above but viewed from the edge of the torus.
    \item Asymmetry~(AS)~~~~~~The line can be fitted by a broad component accompanied by an additional component with arbitrary width and shift of the center wavelength. This again requires a deviation from a pure spherically symmetric ejecta, pointing to the existence of a single dominating blob corresponding to the narrow component, in addition to the bulk distribution representing the broad component. It should be noted that the only difference between NC and AS is the relative shift of the narrow component. Whether NC/AS are distinct populations is not clear. See the statistic results in \S 3.1 and \S 6.3.
\end{itemize}

Most of the objects in the sample can be well fitted by the method applied in this work (see Figure \ref{fig:IIb_fit} to \ref{fig:BL_fit} in the appendix), although some objects, e.g., SN 2006ld and SN 2008aq, possibly require more complicated ejecta geometry.

\begin{figure}
\epsscale{1.2}
\plotone{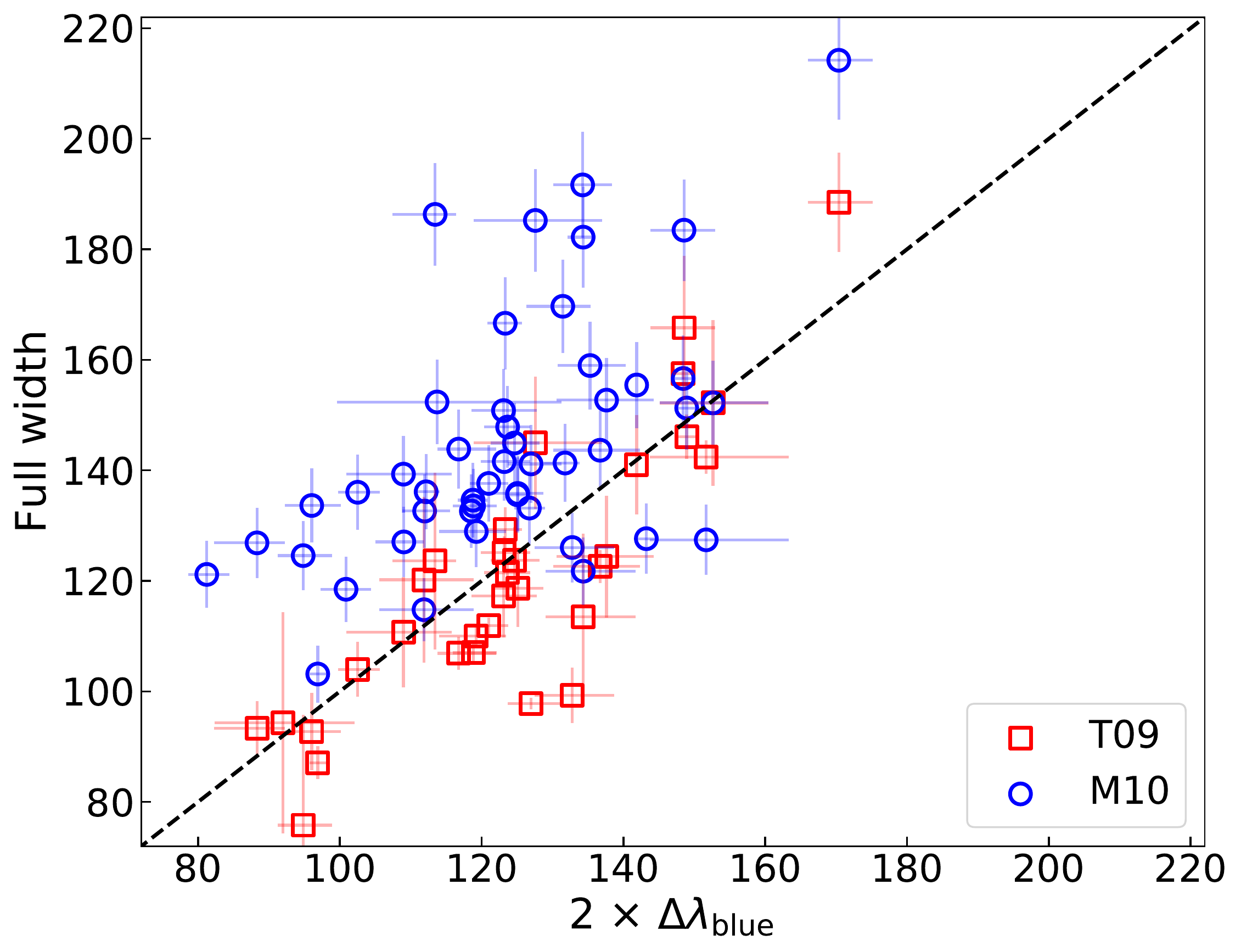}
\centering
\caption{A comparison between the line width measure in this work and previous works. The red squares are for the FWHM from the one-component fit of \citealt{tauben09} (T09). The blue circles represent the line widths transformed from the $v_{\rm 50}$ in \citealt{maurer10} (M10). The uncertainty is set to be 10\%. The black dashed line is for one-to-one correspondence.}
\label{fig:compare_width}
\vspace{4mm}
\end{figure}

\begin{figure}
\epsscale{1.2}
\plotone{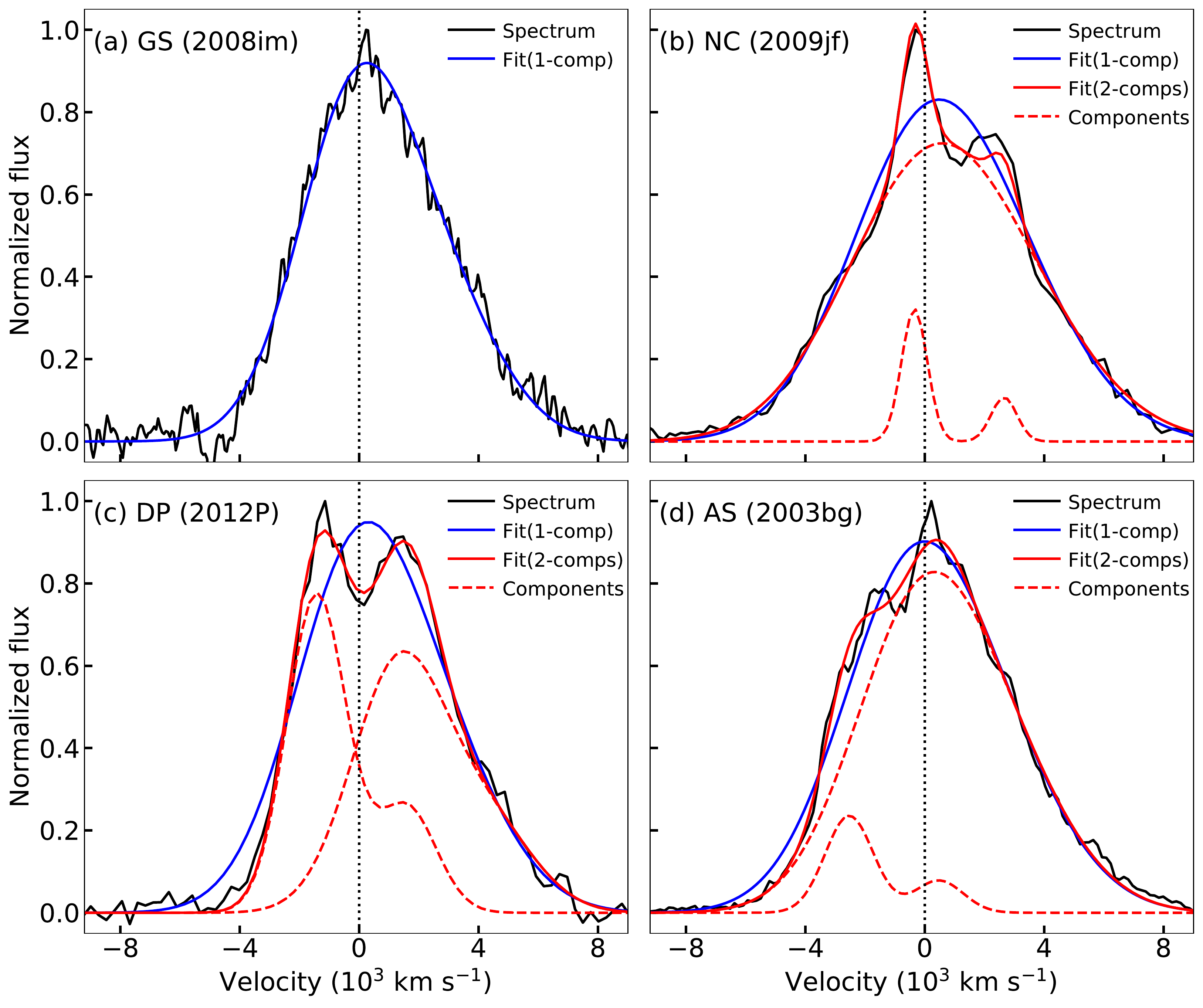}
\centering
\caption{Examples of the four line profiles. Before the fitting procedure, the spectra are subtracted by the background and the symmetric H$\alpha$-like profile as described in the text. The red solid line is the result of the two components fit, and the red dashed lines are the corresponding components. The blue solid line represents the one component fit, and is also plotted in panel (b), (c) and (d) for comparison. In this work, we use the same classification scheme as \citet{tauben09}.}
\label{fig:morpho_example}
\vspace{4mm}
\end{figure}

\section{Statistics of [O {\sc i}] profile}
The profile of the emission line is a useful tracer of the geometry of the ejecta (e.g. \citealt{tauben09}). Although it is not possible to recover the full 3D distribution of the emitter, the measurement in this work can still provide some information on any possible deviation from spherical symmetry. The classifications of [O {\sc i}] line profiles are listed in Table \ref{tab:listIIb} to Table \ref{tab:listIbc}. 

\subsection{Quantitative classification}
\begin{figure}[!t]
\epsscale{1.1}
\plotone{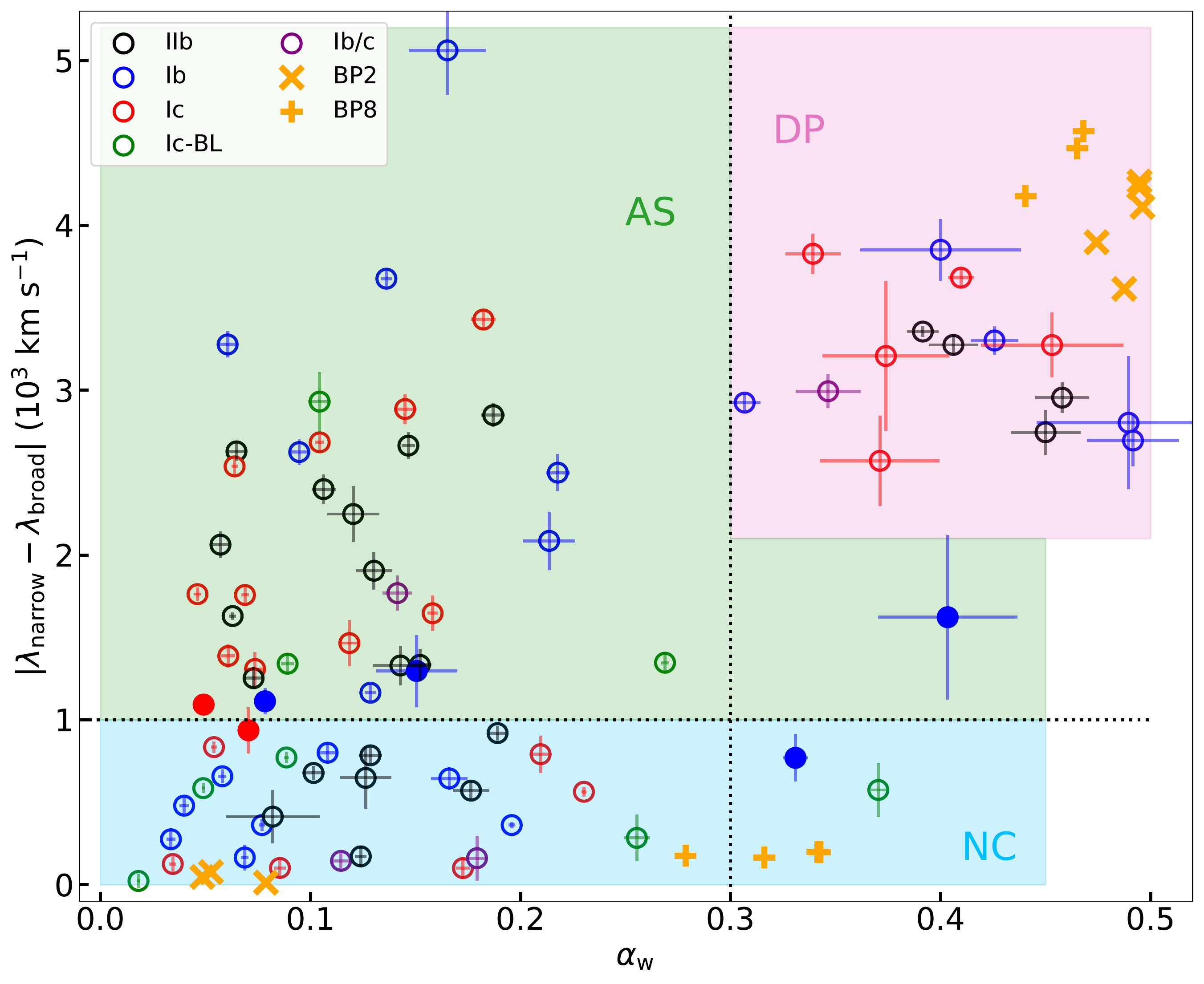}
\centering
\caption{The fractional flux of the secondary component is plotted against the central wavelengths separation. Different [O {\sc i}] profile classes are well separated by $\alpha_{\rm w}$ = 0.3 and $|\lambda_1~-\lambda_2|$ = 1000 km s$^{-1}$ (the dotted lines). Objects of different profile classes are labeled by different colors and markers. BP2 and BP8 are the bipolar explosion models in \citealt{maeda08} with different degree of axisymmetry. We apply the same fitting procedure as described in \S 2.4 to the theoretical spectra. See \S 5.1 for detailed descriptions on the BP models.}
\label{fig:profile_class_quan}
\vspace{4mm}
\end{figure}

To quantify the difference between the classifications, for objects fitted by two components ($N$ = 82), in Figure \ref{fig:profile_class_quan}, the fractional flux of the secondary component $\alpha_{\rm w}$, which is defined to be the component with the smaller flux, is plotted against the absolute central wavelengths offset between the two components. Similarly to Figure 6 of \citet{tauben09}, objects of different line profile classes, by definition, occupy different regions in the plot and are well separated. NC objects are characterized by a narrow strip located at the lower-left region ($\alpha_{\rm w}\lessapprox$ 0.4, $|\lambda_1~-\lambda_2|\textless$ 1000 km s$^{-1}$). The DP and AS objects have wider central wavelength separation, and are separated at $\alpha_{\rm w}$ $\sim$ 0.3. It should be emphasized that the boundary between NC and AS is changeable. In this work, we choose the same criterion as \citet{tauben09}, i.e., offset = 1000 km s$^{-1}$ ($\sim$ 22 $\rm \AA$). Moreover, the uncertainty of the fitting allows some objects, especially those near the boundary, to be re-classified to the other category. Objects with non-negligible probability ($\textgreater$ 0.05) of shifting to the other category are labeled by filled markers in Figure \ref{fig:profile_class_quan}.

According to the classification of \citet{tauben09}, objects with narrow component shift smaller (or larger) than 1000 km s$^{-1}$ are classified as NC (or AS). However, as shown in Figure \ref{fig:profile_class_quan}, the narrow component shift has a continuous distribution, and it is questionable whether NC and AS are two distinct populations. A more detailed discussion on the classification of NC/AS is left to \S 6.3.

\subsection{Statistical evaluation}
The fractions of the line profiles are shown in Table~\ref{tab:Profile}. In the sample of this work, the fractions of GS, NC, DP and AS objects are: 0.20 ($N$ = 21), 0.29 ($N$ = 30), 0.16 ($N$ = 16) and 0.35 ($N$ = 36), respectively. The large fraction of AS/DP objects suggests that the deviation from spherical symmetry is common for the ejecta of SESNe; these two categories require the deviation from spherical symmetry, and thus places an lower limit of $\sim 50\%$ for SESNe having non-spherical ejecta (note that the other two categories, GS/NC, can be explained by, but does not require, spherically symmetric ejecta; \S 2.4). This finding is consistent with previous studies (\citealt{maeda08, modjaz08, tauben09, milisa10}). The line profile fractions are generally in good agreement with the results of \citealt{tauben09}. Given that the fractions show no significant variation after the sample is enlarged by a factor of 2.5 (39 objects in \citealt{tauben09} and 103 objects in this work), we conclude that the distribution of [O {\sc i}] profiles, which is directly linked to the ejecta geometry, is already statistically well determined, and can be a potential constrain on the explosion mechanism.

\begin{deluxetable}{lcccccc}
\caption{Statistics of [O {\sc i}] profile}
\label{tab:Profile}
\tablehead{
\colhead{Types} & \colhead{Full} & \colhead{IIb} & \colhead{Ib} & \colhead{Ic}&\colhead{Ib/c}& \colhead{Ic-BL}\\
\multicolumn{6}{c}{$N$ (fraction)}
}
\startdata
GS&21(0.20)&3(0.12)&7(0.23)&9(0.28)&1(0.20)&1(0.11)\\
NC&30(0.29)&7(0.27)&9(0.29)&7(0.22)&2(0.40)&5(0.56)\\
DP&16(0.16)&4(0.15)&5(0.16)&5(0.16)&1(0.20)&0(-)\\
AS&36(0.35)&12(0.46)&10(0.32)&11(0.34)&1(0.20)&3(0.33)\\
\enddata
\end{deluxetable}

\begin{figure}[!t]
\epsscale{1.2}
\plotone{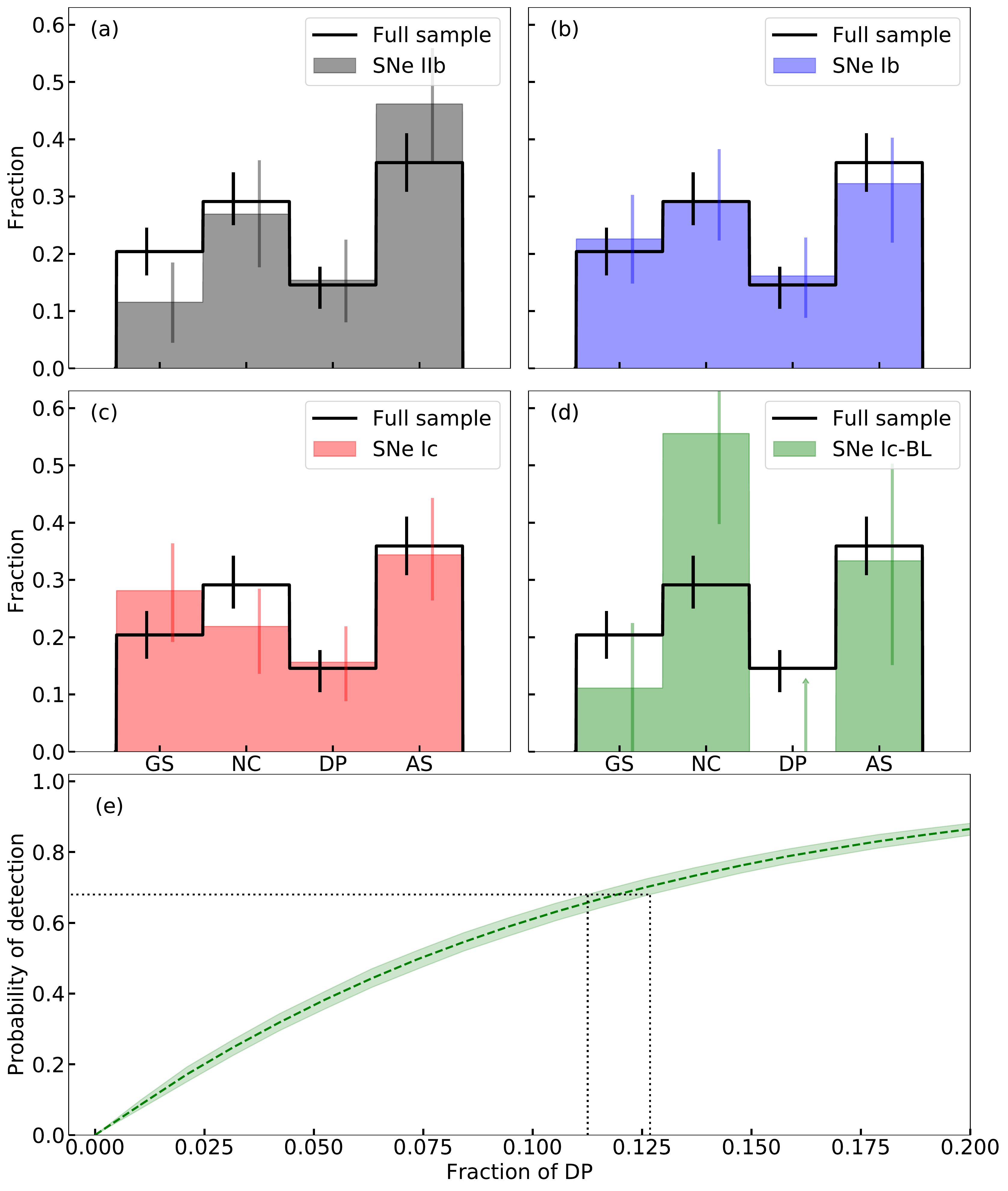}
\centering
\caption{$Panel~(a)~to~(d)$: The distributions of the [O {\sc i}] profile of different SN sub types, which are shown in different colors and panels. The histogram plotted by black-solid line is the distribution of the full sample. The error bars are estimated by the bootstrap Monte Carlo method described in the main text. The $1 \sigma$ upper limit of the DP fraction of SNe Ic-BL is marked by the arrow in panel (d). In $Panel~(e)$, the probability for DP detection as a function of intrinsic DP fraction for SNe Ic-BL is shown. The shaded region represents the 95\% CI. The vertical dotted lines mark the range of $f_{\rm DP}$ such that the probability of detection is equal to 0.68.}
\label{fig:profile_class_distribute}
\vspace{4mm}
\end{figure}

The distributions of the line profiles of different SN sub types, along with the full sample for comparison, are shown in  Figure~\ref{fig:profile_class_distribute}. 
In general, the line profiles distributions of the canonical SESNe (IIb/Ib/Ic) are quite similar with each other.
The uncertainties of the fractions of different line profiles in different SN sub types are estimated by a bootstrap-based Monte Carlo method. We run 10$^4$ simulations. In each trial, the SNe sample is re-sampled with replacement. For NC and AS objects, the probability of re-classification into the other category is also included (see \S 3.1). The fractions of different line profiles of the new sample and the different SN sub types are then calculated. The 16\% and 84\% percentage of the 10$^4$ trials is employed as the lower and upper limits of the line profile fractions.

\citet{tauben09} suggested the objects with an extended envelope tend to be more aspherical, as the SNe Ib in their sample mainly belong to the AS category. The results in this work do not support their finding. Although the fraction of AS objects in the SNe IIb sample is slightly larger than the average, we find no significant difference in the line profile distributions among SNe IIb/Ib/Ic. The similarity likely indicates an limited effect of the presence of the helium layer or the residual hydrogen envelope on the ejecta geometry. For each sub type, at least 50\% (and likely more) objects cannot be interpreted by the spherical symmetric ejecta, and such deviation is commonly seen for all types of canonical SESNe.

Some differences of SNe Ic-BL when compared with the average behavior can be discerned: (1) large fraction of NC objects and (2) lack of DP objects. However, in this work, the number of SNe Ic-BL is small ($N$=9). The lack of DP objects can be the result of small-sample statistics. We therefore need to estimate the upper limit of the intrinsic DP fraction above which the non-detection is statistically significant. For this purpose, we run 10$^4$ simulations. In each trial, the GS, NC and AS fractions ($f_{\rm GS}$,$f_{\rm NC}$,$f_{\rm AS}$) are randomly drawn from the full sample with the bootstrap-based Monte Carlo method introduced above. The intrinsic DP fraction $f_{\rm DP}$ is varied from 0 to 0.2, with the ratio of $f_{\rm GS}$,$f_{\rm NC}$ and $f_{\rm AS}$ kept fixed. For the fixed $f_{\rm DP}$, 10$^3$ samples (size $N$=9), are generated according to the current line profile distribution. The rate of the samples with DP detected is then calculated. The relation between the DP fraction and the detection probability are shown by the green dashed line in the panel (e) of Figure \ref{fig:profile_class_distribute}. The shaded region is the 95\% confidence interval (CI) of the 10$^4$ simulations. When $f_{\rm DP}$=0, no DP object can be detected in all trials by definition. As $f_{\rm DP}$ increases, the probability of detection increases as expected. The upper limit of $f_{\rm DP}$ is defined to be the value such that detection probability $p_{\rm detect}$=0.68 (or non-detection probability = 0.32). This $1 \sigma$ upper limit ranges from 0.112 to 0.126 (mean value = 0.119), as indicated by the vertical dotted lines in the panel (e) of Figure \ref{fig:profile_class_distribute}. The conservative value 0.126 is employed as the upper limit of the DP fraction of SNe Ic-BL, which is still smaller then the DP fraction (0.155) of the full sample, but slightly larger than its lower limit (0.120). Therefore, there is an indication, at a confidence level of about $1\sigma$, that the lack of double-peaked SNe Ic-BL is an intrinsic feature rather than statistics effect.

A hint that the distribution of the line profiles of SNe Ic-BL is different from those of the canonical SESNe can thus be discerned, which suggests difference in ejecta geometry. From early-phase observation, SNe Ic-BL are already found to be distinct from other SESNe with their extreme nature. The finding in this work further extends such distinction in the nebular phase.

The full sample is large enough for statistical evaluation. However, the size of each SNe sub type is still limited, especially lacking SNe Ic-BL. Inferences made based on the fractions of small samples are uncertain (\citealt{park06}). To reliably investigate the dependence of the line profiles on SNe sub types, an even larger sample is required.

\section{[O {\sc i}]/[Ca {\sc i}{\sc i}] and [O {\sc i}] width}
\begin{figure*}[!t]

\plotone{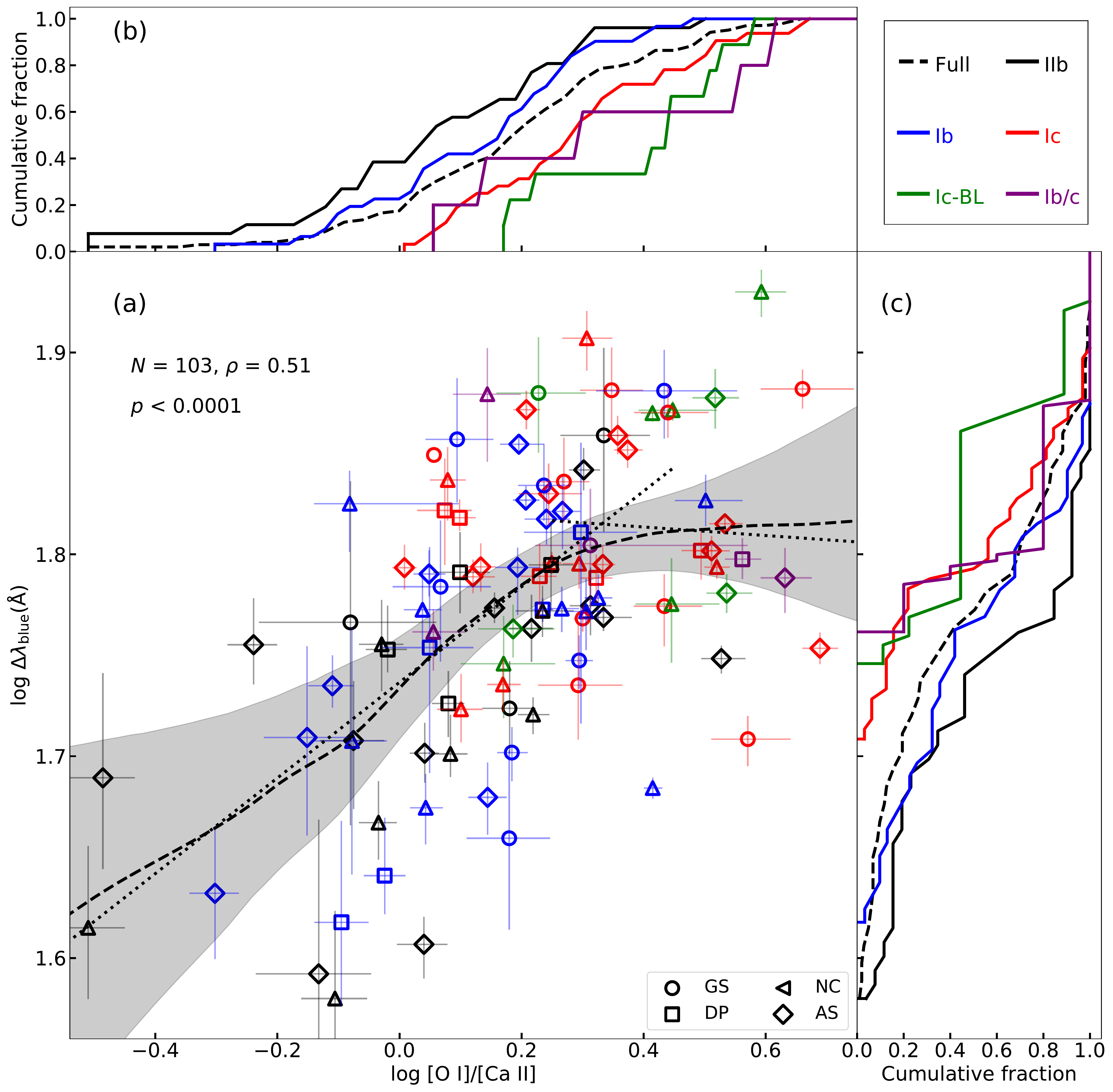}
\centering
\caption{$Panel (a)$ The relation between the [O {\sc i}]/[Ca {\sc i}{\sc i}] ratio and the [O {\sc i}] width $\Delta\lambda_{\rm blue}$. SESNe of different SN sub types and line profile classes are labeled by different colors and markers. The black dashed line is the result of the local non-parametric regression for the full sample, and the shaded region represents the 95\% CI. The black dotted lines are the results of linear regression performed for the objects with log [O {\sc i}]/[Ca {\sc i}{\sc i}] $\textless$ 0.4 and log [O {\sc i}]/[Ca {\sc i}{\sc i}] $\textgreater$ 0.4, respectively. (b) Cumulative fraction of the [O {\sc i}]/[Ca {\sc i}{\sc i}] ratio. (c) Cumulative fraction of the [O {\sc i}] width $\Delta\lambda_{\rm blue}$.}
\label{fig:main}
\end{figure*}

The individual measurements of the [O {\sc i}]/[Ca {\sc i}{\sc i}] ratio and the [O {\sc i}] width $\Delta\lambda_{\rm blue}$ for each object in the sample is plotted in the panel (a) of Figure \ref{fig:main}. The cumulative distributions of these two quantities are plotted in the panels (b) and (c), respectively, where the objects of different SN sub types are labeled by different colors and the cumulative fraction of the full sample is labeled by the black dashed line. Objects of different line profile classes (i.e., GS, NC, DP and AS. See the previous section for details) are distincted by different markers.

\subsection{Statistical evaluation}
Similarly to the result in \citet{fang19}, for [O {\sc i}]/[Ca {\sc i}{\sc i}], an increasing sequence is discerned; SNe IIb/Ib $\rightarrow$ SNe Ic $\rightarrow$ SNe Ic-BL. Although compared with the results in \citet{fang19}, SNe Ib seem to have slightly larger average [O {\sc i}]/[Ca {\sc i}{\sc i}] ratio than SNe IIb, still the hypotheses that the SNe IIb/Ib have the same [O {\sc i}]/[Ca {\sc i}{\sc i}] distribution cannot be rejected at the significance level $p~\textgreater$ 0.25, based on the two-sample Anderson-Darling~(AD) test. 
For SNe Ic, the difference is significant when compared with He-rich objects (SNe IIb + Ib), with $p~\textless~0.001$. Similarly, the [O {\sc i}]/[Ca {\sc i}{\sc i}] of SNe Ic-BL is significantly larger than SNe IIb/Ib ($p~\textless$ 0.001 when compared with both IIb and Ib), but the distribution is indistinguishable from SNe Ic ($p~\approx$ 0.23). These findings are consistent with \citet{fang19}.

From the panel(c) of Figure~\ref{fig:main}, a possible [O {\sc i}] width sequence is also discerned; SNe IIb $\rightarrow$ SNe Ib $\rightarrow$ SNe Ic $\rightarrow$ SNe Ic-BL. Unlike the case of [O {\sc i}]/[Ca {\sc i}{\sc i}], the differences between SNe IIb/Ib/Ic are significant, showing an increasing trend ($p~\approx ~0.09$ for SNe IIb versus SNe Ib and $p~\approx ~0.04$ for SNe Ib versus SNe Ic). While SNe IIb and SNe Ic are limited to a narrow range, occupying the low- and high-ends of $\Delta\lambda_{\rm blue}$ respectively, the range of the [O I] width of SNe Ib is rather large.

In the early-phase spectra, the SNe Ic-BL show evidence of fast-expanding ejecta. The average photospheric velocity of SNe Ic-BL, measured near light curve peak, is about 20000 km s$^{-1}$, much larger than that of the canonical SNe ($\textless$ 10000 km s$^{-1}$, see \citealt{lyman16}). Surprisingly, the [O {\sc i}] width distribution of SNe Ic-BL is not statistically different from normal SNe Ic. The null hypothesis can be rejected only at the significance level $p~\approx$ 0.21 from AD test when compared with SNe Ic. The AD significance level $p$ reduces to 0.012 when the [O I] width distribution of SNe Ic-BL is compared with the canonical SNe (IIb + Ib + Ic). If the [O I] width is transformed to velocity as
\begin{equation}
   \frac{v}{c} ~ = ~ \frac{\Delta\lambda}{6300 ~\rm{\AA}},
\end{equation}
the average velocity of SNe Ic-BL is about 3300 km s$^{-1}$, slightly larger than that of the canonical SNe (about 2900 km s$^{-1}$) and SNe Ic (about 3100 km s$^{-1}$). The difference of the velocity scales of the innermost ejecta between SNe Ic-BL and the canonical objects is not as striking as the photospheric velocities around the light curve peak, which measure the expansion velocities of the outermost ejecta. 

\begin{figure}[!t]
\epsscale{1.2}
\plotone{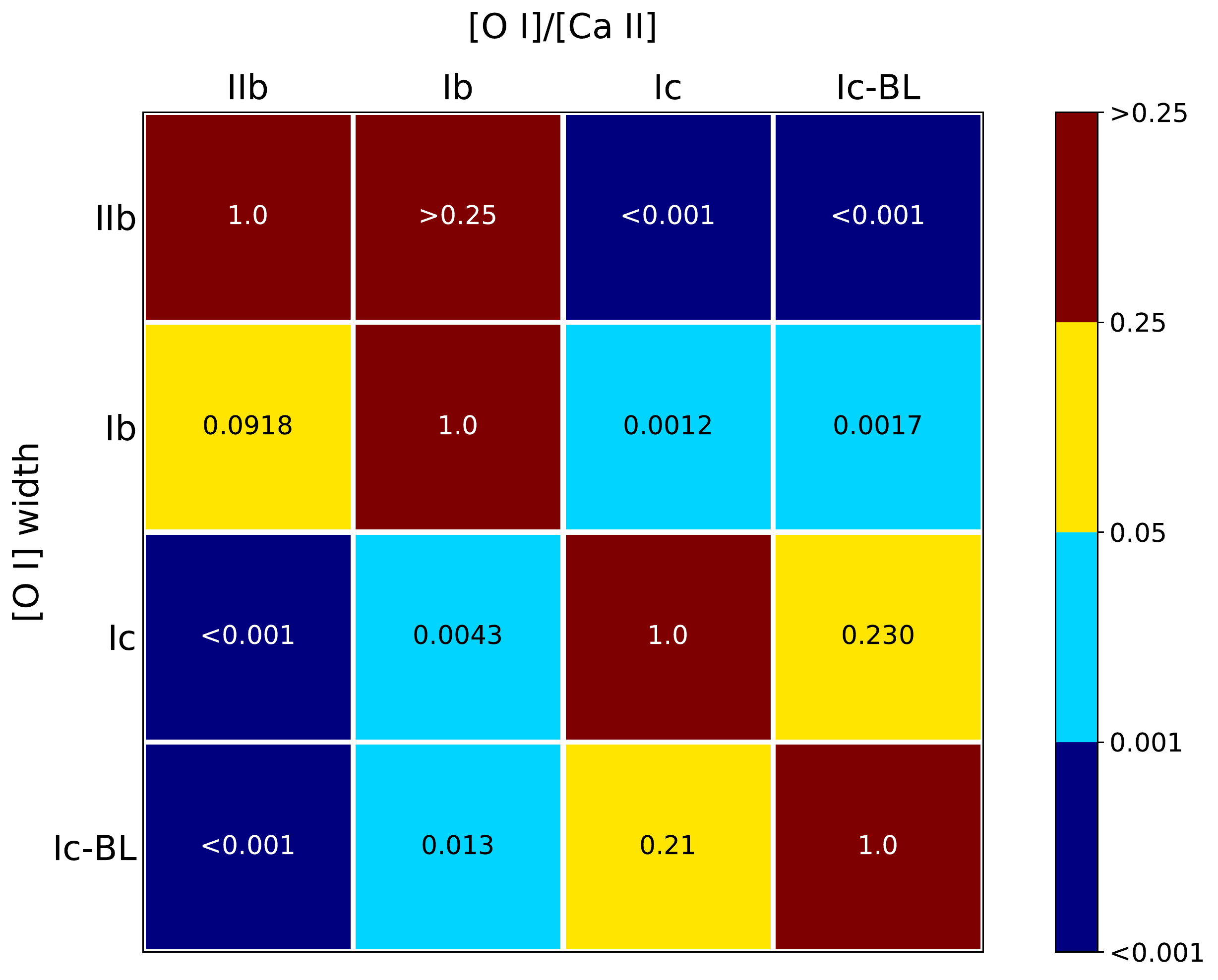}
\centering
\caption{Matrix of AD test significance level when the [O {\sc i}]/[Ca {\sc i}{\sc i}] and [O {\sc i}] width distributions of different SN sub types are compared. The upper right region is for [O {\sc i}]/[Ca {\sc i}{\sc i}] ratio and the lower left region is for [O {\sc i}] width $\Delta\lambda_{\rm blue}$. The colour bar indicates the probability that the samples are drawn from the same distribution, and the blue end indicates significant differences.
\label{fig:KS_sub type}}
\vspace{4mm}
\end{figure}

For both [O {\sc i}]/[Ca {\sc i}{\sc i}] and [O {\sc i}] width, it is clear from Figure \ref{fig:main} (b) (c) that SNe IIb/Ib are lower than the average (black dashed line), while SNe Ic and Ic-BL are higher. The above discussions are summarized in Figure~\ref{fig:KS_sub type}.

\subsection{{\rm [O {\sc i}]/[Ca {\sc i}{\sc i}]-[O {\sc i}]} width correlation}
In Figure~\ref{fig:main} (a), the [O {\sc i}]/[Ca {\sc i}{\sc i}] ratio is plotted against the [O {\sc i}] width for comparison. The objects with small [O {\sc i}]/[Ca {\sc i}{\sc i}] tend to have narrow [O {\sc i}]. The two quantities are moderately correlated (Spearman correlation coefficient $\rho$ = 0.51), and the correlation is significant, with $p\textless~$0.0001 for the sample size of 103 objects. 

To further investigate the dependence of the [O {\sc i}] width on the [O {\sc i}]/[Ca {\sc i}{\sc i}] ratio, in Figure~\ref{fig:main}, the local non-parametric regression is performed to the full sample (black dashed line). To estimate the uncertainties, we run 10$^{4}$ simulations. In each trial, the sample is re-sampled with replacement, and for each object in the new sample, its [O {\sc i}]/[Ca {\sc i}{\sc i}] ratio and the [O {\sc i}] width are added by the errors, which are assumed to follow Gaussian distribution. Then local non-parametric regression is applied to the new sample. The 97.5\% and 2.5\% percentages of the results from 10$^4$ simulations are defined to be the boundaries of the 95\% confidence interval (CI) of the regression, as labeled by the grey shaded region in Figure \ref{fig:main}.

The linear regression is performed to the full sample, because analytical form could be useful for further study. The best-fit result gives
\begin{equation}
\begin{split}
  {\rm log}\frac{\Delta\lambda_{\rm blue}}{\rm \AA}~=~(0.16\pm0.03)\times {\rm log}[{\rm O~I}]/[{\rm Ca~II}] \\
  +~(1.74\pm0.01). 
\end{split}
\end{equation}
From the result of local non-parametric regression, the increasing tendency stops at roughly log[O {\sc i}]/[Ca {\sc i}{\sc i}] = 0.4 (or [O {\sc i}]/[Ca {\sc i}{\sc i}] = 2.5). If the line regression analysis is restricted to the objects with log[O {\sc i}]/[Ca {\sc i}{\sc i}] $\textless$ 0.4 ($N$ = 82), the correlation becomes significant with $\rho$ = 0.56 and $p~\textless$ 0.0001. For objects with log[O {\sc i}]/[Ca {\sc i}{\sc i}] $\textless$ 0.4, the best linear regression gives
\begin{equation}
\begin{split}
  {\rm log}\frac{\Delta\lambda_{\rm blue}}{\rm \AA}~=~(0.22\pm0.04)\times {\rm log}[{\rm O~I}]/[{\rm Ca~II}] \\
  +~(1.74\pm0.01), 
\end{split}
\end{equation}
while for the rest (log[O {\sc i}]/[Ca {\sc i}{\sc i}] $\textgreater$ 0.4, $N$ = 21), $\rho$ reduces to -0.07 and $p~\textless$ 0.77, indicating no correlation exists. For this range, 
\begin{equation}
\begin{split}
  {\rm log}\frac{\Delta\lambda_{\rm blue}}{\rm \AA}~=~(-0.02\pm0.18)\times {\rm log}[{\rm O~I}]/[{\rm Ca~II}] \\
  +~(1.82\pm0.09).
\end{split}
\end{equation}

\begin{figure}[!t]
\epsscale{1.2}
\plotone{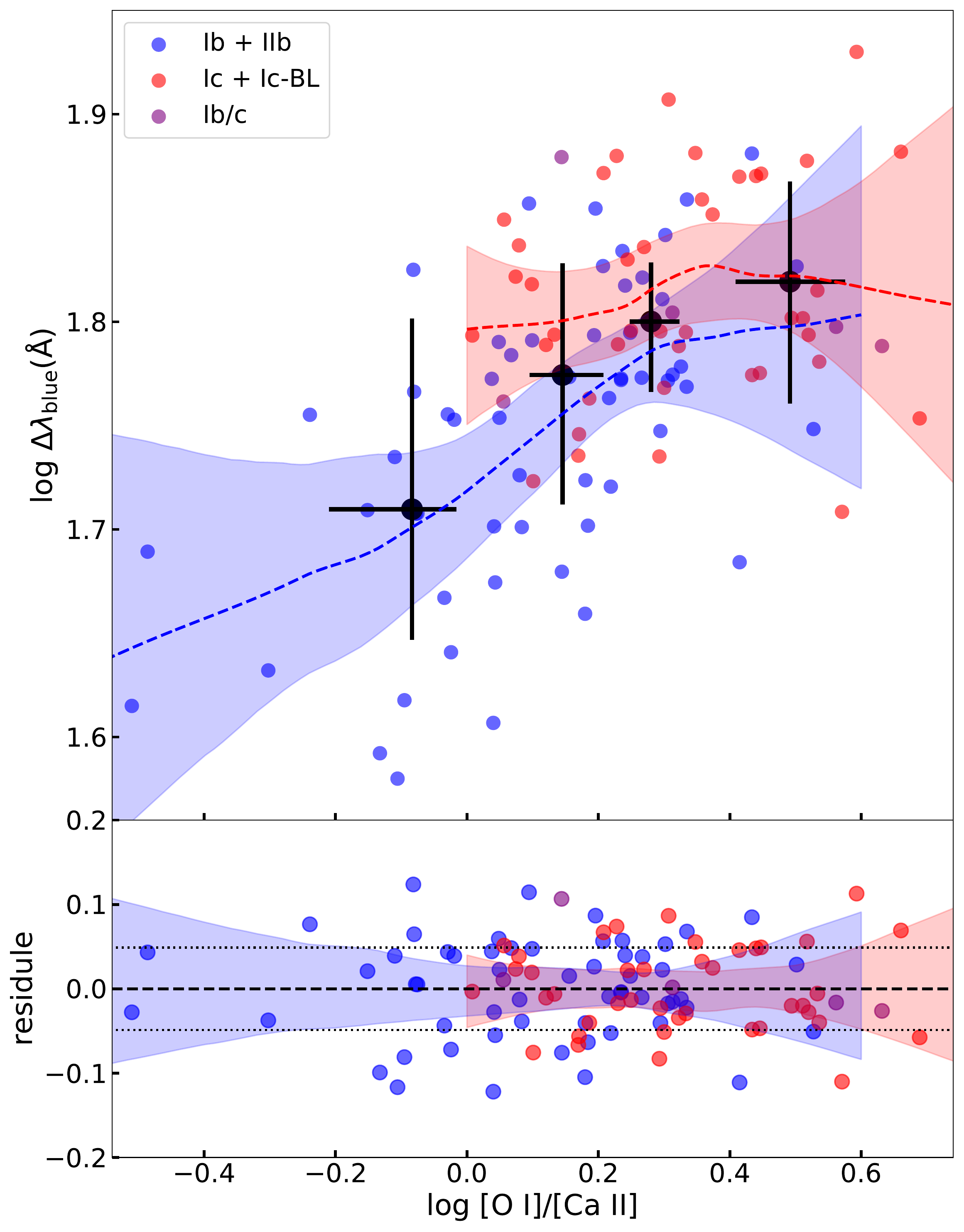}
\centering
\caption{$Upper ~panel$: The relation between the [O {\sc i}]/[Ca {\sc i}{\sc i}] ratio and the [O {\sc i}] width $\Delta\lambda_{\rm blue}$, with SNe of different origins (He star and CO core) labeled by different colors. The 103 objects are divided into 5 groups ($\sim$ 20 objects in each group) according to the range of the [O {\sc i}]/[Ca {\sc i}{\sc i}] ratio, and the black dots represent the mean value of each group. The blue dashed line is the result of the local non-parametric regression to helium-rich SNe (IIb + Ib), and the red dashed line is for helium-deficient SNe (Ic + Ic-BL). The shaded regions are the 95\% CI (see main text for details). A clear increasing trend can be discerned; $Lower ~panel$: The residual of the fitting. The dotted lines represent the standard deviation of the residual, which is about 0.06 dex ($\sim$ 15$\%$ in linear scale).}
\label{fig:main_fitting}
\end{figure}

The significance of the correlation between [O {\sc i}]/[Ca {\sc i}{\sc i}] and [O {\sc i}] width may be affected by SN sub types and line profile classes as follows;

\begin{figure*}[!t]

\plotone{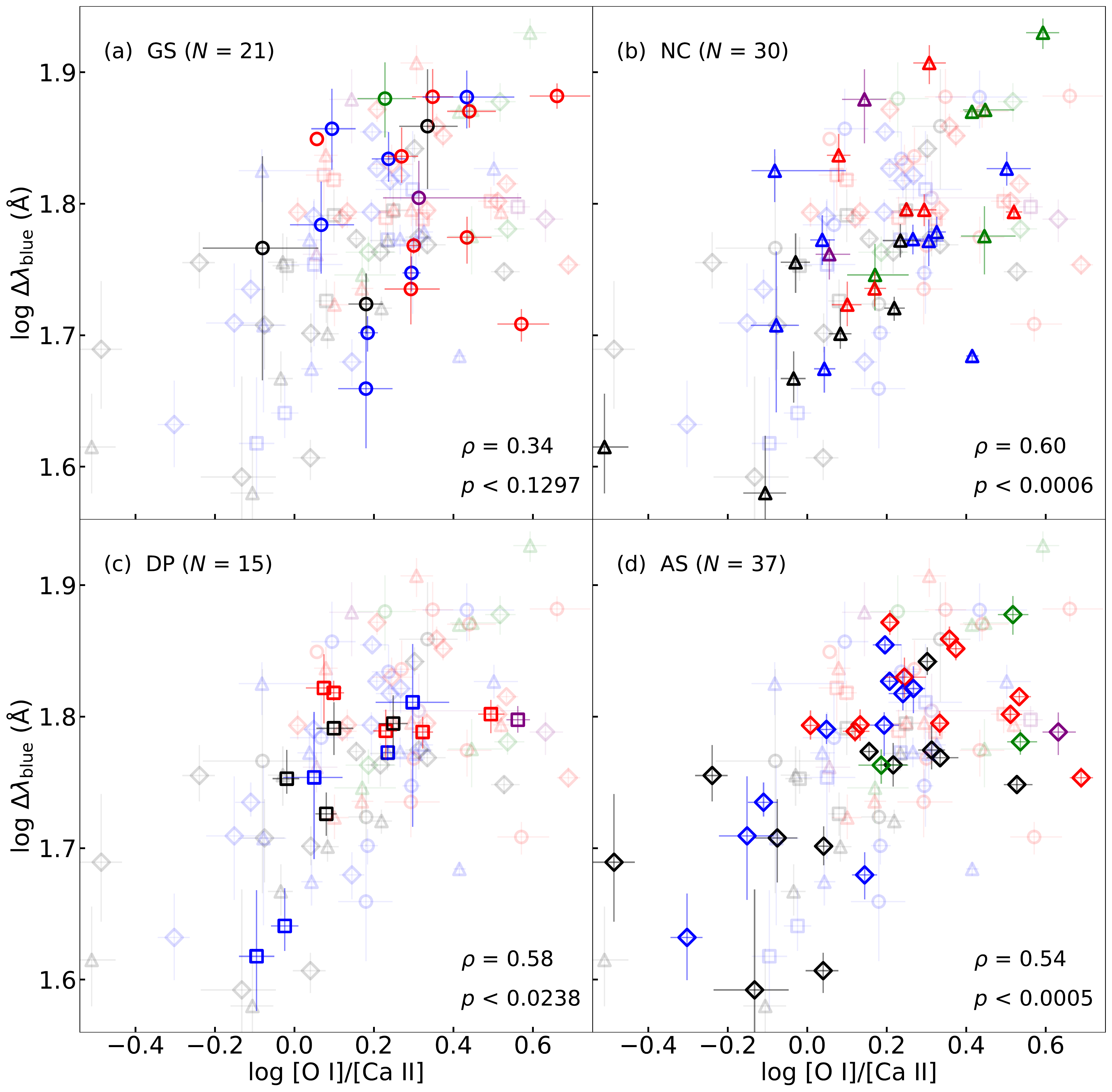}
\centering
\caption{The correlations between the [O {\sc i}]/[Ca {\sc i}{\sc i}] ratio and [O {\sc i}] width of different line-profile classes, and the comparison with other line profiles, are plotted in different panels. The meanings of the different colors and markers are same as Figure \ref{fig:main}. It is clear that NC objects show the tightest correlation, followed by DP and AS. The correlation of GS objects is weak.}
\label{fig:profile_correlation}

\end{figure*}

\begin{itemize}

    \item SN~sub~type.~~~ Objects of different SN sub types (labeled by different colors) behave differently in Figure \ref{fig:main_fitting}. It is clear that the helium-rich objects (SNe IIb + Ib) show increasing tendency ($\rho$ = 0.53, $p~\textless$~0.0001). The local non-parametric regression technique is applied to the helium-rich SNe, with the same bootstrap-based uncertainties introduced above. The result and the 95\% CI are shown by the blue dashed line and the blue shaded region in Figure~\ref{fig:main_fitting}.
    
    However, the [O {\sc i}] width of the helium-deficient SNe (SNe Ic + Ic-BL) remains (almost) constant as [O {\sc i}]/[Ca {\sc i}{\sc i}] increase, showing large scatter and no correlation can be discerned ($\rho$ = 0.10, $p~\textless~$0.54). This is consistent with the result of the local non-parametric regression, as shown by the red dashed line and the red shaded region (95\% CI) in Figure~\ref{fig:main_fitting}.

    \item Line~profile.~~~ 
    The [O {\sc i}]/[Ca {\sc i}{\sc i}]-[O {\sc i}] width correlation separately shown for different line-profile classes are plotted in Figure \ref{fig:profile_correlation}. The NC objects have the tightest correlation ($\rho$ = 0.60 with $p~\textless$ 0.0006), followed by DP and AS ($\rho$ = 0.58 and 0.54, with $p~\textless$0.0238 and 0.0005, respectively). For GS objects, the correlation is weak and not significant ($\rho$ = 0.34 with $p~\textless$0.1297).

\end{itemize}
The above discussions are summarized in Table \ref{tab:affect_factors}.

\begin{deluxetable}{l|cc}
\caption{Factors affect the [O {\sc i}]/[Ca {\sc i}{\sc i}]-[O {\sc i}] width correlation}
\label{tab:affect_factors}
\tablehead{
\nocolhead{} & \colhead{$\rho$} & \colhead{$p$}
}
\startdata
\multicolumn{3}{c}{log[O {\sc i}]/[Ca {\sc i}{\sc i}]}\\
 \hline
$\textless$ 0.4&0.56&$\textless$0.0001\\
$\textgreater$ 0.4&-0.07&0.7706\\
 \hline
\multicolumn{3}{c}{Line profile}\\
\hline
GS &0.34&0.1297\\
NC &0.60&0.0006\\
DP &0.58&0.0238 \\
AS &0.54&0.0005\\
AS + DP&0.56&$\textless$0.0001\\
GS + NC&0.50&0.0003\\
\hline%
\multicolumn{3}{c}{SN sub types}\\
 \hline
IIb&0.67&$\textless$0.0002\\
Ib &0.48&0.0064\\
IIb + Ib &0.58&$\textless$0.0001\\
Ic &0.02&0.8948\\
Ic-BL&0.56&0.1108\\
Ic + Ic-BL &0.14&0.3862\\
\enddata
\end{deluxetable}

Spectral phase is also a possible factor that affects the correlation, as both [O {\sc i}]/[Ca {\sc i}{\sc i}] and $\Delta\lambda_{\rm blue}$ are time dependent, although not sensitive (\citealt{maurer10}, \citealt{fang19}). In \S 6.1, the effect of spectral evolution will be discussed.

\subsection{{\rm [O {\sc i}]/[Ca {\sc i}{\sc i}]} and line profile}

\begin{figure}[!t]
\epsscale{1.1}
\plotone{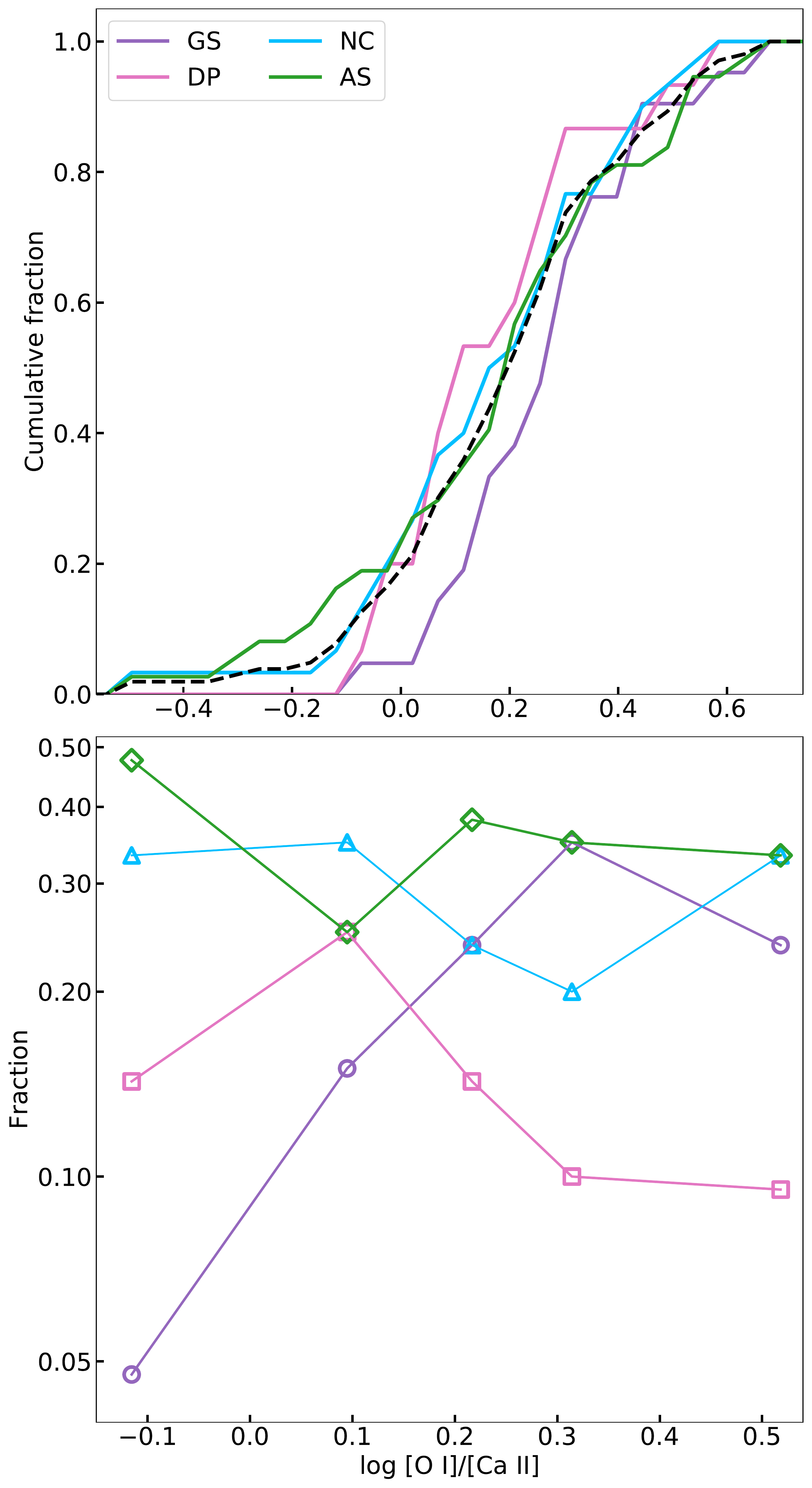}
\centering
\caption{$Upper~ panel$: The cumulative fractions of log [O {\sc i}]/[Ca {\sc i}{\sc i}] of the objects with different line profiles. The black dashed line is the distribution of the full sample. $Lower~ panel$: The fractions of line profile as functions of log [O {\sc i}]/[Ca {\sc i}{\sc i}] are plotted by different color solid lines and different markers.}
\label{fig:profile_oxygen_distribute}

\end{figure}

The cumulative fractions of [O {\sc i}]/[Ca {\sc i}{\sc i}] in terms of the line profiles are shown in the upper panel of Figure~\ref{fig:profile_oxygen_distribute}. The GS objects tend to have the largest [O {\sc i}]/[Ca {\sc i}{\sc i}] on average, followed by NC/AS, then DP. However, such difference is not significant, possibly except for the difference between DP and GS, where the null hypothesis can be rejected at the significant level $p~\approx$ 0.08 from AD test. The distributions of AS and NC objects are remarkably similar, and the [O {\sc i}]/[Ca {\sc i}{\sc i}] distributions of all line profiles are indistinguishable from the average ($p~\textgreater$ 0.25 from AD test).

To investigate how the distributions of the line profiles change as the [O {\sc i}]/[Ca {\sc i}{\sc i}] ratio increases, the full sample is binned into 5 groups with equal number of members ($N$=20 or 21) according to the [O {\sc i}]/[Ca {\sc i}{\sc i}]. In each group, the fractions of each line profile is calculated. The results are plotted by the color solid lines in the lower panel of Figure \ref{fig:profile_oxygen_distribute}.

It is clear that there is a systematic trend where the fraction of GS objects increases as the [O {\sc i}]/[Ca {\sc i}{\sc i}] ratio increases, and then becomes saturated at log [O {\sc i}]/[Ca {\sc i}{\sc i}] $\sim$ 0.3 ($\rho$ = 0.82, $p~\textless$ 0.09). For DP objects, the trend goes to the opposite direction ($\rho$ = -0.82, $p~\textless$ 0.06). Another interesting feature is the fractions of NC and AS objects are fluctuating around 0.3 and no significant dependence on [O {\sc i}]/[Ca {\sc i}{\sc i}] can be discerned ($\rho$ = -0.41, $p~\textless$ 0.49 for NC and $\rho$ = -0.40, $p~\textless$ 0.51 for AS).

\section{Physical implication}
In \S 3 and \S 4, the statistical properties of the [O {\sc i}] profile, the [O {\sc i}]/[Ca {\sc i}{\sc i}] ratio and the [O {\sc i}] width $\Delta\lambda_{\rm blue}$, along with their mutual relations, are investigated. In this section, the possible physical implications behind the statistics are discussed.
\subsection{Constrains on the ejecta geometry}

As introduced in \S 2.4, different ejecta geometry will lead to different line profile. To further constrain the configuration of the ejecta, it is useful to compare the observational data with some models. For this purpose, a specific bipolar explosion model(s) from \citet{maeda06} is employed, as this model has been frequently referred to in the previous works to study the ejecta kinematics through the [O {\sc i}] profile. Note that the model prediction should not be over-interpreted, given various assumptions under which the model is constructed. For example, the models are assumed to be perfectly axisymmetric and the two hemispheres are symmetric, which are probably too simplified. Indeed, not only the consistency but the inconsistency between the data and the model are important; the latter will be useful to clarify what are still missing in the model, by investigating what assumption is a potential cause of the inconsistency. 

To compare the observational data to the theoretical predictions, the multi-Gaussian fit procedure is applied to the synthetic spectra of the bipolar explosion models (\citealt{maeda06}) in the same way as applied to the observational data. In this model sequence, oxygen-rich materials are distributed in a torus-like structure surrounding the bipolar jets that convert the stellar material (e.g., oxygen) into the Fe-peak elements (\citealt{maeda02}, \citealt{maeda03}, \citealt{maeda06}, \citealt{maeda08}). The [O {\sc i}] profiles of the models depend on the degree of asphericity and the viewing angle. In this work, two representative models in \citet{maeda06}, the mildly aspherical model (BP2) and the extremely aspherical one (BP8), are employed.

A basic assumption of the SN ejecta kinematic is homologous expansion, i.e., $v(r,t)=r/t$, where $v(r,t)$ is the velocity of the point located at radial coordinate $r$ at time $t$. For a photon emitted from $\vec{r}$, the Doppler shift of its wavelength is $\Delta\lambda$=-$\lambda_0$($v_{\parallel}$/c), where $\lambda_0$ is the intrinsic wavelength and $v_{\parallel}$ is the line of sight velocity toward the observer. For the homologously expanding ejecta, $\Delta\lambda\propto d$, where $d$ is the projection of $\vec{r}$ onto the direction of line of sight. At late phases, the photons emitted from the same plane, which is perpendicular to the line of sight, have the same observed wavelength. The line profile therefore provides the `scan` of the integrated emissions on these planes. The readers may refer to \citet{maeda08} and \citet{jerk17} for more detailed discussions on the formation of nebular line profile.

For the BP models, the O-rich material is distributed in a torus. When the ejecta is viewed from the on-edge direction, the integrated emission on the scan plane increases as it moves from the outer edge toward the inner edge of the hole, then decreases as it further moves to the center, where the integrated emission reaches its minimum. The [O {\sc i}] is therefore expected to have a horn-like profile. If the ejecta is viewed from the on-axis direction, i.e., along the jet, the integrated emission monotonically increases as the scan plane moves toward the center. Most of the O-rich materials are distributed on the equatorial plane, therefore contribute to the flux at $v\approx$0 km s$^{-1}$, giving rise to the narrow-core [O {\sc i}] profile.

\begin{figure*}[!t]
\epsscale{1.2}
\plotone{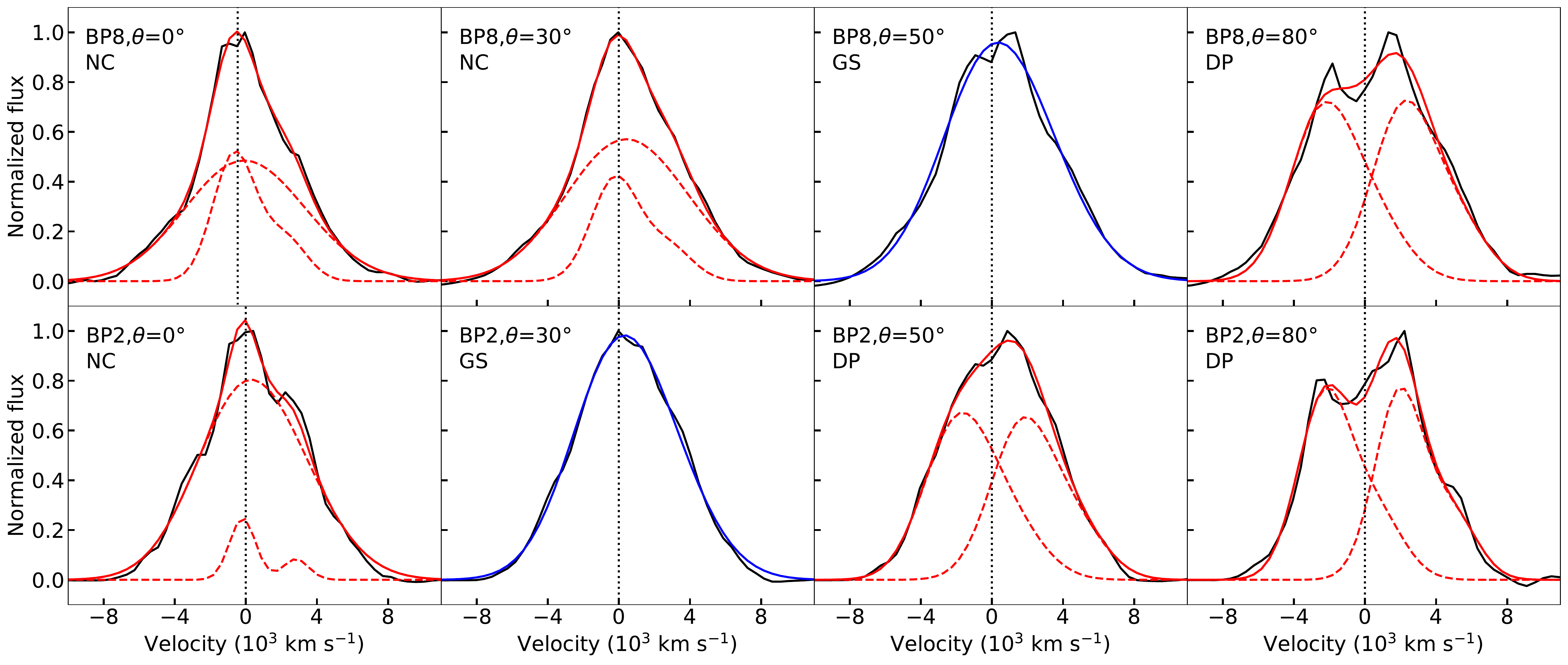}
\centering
\caption{The fitting results of the bipolar explosion models with different degrees of axisymmetry and different viewing angles from the direction of the poles. The black solid lines are the model spectra. The blue and red solid lines are the results of one-component and two-component fit. The red dashed lines are the corresponding components. The spectra are plotted in velocity space. The dotted vertical lines represents zero velocity (6300 $\rm \AA$).}
\label{fig:fit_BP}

\end{figure*}

Applying the same multi-Gaussian fit procedure, the [O {\sc i}] of the extremely aspherical model (BP8) is classified into the NC and DP profiles if the viewing angles from the jet axis are 0$\degree$...30$\degree$ and 70$\degree$...90$\degree$, respectively. For the mildly aspherical model (BP2), the corresponding viewing angles change to 0$\degree$...20$\degree$ and 50$\degree$...90$\degree$. Some examples of the fitting results are shown in Figure \ref{fig:fit_BP}. If the viewing angles are just randomly distributed without any preference, the fractions of different line profiles can be estimated by 
\begin{equation}
    f~=~\frac{1}{2\pi}\int_{\theta_0}^{\theta_1}2\pi{\rm sin}\theta {\rm d}\theta,\\
\end{equation}
where $\theta_0$ and $\theta_1$ are the lower and upper limits of the viewing angle described above. The occurrence rates of the DP objects for the bipolar explosion are 34\% (BP8) and 68\% (BP2). Using the same method, the corresponding NC fractions are 13\% and 6\%. The results are summarized in Table \ref{tab:model_para}. 

It should be noted that the assumption of randomly-distributed viewing angle may not be valid for SNe Ic-BL, as these events are frequently accompanied by the occurrence of GRB and may favor on-axis direction. However, the number of these events is small in this sample ($N$ = 9 out of 103). The following analysis will be restricted to the canonical SESNe (SNe Ic-BL excluded), and SNe Ic-BL will be discussed separately.

The line profile fractions of canonical SESNe are:21\% (GS), 27\% (NC), 16\% (DP) and 36\% (AS). Based on the bipolar explosion models, the observed fraction of DP objects suggests that the fraction of the bipolar supernovae is $\sim$ 25\% (BP2) to 48\% (BP8), if the sample is assumed to be unbiased in orientation and all the DP objects are originated from the oxygen-rich torus viewed from the on-edge direction. 
The relatively low fraction of bipolar supernovae also implies that most of the NC objects can not be interpreted by the same configuration but viewed on-axis. Using the estimated occurrence rates of the NC objects (6\% and 13\% for BP2 and BP8), the expected NC fraction arising from this configuration is only about 1.5\% (BP2) to 6.2\% (BP8) of the full canonical SESNe sample, much less than the observed NC fraction (27\%). Therefore more than $\sim$80\% of the NC objects cannot be interpreted by the bipolar explosion. It may leave a massive oxygen blob moving perpendicular to the line of sight or the enhanced core density as more plausible scenarios.

However, as stated above, the model should not be over-interpreted. The classification of the model [O I] profiles into the NC and GS categories is one issue; this is very sensitive to the detailed density distribution, which might be affected by the details of the model construction (e.g., the treatment of the boundary condition in the explosion model). Conservatively, we may thus consider a combination of the NC and GS profiles as the `single-peak' category. If we allow this combined classification, then the single-peak fraction expected in the model is 36\% (BP2) to 66\% (BP8). Taking into account the fraction of the bipolar model as constrained by the BP fraction (i.e., 26\% in BP2 or 50\% in BP8), the expected fraction of the single-peak objects is $\sim $ 9.4\% (BP2) or 33\% (BP8). The fraction of the single-peak objects in the canonical SESNe sample is 48\%, and thus the bipolar configuration can explain up to 70\% of the NC/GS objects in this case. 

Another issue is the classification of the AS and DP profiles as individual classes, for the following two reasons: (1) the classification of the AS and DP objects in the fitting procedure is not very strict, and for some objects the classification is found to be interchangeable (see \S 6.3 for further discussion). (2) There is indeed no `AS' profile predicted in the model, and this stems from the two strong assumptions in the model; perfect axisymmetry plus symmetry in the two hemispheres, from which only the line profile symmetric with respect to the line rest wavelength is predicted. In reality, these two assumptions are probably too strong; for example, the observed neutron star kick naturally indicates there must be some overall shift in the momentum distribution within the ejecta (\citealt{holland17},\citealt{katsuda18}). Therefore, we may consider the AS and DP collectively as the `non-single profile' category and compare it to the model DP fraction. As combined with the above caveat on the classification between the NC and GS categories, we may then compare the fractions of the `single-peaked category' (NC and GS) and the `non-single profile' category (AS and DP). Then, the observed fraction of the non-single profile category is 52\%, while this is 64\% (BP2) or 34\% (BP8). The single-peak category accounts for 48\% of the canonical SESNe sample, and its fraction is 34\% (BP2) or 66\% (BP8). Therefore, the bipolar-like model could account for the full canonical SESNe sample, once one allows the deviation from either the axisymmetry or the symmetry between the two hemispheres to some extent. In other word, the above analysis suggests that (1) the deviation from spherical symmetry could be a common feature in the SN explosion, (2) most of the SN explosion would also have a specific direction, (3) the configuration having negligible deviation from the axisymmetry and the two-hemispheres symmetry could account only up to one third of the canonical SESNe.

The leading scenario for GRBs includes two components: a narrow and relativistic jet for an high-energy GRB emission and a quasi-spherical (but perhaps with a substantial asphericity) component for an optical SN emission. For those associated with GRBs, there could indeed be a preferential viewing direction (\citealt{maeda06b}). In the sample of 9 SNe Ic-BL, two are definitely associated with GRBs (SNe 1998bw and 2006aj). SN 1997ef might also have been associated with a GRB, and there could also be a bias in the viewing direction for SN 2012ap given its strong radio emission. Therefore, up to $\sim 45$\% of the SNe Ic-BL in this sample may indeed suffer from an observational bias in the viewing direction. If we would take this fraction in the model prediction (Table \ref{tab:model_para}), then the NC, GS, DP fractions expected in the model would change to (as the most extreme case) 48\%:17\%:35\% (BP2) or 52\%:29\%:19\% (BP8). This is indeed compatible to the observed fractions of the NC (56\%) and GS (11\%) objects, or the sum of the NC and GS fractions (67\%; see above for the uncertainty associated with the NC/GS classifications) among the SN Ic-BL sample. 

While the specific model used here would not allow quantitative discussion on the difference between the NC and GS categories (see above), qualitative comparison between different SN sub types may still be possible; a larger degree of asphericity leads to a larger ratio of the NC objects to the GS object. This may partly explain a larger fraction of the NC objects in SNe Ic-BL than the other sub types, together with the effect of a possible bias in the viewing direction as stated above. A lack of the DP objects in SNe Ic-BL is puzzling\footnote{SN 2003jd is a prototype of the DP object (\citealt{mazzali05},\citealt{tauben09}). However, its publicly available spectra do not meet the wavelength range required in this work, so it is not included in our sample.}. As one possibility, this may indicate that SNe Ic-BL may tend to have a specific direction in the explosion and the deviation from the axisymmetry and/or two-hemispheres symmetry is more important than in the other SESN sub types. This might further be related to the larger asphericity indicated by a large fraction of the NC objects in SNe Ic-BL. Further investigation focusing on the difference of nebular behaviors between SNe Ic-BL with and without GRB association, based on a larger sample, is required. 

\begin{deluxetable}{l|l|c|c}
\caption{Properties of the BP2 and BP8 models}
\label{tab:model_para}
\tablehead{
\nocolhead{} & \nocolhead{} & \colhead{BP2}& \colhead{BP8}
}
\startdata
\multirow{3}{*}{NC} & angle\tablenotemark{*}
 & $\leqslant20\degree$& $\leqslant30\degree$\\
 & fraction &  0.06   &   0.13  \\
                    & scatter of [O {\sc i}] width&0.003&0.004\\ \hline
\multirow{3}{*}{GS} & angle    & 30$\degree$...40$\degree$    & 40$\degree$...60$\degree$ \\
                    & fraction &  0.30   &   0.53  \\
                    & scatter of [O {\sc i}] width&0.050&0.014\\ \hline
\multirow{3}{*}{DP} & angle    &  50$\degree\leqslant$   & $70\degree\leqslant$ \\
                    & fraction &  0.64   &   0.34  \\
                    & scatter of [O {\sc i}] width&0.012&0.004\\ \hline
\enddata
\tablenotetext{*}{The dividing angles for DP objects of BP2 and BP8 models are different from that in \citet{maeda08}. This is because we employ a different definition of DP in this work, which is based on the fitting procedure described in \S 2. To avoid confusion, throughout the paper, we will adhere to this criterion.}
\end{deluxetable}

\subsection{{\rm [O {\sc i}]/[Ca {\sc i}{\sc i}]}-{\rm [O {\sc i}]} width correlation}

In \S 4.2, using the thus far largest spectral sample of nebular SESNe, a correlation between the [O {\sc i}]/[Ca {\sc i}{\sc i}] ratio and the [O {\sc i}] width is discerned. In the computed nebular spectra of SESNe, the [O {\sc i}]/[Ca {\sc i}{\sc i}] ratio is found to be positively correlated with the progenitor CO core mass (\citealt{fransson89, jerk15, dessart21a}), and is therefore routinely employed as the indicator of this very important quantity (\citealt{kun15, maeda15, fang19}).  Based on this \emph{assumption} (its validity will be discussed in \S 5.4), the correlation implies that the ejecta of SN with a larger CO core tends to expand faster. The typical velocity of the ejecta can be estimated as:
\begin{equation}
\label{eq:0}
v^2~\sim~\frac{E_{\rm K}}{M_{\rm ejecta}}~=~\frac{E_{\rm K}}{M_{\rm pre-SN}-M_{\rm NS}}.\\
\end{equation}
Within each sub type, a more massive progenitor will thus tend to have larger ejecta mass. If the kinetic energy of the ejecta is a constant, for example, 10$^{51}$ erg, the velocity of the ejecta would be expected to be anti-correlated with the progenitor ZAMS mass or the CO core mass, which contradicts the result in this work. The positive correlation of the [O {\sc i}] width and [O {\sc i}]/[Ca {\sc i}{\sc i}] ratio implies that the SN with a progenitor possessing a more massive CO core will tend to have larger kinetic energy. Assuming that the kinetic energy is a function of the CO core mass, i.e., $E_{\rm K}=E_{\rm K} (M_{\rm CO})$, the observational tendency in this work can be qualitatively reproduced. 

For SNe Ic/Ic-BL, the typical velocity can be estimated as:
\begin{equation}
\label{eq:1}
v^2~\sim~\frac{E_{\rm K}}{M_{\rm ejecta}}~=~\frac{E_{\rm K}(M_{\rm CO})}{M_{\rm ejecta}}~\approx~\frac{E_{\rm K}(M_{\rm O})}{M_{\rm O}},\\
\end{equation}
where $M_{\rm CO}$ is the CO core mass and $M_{\rm O}$ is the mass of the oxygen in the ejecta. Since $M_{\rm O}$ is tightly correlated with $M_{\rm CO}$, $E_{\rm K}(M_{\rm CO})$ can also be written as $E_{\rm K}(M_{\rm O})$. We assume $M_{\rm ejecta}\approx M_{\rm O}$, as the oxygen-rich material makes up a significant part of the ejecta of SNe Ic/Ic-BL. If the dependence of $E_{\rm K}$ on $M_{\rm O}$ is in the form of power law, i.e., $E_{\rm K}\propto M_{\rm O}^{\alpha}$, and the power index ${\alpha}$ is close to unity, the typical velocity of SNe Ic/Ic-BL will be a constant. 

For SNe IIb/Ib, if the residual hydrogen envelope of SNe IIb is neglected ($\sim 0.1 M_{\odot}$), Equation (\ref{eq:1}) becomes
\begin{equation}
\label{eq:2}
v^2~\sim~\frac{E_{\rm K}(M_{\rm O})}{M_{\rm O}+M_{\rm He}}~=~\frac{E_{\rm K}(M_{\rm O})}{M_{\rm O}}\frac{1}{1 + \frac{M_{\rm He}}{M_{\rm O}}},\\
\end{equation}
where $M_{\rm He}$ is the mass of the helium in the ejecta. The quantity $M_{\rm He}$/$M_{\rm O}$ is a decreasing function of CO core mass, as the He burning is efficient for large $M_{\rm ZAMS}$ (\citealt{dessart20}). Therefore the typical velocity of SNe IIb/Ib is an increasing function of $M_{\rm CO}$ if $\alpha \sim 1$, which explains the behaviors of SNe IIb/Ib in Figure~\ref{fig:main_fitting}.

The gravitational binding energy of pre-SN progenitor is $E_{\rm g}$ $\sim M^2/R$, where $M$ is its mass and $R$ is the radius. The above qualitative analysis gives $E_{\rm K}\propto M_{\rm O}$. Based on the helium star models in \citet{dessart20} (the parameters are listed in their Table 1), we derive the scaling relation $E_{\rm K}\propto E_{\rm g}^{0.60}$ to explain the observed correlation. However, the above discussion is greatly simplified, and highly dependent on the stellar evolution and the mass-loss scheme. A more detailed treatment of the quantitative relation between the kinetic energy and the progenitor CO core mass will be presented in a forthcoming work (Fang et al., in preparation).

Another interesting feature is the dependence of the correlation on the line profile. In Figure~\ref{fig:profile_correlation}, if only NC objects are included, the [O {\sc i}]/[Ca {\sc i}{\sc i}] ratio and the [O {\sc i}] width have the tightest correlation, followed by AS and DP objects. If the NC objects are originated from the oxygen-rich torus viewed from the on-axis direction, then the difference in the velocity projection can be neglected, because the viewing angle is restricted to a small range. The effect of the viewing angle can thus be a potential origin of the relatively large scatter seen in GS objects.

To test how the viewing angle affects the scatter level, the same [O {\sc i}] width measurement is applied to the BP2 and BP8 model spectra. As shown in \S5.1, the range of the viewing angle relative to the on-jet direction will affect the emission line profile. We measure the [O {\sc i}] width of the models, and calculate the standard deviation in each line profile group. The results are summarized in Table~\ref{tab:model_para}. In general, the scatter levels of the models are much smaller than observation (about 0.06 dex, see the lower panel of Figure~\ref{fig:main_fitting}), but both the BP2 and BP8 models give the correct tendency; the scatter levels of the NC and DP types are relatively small compared with the GS type, as the viewing angle of the NC and DP models are restricted to a narrow range where the effect of velocity projection can be neglected.

\subsection{{\rm [O {\sc i}]/[Ca {\sc i}{\sc i}]}-line profile correlation}
The relation between the [O {\sc i}]/[Ca {\sc i}{\sc i}] ratio and the line profiles (\S 4.3) can be summarized as follows: (1) the GS objects have the largest average [O {\sc i}]/[Ca {\sc i}{\sc i}], followed by AS/NC, then DP; (2) the fraction of GS objects increases with [O {\sc i}]/[Ca {\sc i}{\sc i}]; (3) the fraction of DP objects decreases with [O {\sc i}]/[Ca {\sc i}{\sc i}] and (4) the fractions of NC and AS objects are not monotonic functions of [O {\sc i}]/[Ca {\sc i}{\sc i}].

The relation between the [O {\sc i}] profile distribution and the [O {\sc i}]/[Ca {\sc i}{\sc i}] ratio suggests the geometry of the O-rich ejecta probed by the [O {\sc i}] profile is a function of the progenitor CO core mass, which is \emph{assumed} to be measured by the [O {\sc i}]/[Ca {\sc i}{\sc i}] ratio. The interpretation of this relation is uncertain. In the classification scheme of \citet{tauben09}, the geometry origins of GS/NC/AS objects are degenerated. Meanwhile, the DP objects are unambiguously related to the O-rich torus, therefore can be an useful indicator of bipolar explosion. However, the fraction of DP objects is affected by two factors, i.e., the occurrence rate and the (average) degree of asymmetry of the bipolar explosion. Two extreme cases will be discussed in the following, which account for the effects of (A) the bipolar explosion rate and (B) the degree of asymmetry on the interpretation of the [O {\sc i}]/[Ca {\sc i}{\sc i}]-[O {\sc i}] profile relation.

\begin{figure}[!t]
\epsscale{1.2}
\plotone{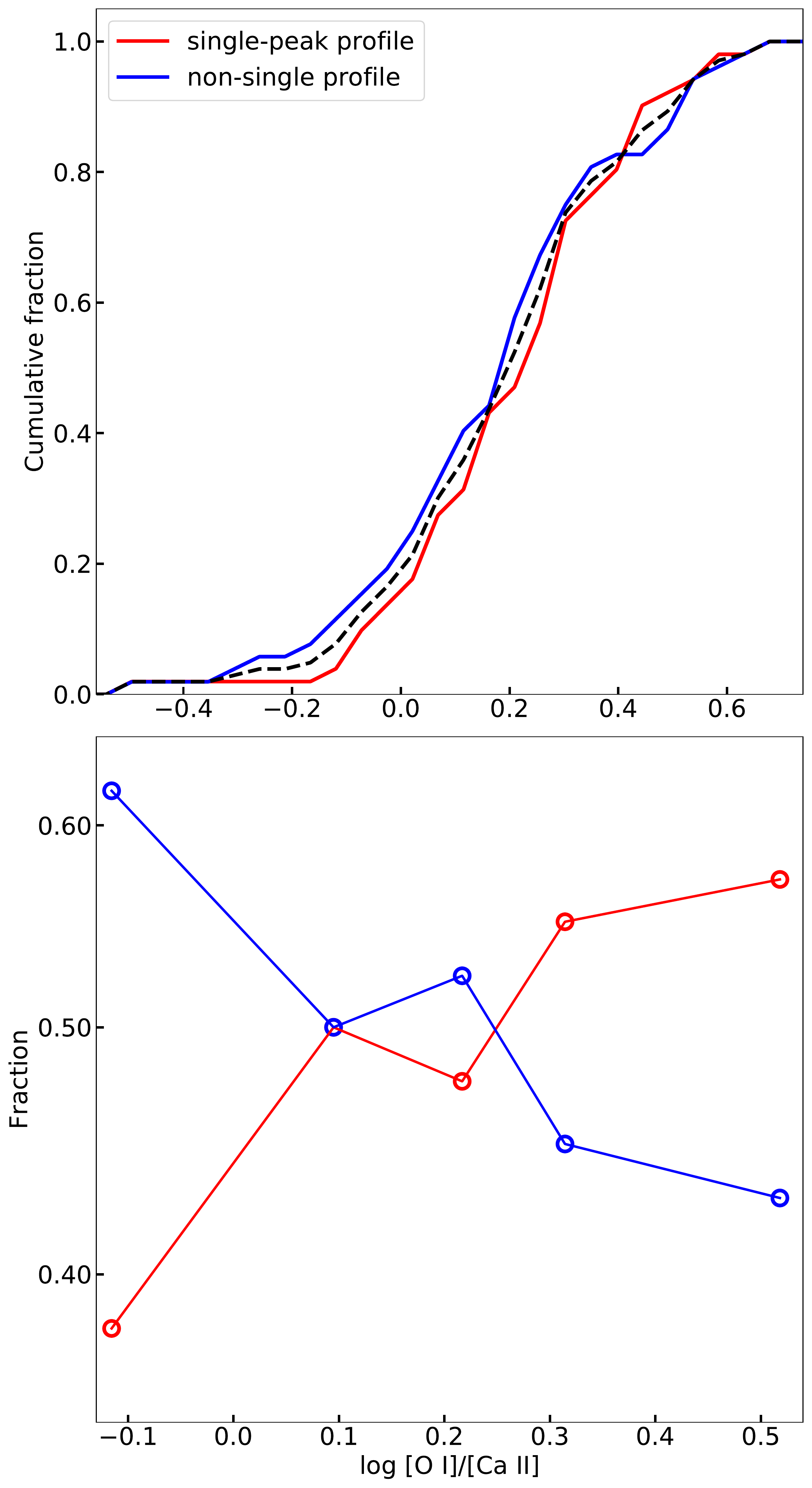}
\centering
\caption{$Upper~ panel$: The cumulative fractions of log [O {\sc i}]/[Ca {\sc i}{\sc i}] of the objects with single-peak (NC + GS) and non-single (DP + AS) profiles. The black dashed line is the distribution of the full sample; $Lower~ panel$: The same as the lower panel of Figure \ref{fig:profile_oxygen_distribute} but based on the classification scheme of single-peak and non-single profile.}
\label{fig:combinedprofile_oxygen_distribute}
\end{figure}

\begin{itemize}
    \item Case A.~~~The global geometry of the ejecta is assumed to be either spherically symmetric (a broad GS base, possibly plus a moving blob to account for the AS and NC objects) or have an axisymmetric bipolar configuration with the \emph{fixed} degree of asymmetry. In this case, the fraction of the DP objects can be an indicator of the occurrence rate of the bipolar explosion. The decreasing trend of DP fraction in Figure \ref{fig:profile_oxygen_distribute} implies the rate of this configuration is anti-correlated with the progenitor CO core mass. Therefore, the ejecta of SESN with a more massive progenitor will tend to be spherical symmetric. This is also consistent with the increasing trend of GS fraction.
    
    By assuming no spatial preference in the viewing angle, only a small fraction of NC objects originate from the bipolar explosion model viewed from the jet-on direction. The NC/AS objects are characterized by globally spherical symmetry plus a narrow component, which can be interpreted as the massive moving blob or enhanced core density. The insensitivity of the fractions of the AS/NC objects on the [O {\sc i}]/[Ca {\sc i}{\sc i}] suggests that the CO core mass is not responsible for the occurrence of these local clumpy structures.
    
    \item Case B.~~~The SESNe in this sample are all assumed to be originated from bipolar explosions (i.e., the occurrence rate is fixed to be 100\%) with different degrees of asymmetry, which are reflected by the fractions of the DP objects (Table~\ref{tab:model_para}). The bipolar explosions are allowed to be non-axisymmetric to account for the AS objects (see the discussion in \S 5.1). As already discussed in \S 5.1, if the GS and NC profiles are combined as `single-peak profile', and the AS and DP profiles are combined as `non-single profile' (the assumption of perfect axisymmetry is discarded), the bipolar explosion models could account for the line profile distribution of the full sample. If this is the case, the dependence of the single-peak/non-single profiles on the [O {\sc i}]/[Ca {\sc i}{\sc i}] may provide important constrain on the development of the bipolar configuration of SNe.
    
    The cumulative fractions of log [O {\sc i}]/[Ca {\sc i}{\sc i}] of the objects with single-peak and non-single profiles are plotted in the upper panel of Figure \ref{fig:combinedprofile_oxygen_distribute}. Although the average log [O {\sc i}]/[Ca {\sc i}{\sc i}] of single-peak objects is slightly larger than those with non-single profile, the difference is not significant, and the [O {\sc i}]/[Ca {\sc i}{\sc i}] distributions are indistinguishable ($p~\textgreater$ 0.25 based on the two-sample AD test). The relation between the [O {\sc i}]/[Ca {\sc i}{\sc i}] ratio and the distribution of line profile is derived using the same method as \S 4.3. As shown in the lower panel of Figure \ref{fig:combinedprofile_oxygen_distribute}, the trends where the fraction of the single-peak objects increases as log [O {\sc i}]/[Ca {\sc i}{\sc i}] increases, while the fraction of their non-single counterparts decreases, can be discerned ($p$=$\pm 0.90$ respectively and $\rho\textless0.03$).
    
    The discussion in \S 5.1 shows that the BP2 model has a smaller fraction of single-peak than the BP8 model (see Table~\ref{tab:model_para}). With [O {\sc i}]/[Ca {\sc i}{\sc i}] being a measurement of the CO core mass, the statistics evaluation is qualitatively consistent with the scenario where the ejecta geometry develops as the progenitor CO core mass increases, gradually converting from the mildly aspherical BP2 cases to the extremely aspherical BP8 cases, i.e.,  the deviation of the explosion from spherical symmetry develops as the CO core mass (or ZAMS mass of the progenitor) increases.
    
    Comparison of the data using the specific bipolar model is just for demonstration purpose. In reality, the ejecta structure can be more complicated, and the full SESNe samples may not be represented by a single model sequence, we thus limit ourselves to discuss the general tendency using this specific models.
\end{itemize}

The investigation on the physics that governs the dependence of the ejecta geometry on the progenitor CO core mass is related to the development of the asphericial explosion, which may put important constrain on the explosion mechanism of SESNe. However, the interpretation of this dependence can be different (or even opposite) when different assumptions are made, as exemplified by the two extreme cases discussed above. In reality, the situation may be the mixture of the two cases, or even more complicated. To firmly interpret the relation between the [O {\sc i}]/[Ca {\sc i}{\sc i}] ratio and the distribution of line profile, we thus need another tool, which should be independent from the [O {\sc i}] profile, to probe the geometry of the ejecta. The investigation on this topic will be presented in a forthcoming work (Fang et al. in preparation).

\subsection{{\rm [O {\sc i}]/[Ca {\sc i}{\sc i}]} as measurement of progenitor M$_{\rm ZAMS}$}
The discussion in \S 5.2 and \S 5.3 is largely based on the assumption that [O {\sc i}]/[Ca {\sc i}{\sc i}] ratio is positively correlated with the progenitor CO core mass, and thus its ZAMS mass. This is the case for the currently available models (\citealt{fransson89, jerk15, jerk17, dessart21a}). However, whether this diagnostics is robust remains uncertain (\citealt{jerk17}); the [O {\sc i}]/[Ca {\sc i}{\sc i}] ratio is affected by the phase of observation, the expansion velocity of the ejecta (or more specifically, kinetic energy), and the distribution of the calcium. In \S 6.1, we will show that the spectral phase will not affect the above correlation. In this subsection, the latter two points, i.e., the effect of kinetic energy, and the pollution of calcium into the O-rich material, are discussed.

\begin{itemize}
    \item Kinetic~energy.~~~~The [Ca {\sc i}{\sc i}] is emitted from the ash of the explosive burning, the physical properties of which are affected by the explosion energy. The density structure of the ejecta is also related to its expansion velocity, which again affects the [O {\sc i}]/[Ca {\sc i}{\sc i}] ratio (\citealt{fransson89}). We will now investigate whether the effect of the explosion energy alone can account for the wide range of the [O {\sc i}]/[Ca {\sc i}{\sc i}] ratio.
    
    To simplify the discussion, the amount of the newly synthesize elements, including calcium, is assumed to be positively correlated with the kinetic energy of the ejecta (\citealt{woosley02, limongi03}). With this assumption, for a fixed CO core mass, the kinetic energy will affect the [O {\sc i}]/[Ca {\sc i}{\sc i}] in two aspects: (1) SNe with larger kinetic energy will synthesize a larger amount of calcium, which increases the intensity of the [Ca {\sc i}{\sc i}] and decreases the [O {\sc i}]/[Ca {\sc i}{\sc i}] ratio, and (2) with larger kinetic energy, the ejecta will expand faster, which decreases its density and again, decreases the [O {\sc i}]/[Ca {\sc i}{\sc i}] ratio (\citealp{fransson89}).  

    For the same CO core with different kinetic energy injected, the [O {\sc i}]/[Ca {\sc i}{\sc i}] will be expected to be anti-correlated with the expansion velocity of the ejecta, which contradicts the observed correlation in Figure~\ref{fig:main}. The correlation of the [O {\sc i}]/[Ca {\sc i}{\sc i}] ratio and the [O {\sc i}] width suggests that the effect of the kinetic energy is limited and cannot be a main driver of the large range of [O {\sc i}]/[Ca {\sc i}{\sc i}] ($\sim$ 1 dex);
    
    \item Calcium pollution.~~~~The [Ca {\sc i}{\sc i}] is mainly emitted by the newly synthesized calcium from the explosive oxygen burning ash \citep{jerk15}. However, in several CCSN nebular models, if the calcium produced by the pre-SN nucleosynthesis is microscopically mixed into the O-rich layer through shell merger (which may happen during Si burning stage), its contribution to the [Ca {\sc i}{\sc i}] becomes significant (\citealt{dessart21}). The [O {\sc i}]/[Ca {\sc i}{\sc i}] will be dramatically reduced because [Ca {\sc i}{\sc i}] is a very effective coolant (\citealt{dessart20a, dessart21}). In this case, [O {\sc i}]/[Ca {\sc i}{\sc i}] is no longer a monotonic function of progenitor CO core mass. 
    
    Several works (\citealt{collins18,dessart20a}) reported that the occurrence rate of calcium pollution is high for a more massive star. If the progenitor mass is increased, the [O {\sc i}]/[Ca {\sc i}{\sc i}] will be affected by two competing factors along different directions; increased by the CO core mass, but decreased by the higher degree of microscopically mixed calcium. We may consider the most extreme case, in which the effect of the calcium pollution on the progenitor mass is so strong that the correlation between the [O {\sc i}]/[Ca {\sc i}{\sc i}] and CO core mass is inverted, i.e., a small [O {\sc i}]/[Ca {\sc i}{\sc i}] implies large CO core mass. With this assumption, a constant kinetic energy can produce the correlation between [O {\sc i}]/[Ca {\sc i}{\sc i}] and [O {\sc i}] in Figure \ref{fig:main}.

    From the current observation, the degree of calcium pollution is difficult to constrain. However, its effect on [O {\sc i}]/[Ca {\sc i}{\sc i}] is probably not very strong from several observational lines of evidence; (1) The measured progenitor masses of SNe 2011dh, 2013df and iPTF 13bvn are relatively small from pre- or post-SN image (\citealt{maund11, dyk14, cao13}), and their [O {\sc i}]/[Ca {\sc i}{\sc i}] are among the lowest of the full sample. SNe 1998bw and 2002ap are also believed to have massive progenitors, meanwhile their [O {\sc i}]/[Ca {\sc i}{\sc i}] are at the highest end (\citealt{nakamura01, mazzali02}). (2) A correlation between the light curve width and the [O {\sc i}]/[Ca {\sc i}{\sc i}] ratio is reported by \citet{fang19}. The light curve width can be an independent measurement of the ejecta mass. If the [O {\sc i}]/[Ca {\sc i}{\sc i}] is mainly determined by the degree of microscopic mixing, an anti-correlation between the [O {\sc i}]/[Ca {\sc i}{\sc i}] and light curve width would be expected, which contradicts the observation.

\end{itemize}

We have discussed the possible factors that would affect the [O {\sc i}]/[Ca {\sc i}{\sc i}] ratio. However, it should be emphasized that the current understanding on the [O {\sc i}]/[Ca {\sc i}{\sc i}] ratio itself, as well as its relations with the physical properties (CO core mass, kinetic energy, microscopic mixing, etc.) is still limited. To firmly establish the relations between the observables and the ejecta properties, which is crucial to explain the correlation in Figure \ref{fig:main}, a sophisticated nebular SESN model with all the above factors involved is needed.

\section{Discussion}
\subsection{Temporal evolution}
The nebular spectra in this work cover a quite large range of phases (mean value $\textless$phase$\textgreater$ = 213 days, standard deviation $\sigma$ = 61 days). Therefore it is important to investigate whether the phases of the spectra will affect the correlation in Figure~\ref{fig:main}. 

The most straightforward method is to calculate the rate of change of the [O {\sc i}]/[Ca {\sc i}{\sc i}] ratio or [O {\sc i}] width by following the evolution of each object. However, the number of objects with multiple nebular spectra covering a wide range of phases is too small for such investigation. Fortunately, the main focus of this work is on the statistical properties of these two quantities. Unless there is a strong bias in the sample (for example, objects with large [O {\sc i}]/[Ca {\sc i}{\sc i}] tend to be observed in late phases), the average difference of the quantities at different phases can be employed to estimate the effect of the spectral phase on bulk statistics. In this work, two methods are employed to estimate the rates of change of [O {\sc i}]/[Ca {\sc i}{\sc i}] ratio or [O {\sc i}] width; one based on the statistics of the full sample, and the other based on the evolution of individual objects.

\begin{figure*}
\epsscale{1.2}
\plotone{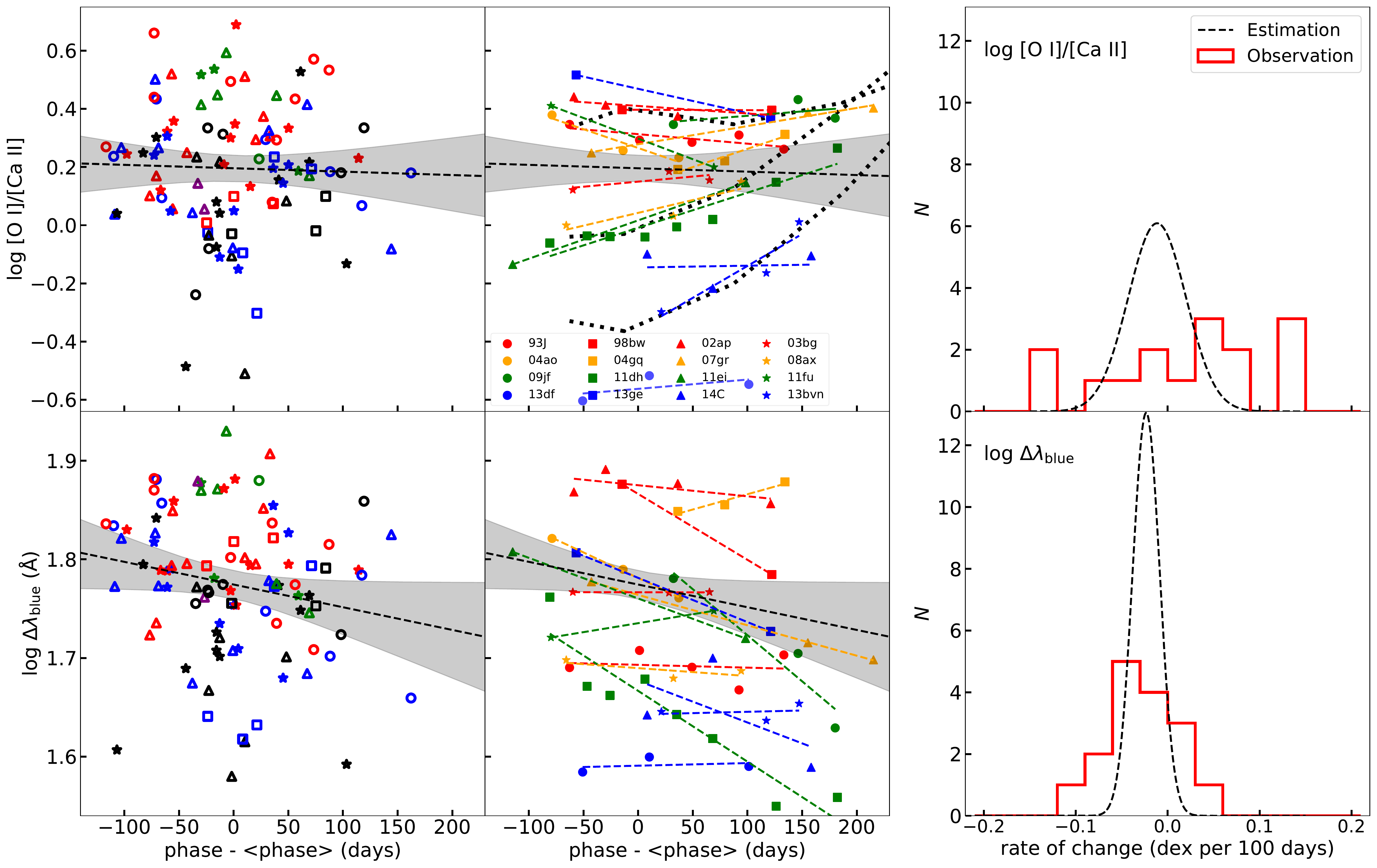}
\centering
\caption{$Left~panels$: The [O {\sc i}]/[Ca {\sc i}{\sc i}] ratio and the [O {\sc i}] width $\Delta\lambda_{\rm blue}$ versus the spectral phase relative to the mean value of the sample ($\textless$phase$\textgreater$ = 213 days). The black dashed lines are the results of the linear regression for the full sample, and the shaded regions are the 95\% CIs estimated from the bootstrap-based Monte Carlo method. The meanings of the different colors and marker are same as Figure \ref{fig:main}; $Middle~panels$: The time evolution of [O {\sc i}]/[Ca {\sc i}{\sc i}] and $\Delta\lambda_{\rm blue}$ of individual objects. Different objects are labeled by different colors and markers. The color dashed lines are the results of linear regression for each object. The fitting results of the full sample are also plotted for comparison. The black dotted lines in the upper-middle panel are the measurements of the model spectra from \citet{jerk15}; $Right~panels$: The distributions of the rates of change of log [O {\sc i}]/[Ca {\sc i}{\sc i}] and log $\Delta\lambda_{\rm blue}$. The red histograms are the observed rates of change of individual objects. The black dashed lines are the expected distributions of the rates of change estimated from the linear regression of the full sample, scaled to $N$=16. The sources of the spectra: SN 1993J (\citealt{barbon95,matheson00,jerk15}); SN 1998bw (\citealt{patat01}); SN 2002ap (\citealt{foley03}); SN 2003bg (\citealt{hamuy09}); SN 2004ao (\citealt{modjaz08,shivvers19}); SN 2004gq (\citealt{maeda08,modjaz14}); SN 2007gr \citealt{shivvers19}; SN 2008ax (\citealt{chornock11,tauben11,modjaz14}); SN 2009jf (\citealt{valenti11,modjaz14}); SN 2011dh (\citealt{shivvers13,ergon15}); SN 2011ei (\citealt{milisa13}); SN 2011fu (\citealt{morales15}); SN 2013df (\citealt{morales14,maeda15}); SN 2013ge (\citealt{drout16}); SN 2014C (\citealt{milisa15b,shivvers19}); iPTF 13bvn (\citealt{fremling16}).}
\label{fig:evolution}
\end{figure*}

The left panels of Figure~\ref{fig:evolution} shows the time dependence of the [O {\sc i}]/[Ca {\sc i}{\sc i}] ratio and the [O {\sc i}] width $\Delta\lambda_{\rm blue}$ of the full sample. In these panels, each data point represents an individual object. The $\Delta\lambda_{\rm blue}$ is weakly correlated with the spectral phase ($\rho$ = -0.29, $p~\textless$ 0.02). The slope from linear regression is -0.023$\pm$0.012 (unit: dex per 100 days. In the following text, unless explicitly mentioned, the units of rates of change of both [O {\sc i}]/[Ca {\sc i}{\sc i}] and [O {\sc i}] width are dex per 100 days). The uncertainty is estimated from 10$^{4}$ bootstrap re-samples, and the 95\% CIs are indicated by the shaded regions. If we attribute this phase dependence to temporal evolution of $\Delta\lambda_{\rm blue}$, averagely, $\Delta\lambda_{\rm blue}$ changes by about -7.7 to -2.5\% per 100 days, which is in good agreement with the decrease rate reported by \citet{maurer10}. The same analysis is performed to the [O {\sc i}]/[Ca {\sc i}{\sc i}] ratio, which in turn shows no evidence of temporal evolution ($\rho$=-0.03, $p\textless$0.79). Linear regression suggests [O {\sc i}]/[Ca {\sc i}{\sc i}] changes by only -0.012$\pm$0.029 dex (or about -9.7 to 4.0\% in linear scale) per 100 days.

To examine the evolution of the [O {\sc i}]/[Ca {\sc i}{\sc i}] ratio and the [O {\sc i}] width of individual objects, we turn to those SNe in the sample with multiple nebular spectra available from the literature, and the maximum phase span is required to be larger than 100 days. The corresponding measurements of these objects are plotted in the middle panels of Figure~\ref{fig:evolution}. The evolution rates are estimated by linear regression. In the SESNe models of \citet{jerk15}, the oxygen element spreads across a wide range of zones. Initially the [O {\sc i}] is dominated by the emission from the outermost region. As the ejecta expands, the contribution from the innermost region becomes larger, which decreases the average velocity of the emitting elements and therefore the width of the emission line. For most objects ($N$=12 out of 16), the [O {\sc i}] width decreases with time, which is consistent with the above picture. The average and standard deviation of the slopes are -0.026$\pm$0.033. 
The distribution of the slopes is also shown in the lower right panel of Figure \ref{fig:evolution}, with a peak around -0.029. This is consistent with the slope estimated from the full sample (-0.023), and can fully explain the overall time dependence of [O {\sc i}] width in the lower left panel of Figure \ref{fig:evolution}.

However, the temporal evolution of [O {\sc i}]/[Ca {\sc i}{\sc i}] depends on the physical conditions of the ejecta. The complexity is also discerned in the observation data; the observed slopes of the [O {\sc i}]/[Ca {\sc i}{\sc i}] ratio spread over a wide range. The average and standard deviation of the slopes are 0.027$\pm$0.094. Unlike the [O {\sc i}] width, the distribution of the evolution rates of [O {\sc i}]/[Ca {\sc i}{\sc i}] lacks a clear peak, which may possibly explain the lack of correlation between the spectral phase and [O {\sc i}]/[Ca {\sc i}{\sc i}]; the different directions of evolution cancel each other out.

It is useful to compare the evolution of [O {\sc i}]/[Ca {\sc i}{\sc i}] with theoretical models.
For the SNe IIb model spectra of \citet{jerk15}, the [O {\sc i}]/[Ca {\sc i}{\sc i}] increases with time (see also Figure 13 in \citealt{jerk17}). Using the same measurement method in this work, the evolution of the [O {\sc i}]/[Ca {\sc i}{\sc i}] of these models is plotted by the black dotted lines in the upper-middle panel of Figure \ref{fig:evolution} for comparison. The [O {\sc i}]/[Ca {\sc i}{\sc i}] and the evolution of the He star model with $M_{\rm ZAMS}=12M_{\odot}$ (M12 for short hereafter) is consistent with iPTF 13bvn. When compared with SN 2011dh and SN 2008ax, the M13 model evolves faster, but the behaviors are qualitatively similar; the change of [O {\sc i}]/[Ca {\sc i}{\sc i}] is mild before $\sim$300 days, while at later phases the slope increases. For M17 model and the objects with large [O {\sc i}]/[Ca {\sc i}{\sc i}], the rates of change are approximately negligible before $\sim$300 days.

\begin{figure}
\epsscale{1.2}
\plotone{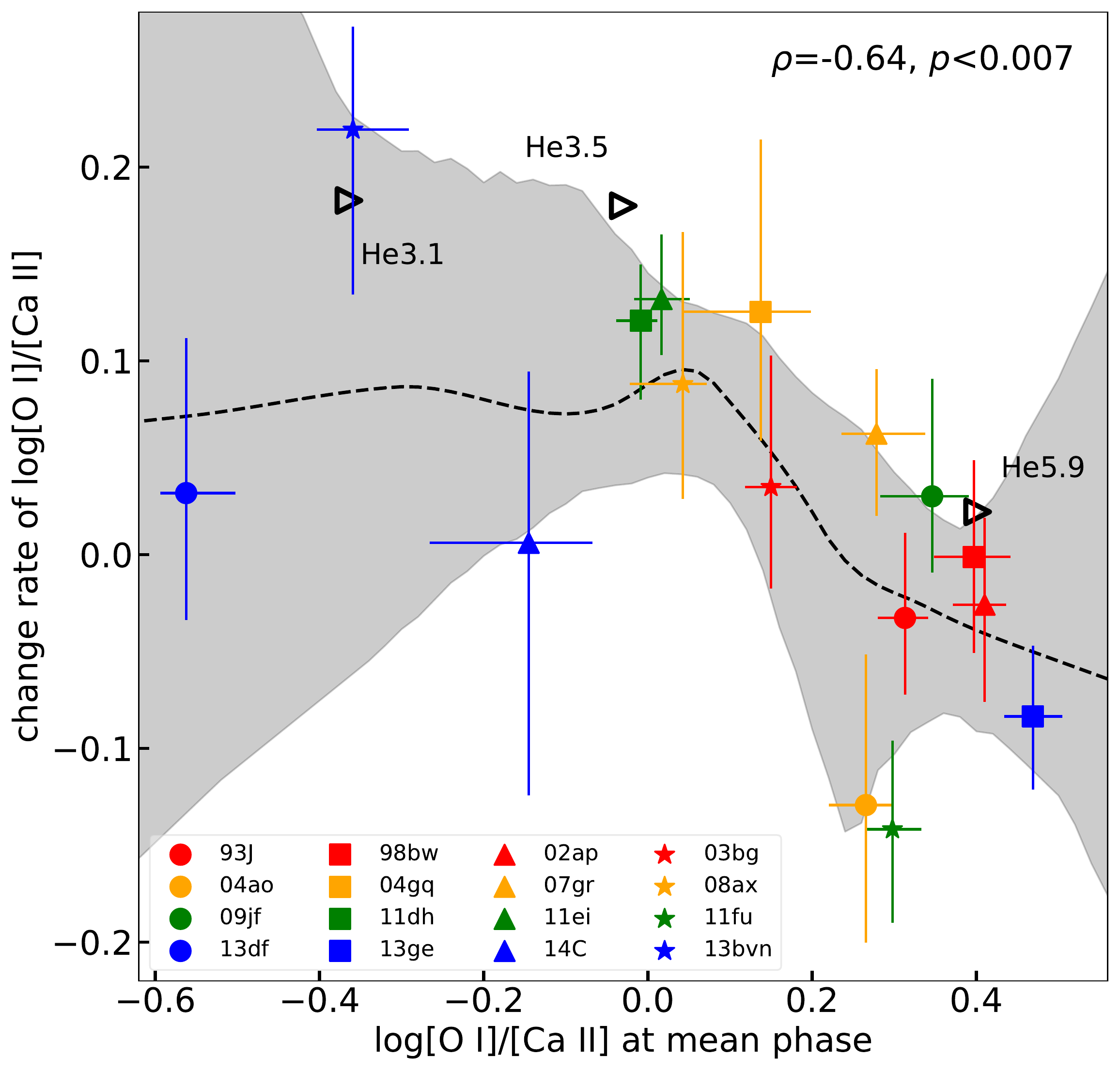}
\centering
\caption{The relation between log [O {\sc i}]/[Ca {\sc i}{\sc i}] and its time evolution rate (unit: dex per 100 days). Different objects are labeled by different colors and markers. The SNe IIb model spectra of \citet{jerk15} are also plotted for comparison, and the evolution rates of the models are calculated from the measurements at 150-400 days. The dashed line is the result of local non-parametric fit and the shaded region is the 95\% confidence interval estimated from the bootstrap-based Monte Carlo method (see main text).}
\label{fig:ratio_slope}
\end{figure}

The above discussion motivates the investigation on the possible dependence of the rate of change of [O {\sc i}]/[Ca {\sc i}{\sc i}] on [O {\sc i}]/[Ca {\sc i}{\sc i}] itself. The He star models of \citet{jerk15} and the observation data suggest objects with large [O {\sc i}]/[Ca {\sc i}{\sc i}] tend to have slowly-evolve [O {\sc i}]/[Ca {\sc i}{\sc i}]. For the 16 SNe with wide spectral phase span and the He star models in \citet{jerk15}, their [O {\sc i}]/[Ca {\sc i}{\sc i}] ratios are corrected to the mean phase (213 days), which are then compared with the slopes estimated from linear regression, as shown in Figure \ref{fig:ratio_slope}. The uncertainties of the corrected [O {\sc i}]/[Ca {\sc i}{\sc i}] and the slopes are estimated from the bootstrap based Monte Carlo method, which includes the uncertainties of the measurement of [O {\sc i}]/[Ca {\sc i}{\sc i}] at different phases. An anti correlation between the slopes and the [O {\sc i}]/[Ca {\sc i}{\sc i}] ratios can be discerned ($\rho=-0.64,p\textless 0.007$), especially for objects with log[O {\sc i}]/[Ca {\sc i}{\sc i}]$\textgreater$0. The relation between the [O {\sc i}]/[Ca {\sc i}{\sc i}] and the slope at low [O {\sc i}]/[Ca {\sc i}{\sc i}] end is hard to constrain because only three objects are available (SNe 2013df, 2014C and iPTF 13bvn) and the scatter is large. The 16 observation data points are then fitted by local non-parametric regression, the result of which is plotted by the black dashed line in Figure \ref{fig:ratio_slope}, and the 95\% CI estimated from the bootstrap-based Monte Carlo method is shown by the shaded region.

Limited by the sample size ($N$=16), the result in this work provides the starting point for the investigation on the dependence of the evolution rate of [O {\sc i}]/[Ca {\sc i}{\sc i}] on [O {\sc i}]/[Ca {\sc i}{\sc i}]. To firmly establish this relation, we need a larger sample of SESNe with nebular spectra covering large ranges of phases, especially later than 300 days.

The direct comparison of the nebular spectra at different phases are presented in Figure \ref{fig:multiphase_example} for some well-observed examples.

To eliminate the effect of spectral evolution, we run 10$^4$ simulations, and in each trial, the rates of change of [O {\sc i}]/[Ca {\sc i}{\sc i}] and [O {\sc i}] width are assigned to each object, which are randomly drawn from (1) the slope estimated from the full sample (the black dashed lines in the right panels of Figure \ref{fig:evolution}), or (2) the distributions of slopes derived from following the evolution of individual objects (the red histograms in the right panels of Figure \ref{fig:evolution}), or (3) the [O {\sc i}]/[Ca {\sc i}{\sc i}]- dependent evolution rate (the shaded region in Figure \ref{fig:ratio_slope}). The [O {\sc i}]/[Ca {\sc i}{\sc i}] and [O {\sc i}] width are then corrected to the mean phase. We find no matter what distributions and combinations are chosen, the two quantities are significantly correlated, with $\rho$ ranges from 0.50 to 0.54 and $p~\textless~0.0001$ for all cases. We therefore conclude that the spectral evolution will not significantly affect the correlation in Figure \ref{fig:main}.

In Figure~\ref{fig:main_fitting}, the helium-rich SNe behave differently from their helium-deficient counterparts. However, the average phases of the SN sub types in this work are similar and no statistical difference can be discerned; 220$\pm$58 days for SNe IIb, 203$\pm$80 days for SNe Ib, 202$\pm$56 days for SNe Ic and 223$\pm$36 days for SNe Ic-BL. Therefore temporal evolution can not be the main reason for the different behaviors of the different SN sub types in both Figure~\ref{fig:main} and Figure~\ref{fig:main_fitting}, which can be another evidence of the limited effect of the spectral phase on the correlation.

\subsection{The effect of Asymmetric H$\alpha$/[N {\sc i}{\sc i}]}
In $\S$2, to derive the `clean' [O {\sc i}] profile, the excess flux at the red wing of [O {\sc i}] is subtracted by assuming it is symmetric with respect to 6563 ${\rm \AA}$. However, this assumption is not necessarily valid and will affect the line width measurement. For example, if the real center of the excess flux is red shifted, assuming an symmetry with respect to 6563 $\rm \AA$ will result in over subtraction of the [O {\sc i}] and under estimation of the line width. It is not always easy to tell whether such asymmetry exists from the nebular spectra, as the [O {\sc i}] and the excess flux are always blended. In this subsection, we will quantitatively estimate how the asymmetry of the H$\alpha$-like structure affect the measurements.

Firstly, an [O {\sc i}] component, which is composed of two Gaussian functions with the same standard deviation ($\sigma$ = 50 $\rm \AA$), is simulated. The central wavelengths are fixed at 6300 and 6364 $\rm \AA$ and the intensity ratio is set to be 3:1 (see \S 2). We then generate a set of excess emissions with detailed profiles listed in Table~\ref{tab:Profile_Halpha} to account for different distributions of the emitters. The half-width at zero intensity (HWZI) of these profiles are fixed to be 220 $\rm \AA$, based on the $\sim$ 10000 km s$^{-1}$ outer edge velocity of the excess profile estimated by \citet{maeda15}. The fluxes of these profiles are set to be 40\% of the [O {\sc i}] emission (about 84 percentage of the full sample). At the same time, we allow the symmetric center $\lambda_{\rm sym}$ move from 6453 to 6673 $\rm \AA$, corresponding to $\lvert v_{\rm shift} \rvert$ $\sim$ 5000 km s$^{-1}$.

\begin{deluxetable}{lcc}
\caption{Excess emission profiles}
\label{tab:Profile_Halpha}
\tablehead{
\colhead{Geometry} & \colhead{Line profile} & \colhead{Notes\tablenotemark{*}}
}
\startdata
Thin shell&Flat-top&d$R_{\rm sh}$ = 0.2$R_{\rm sh}$\\
Thick shell&Flat-top&d$R_{\rm sh}$ = 0.6$R_{\rm sh}$\\
Uniform disk&$\sqrt{1 - (\frac{\lambda~-\lambda_{\rm sym}}{\Delta \lambda})^2}$&$\Delta \lambda$ = 220 $\rm \AA$\\
Uniform sphere&1 - ($\frac{\lambda~-\lambda_{\rm sym}}{\Delta \lambda}$)$^2$&$\Delta \lambda$ = 220 $\rm \AA$\\
\enddata
\tablenotetext{*}{$R_{\rm sh}$ is the maximum radius of the shell, and d$R_{\rm sh}$ is its thickness.}
\end{deluxetable}

\begin{figure}[!t]
\epsscale{1.2}
\plotone{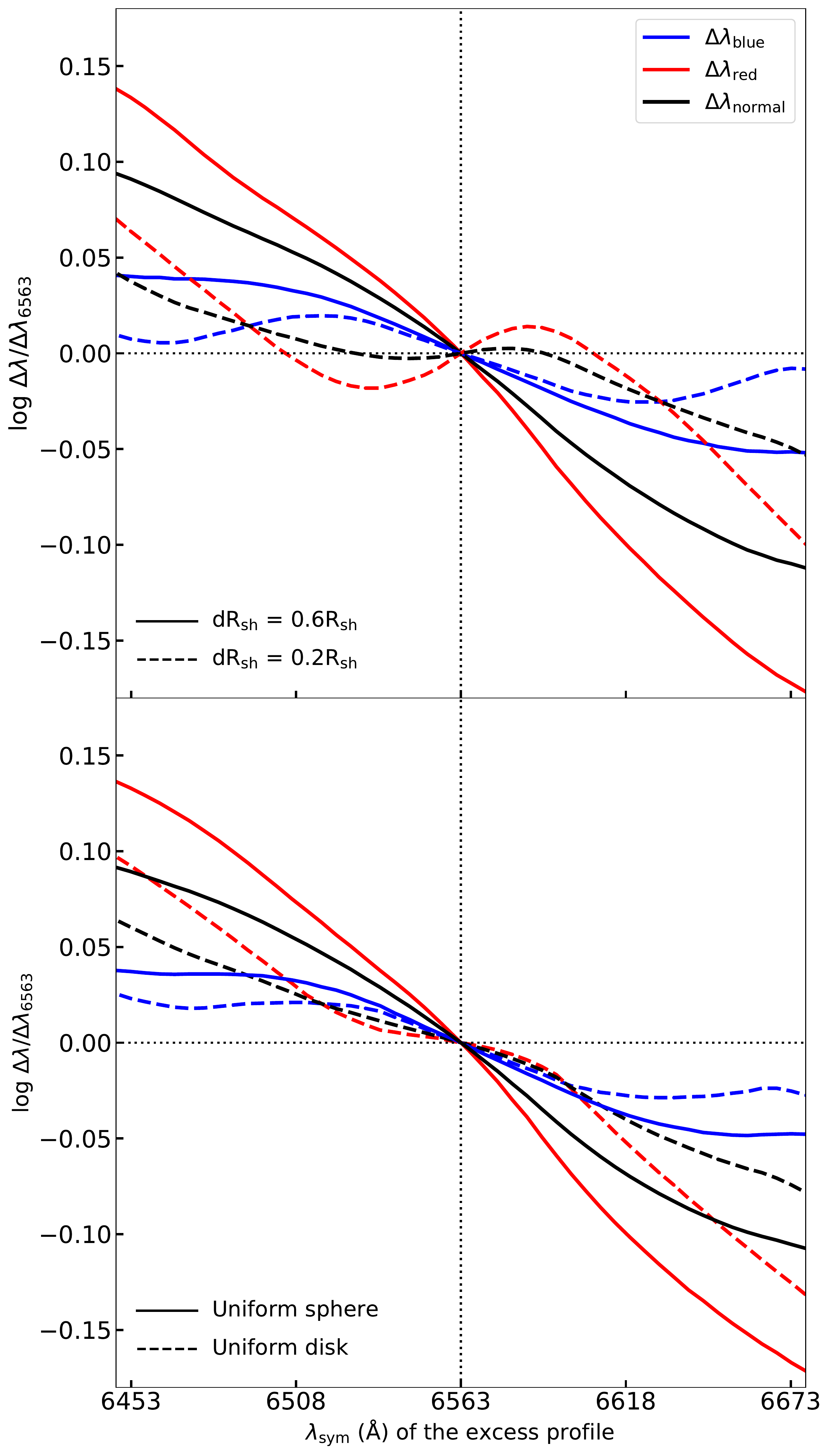}

\centering
\caption{The symmetric center of the excess emission and the deviation of the measured line widths from $\Delta \lambda_{\rm 6563}$ (see main text for definitions). $\Delta\lambda_{\rm blue}$, $\Delta\lambda_{\rm red}$ and $\Delta\lambda_{\rm normal}$ are labeled by different colors and different distributions of the emitters are labeled by different line styles. In most cases, the $\Delta\lambda_{\rm blue}$ is less affected by the asymmetry or the line profile of H$\alpha$-like structure, and is therefore selected to be the measurement of core velocity in this work.}
\label{fig:sim_halpha_effect}
\end{figure}

After adding the [O {\sc i}] profile by the simulated excess emissions, we repeat the measurement in \S 2, assuming the excess flux is symmetric with respect to 6563 $\rm \AA$. The deviation of the measured line widths ($\Delta\lambda_{\rm blue}$, $\Delta\lambda_{\rm red}$ and $\Delta\lambda_{\rm normal}$, see Figure~\ref{fig:example_07Y}) from $\Delta \lambda_{6563}$, which is defined to be the corresponding measured line widths when the excess emission is symmetric with respect to 6563$\rm \AA$, is plotted against the symmetric center $\lambda_{\rm sym}$ in Figure~\ref{fig:sim_halpha_effect}.

It is clear that the asymmetry of the excess flux indeed affects the measured line width. The red-width $\Delta\lambda_{\rm red}$ is sensitive to the distribution of the emitters and the shift of the symmetric center. If the symmetric center is heavily shifted, or the profile is sharply peaked (i.e., thick shell versus thin shell, or disk versus sphere), the deviation will be large and reach to about 15\% to 23\%. However, the blue-width $\Delta\lambda_{\rm blue}$ does not show significant deviation in all cases. The deviation of $\Delta\lambda_{\rm blue}$ changed by about 5\% to 8\%. Even in the most extreme cases, the deviation will not exceed $\sim$ 12\% or 0.05 dex, and cannot account for the 0.3 dex line width difference reported in this work. Given that $\Delta\lambda_{\rm blue}$ is not sensitive to the $\lambda_{\rm sym}$ and the spatial distribution of the emitters, in this work, $\Delta\lambda_{\rm blue}$ is employed as the measurement of line width.

The excess emission can be attributed to shock-CSM induced H$\alpha$, radiative powered [N {\sc i}{\sc i}], or the combination of both (\citealt{patat95, jerk15, fang18}). The insensitivity of the $\Delta\lambda_{\rm blue}$ to the $\lambda_{\rm sym}$ suggests that the identification of the excess emission will make no difference on the measurement.
We note that the conditions tested in this subsection are quite extreme, as most objects do not have excess emission as large as (H$\alpha$ or [N {\sc i}{\sc i}])/[O {\sc i}] = 0.4 (\citealt{fang19}), and the assumption that $\lvert v_{\rm shift} \rvert$ $\sim$ 5000 km s$^{-1}$ does not seem realistic. From the very late phase observation of SN 1993J and SN 2013df, no evidence supports that the boxy profile is significantly asymmetric with respect to H$\alpha$ or [N {\sc i}{\sc i}] (\citealt{maeda15}), so allowing the central wavelength move at a velocity as large as 5000 km s$^{-1}$ ($\sim$ 110 $\rm \AA$) is indeed very conservative. 

Observationally, no significant correlation can be discerned between the central wavelength ($\lambda_{\rm c}$ in Figure \ref{fig:example_07Y}) and the [O {\sc i}] width ($\rho$ = 0.12, $p~\textless$ 0.21), as would be expected if the difference of [O {\sc i}] width is significantly affected by the H$\alpha$-like structure subtraction, which again supports the argument in this section.

\subsection{Narrow-core (NC) and Asymmetry (AS): Doppler-shifted moving blobs?}
As introduced in \S 2.4, the narrow component of the NC profile can be interpreted as a massive oxygen-rich blob moving nearly perpendicular to the line of sight, or enhanced core density. 
Similarly, the AS profile would require a blob moving with non-negligible motion at the direction of line of sight to account for the narrow component. The geometrical origins of NC and AS objects can be unified as ejecta of globally spherical symmery plus (1) Doppler-shifted moving blobs or (2) enhanced core density.
In the enhanced core density scenario, the narrow component is expected to be centered at its rest wavelength. Further, if the direction of the moving massive blob is isotropic, the amount of the red- and blue-shifted narrow components would be similar. In conclusion, the distribution of the narrow component offsets of the AS/NC objects is expected to be symmetric with respect to the Doppler velocity $\mid v_{\rm shift}\mid$ = 0 km s$^{-1}$. Any deviation from such distribution would require an additional effect beyond geometrical effect.

\begin{figure}
\epsscale{1.2}
\plotone{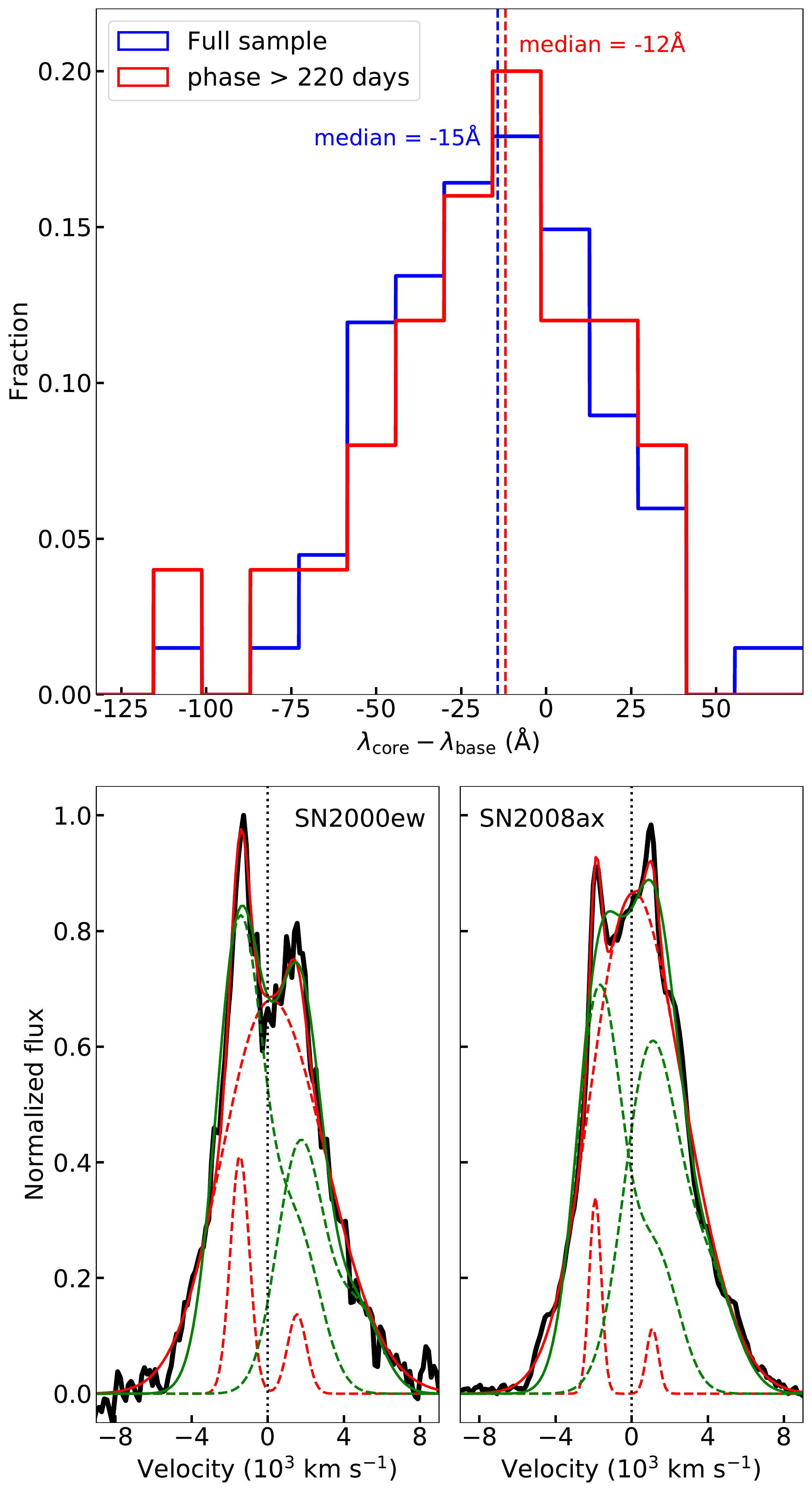}
\centering
\caption{$Upper~panel$:The histogram of the central wavelength offset of the narrow core with respect to the broad base. The blue histogram refers to the full NC + AS sample, while the red one is restricted to the objects observed later than 220 days. The dashed lines indicate the median values; $Lower~panels$: Alternative fits to the SNe 2000ew and 2008ax. The red solid lines are the results of the two components fit with initial guess of case (4), as described in \S 2.4, and the green solid lines are the results where the two components have similar intensity and are forced to blue- and red-shifted. The dashed lines are the corresponding components.}
\label{fig:relative_shift}

\end{figure}

For the combined sample of NC and AS objects, the shift of the broad base is {\rm -5.9} $\rm{\AA}$. The standard deviation is {\rm 9.9} $\rm{\AA}$, which is comparable to the resolution 12.6 $\rm{\AA}$ ($R~\sim$ 500). The low velocity of the broad base is in good agreement with the Gaussian distributed emitter and global spherical symmetry. The histogram of the central wavelength offsets of the narrow component with respect to the broad base is shown in the upper panel of Figure \ref{fig:relative_shift}. On average, the narrow component is blue shifted ({\rm -14}$\rm {\AA}$), which is comparable to its typical width ($\sim$16 $\rm {\AA}$), too large for the enhanced core density scenario. The amount of the red-shifted objects is only about half of the blue-shifted ones ($N_{\rm{red}}$/$N_{\rm{blue}}$ $\sim$ 0.46). This is also not expected if the narrow core originates from the massive moving blob.

\citealt{milisa10} already noticed the [O {\sc i}] with double horns can be classified into two classes:  the two horns are symmetric with respect to zero velocity, or one of the horn is located close to 6300 $\rm{\AA}$ and the other one blue-shifted. These two types can both be fitted by a broad base plus a blue-shifted narrow component. The result in this work suggests the lack of the red-shifted narrow component is a statistically significant phenomenon.

\begin{figure}
\epsscale{1.2}
\plotone{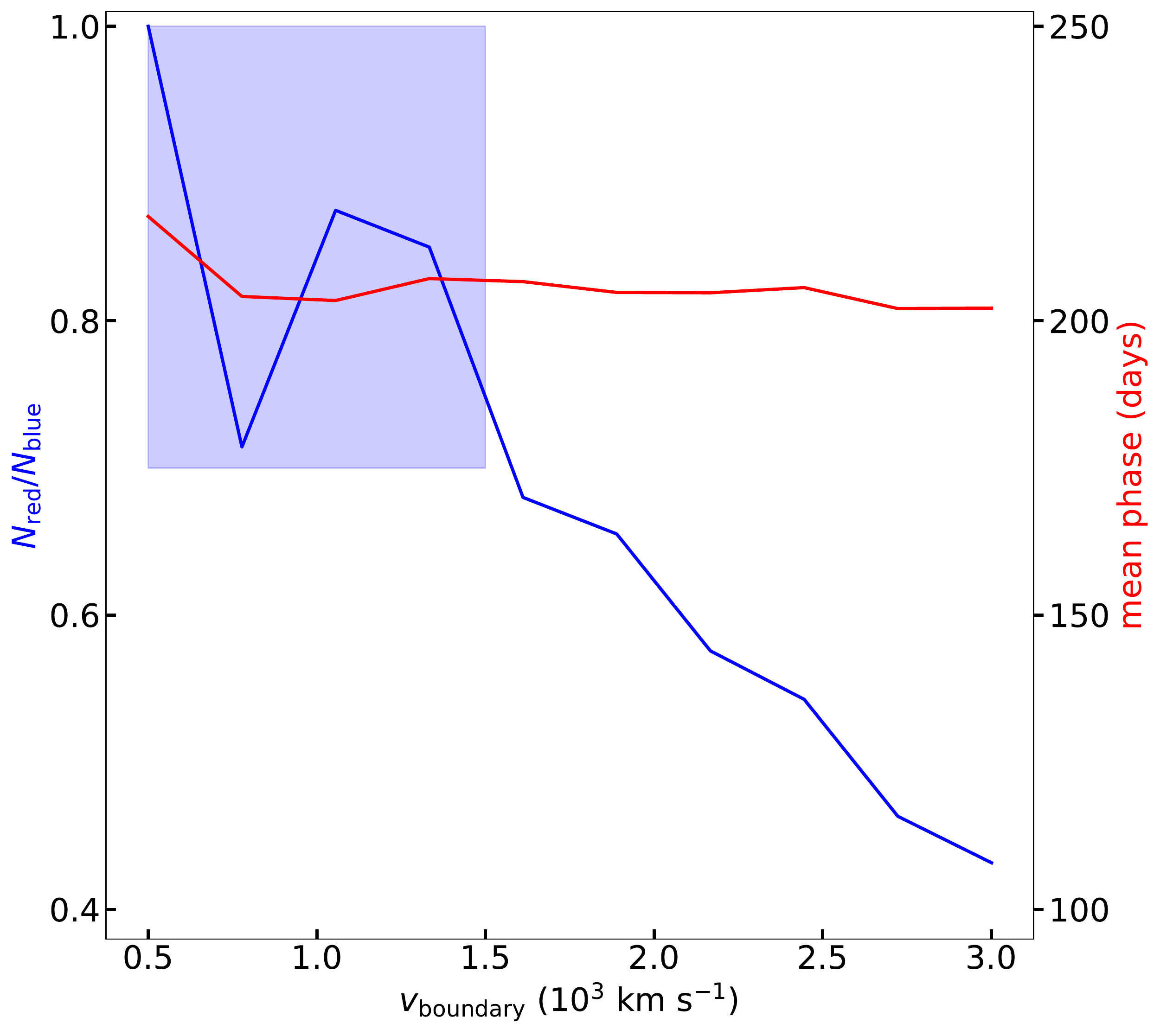}
\centering
\caption{The blue solid line shows the relation between the $N_{\rm red}$/$N_{\rm blue}$ ratio of the NC objects and the $v_{\rm boundary}$, which is defined to be the boundary velocity shift between NC/AS (by default 1000 km s$^{-1}$). The shaded region indicates ratio from 0.7 to 1.0. The red solid line shows the relation between the mean phase of NC objects and $v_{\rm boundary}$, labeled by the right y-axis.}
\label{fig:boundary_ratio}

\end{figure}

The imbalance of $N_{\rm blue}$ and $N_{\rm red}$ is mainly driven by the objects with large narrow component offsets. If the analysis is restricted to NC objects, we find $N_{\rm red}$=14 and $N_{\rm blue}$=16, while for AS objects, $N_{\rm red}$=7 and $N_{\rm blue}$=30. The above statement is not affected by the boundary of AS/NC. To be specific, the boundary velocity between AS/NC $v_{\rm boundary}$, which is by default 1000 km s$^{-1}$, is allowed to vary from 500 to 3000 km s$^{-1}$. The $N_{\rm{red}}$/$N_{\rm{blue}}$ ratio of the NC objects (narrow component offset within -$v_{\rm boundary}$ to $v_{\rm boundary}$) is then calculated. The result is shown as a function of $v_{\rm boundary}$ by the blue solid line in Figure \ref{fig:boundary_ratio}. When $v_{\rm boundary}$ varies from 500 to 1500 km s$^{-1}$, $N_{\rm{red}}$/$N_{\rm{blue}}$ fluctuates between 0.7 and 1.0, consistent with the moving blob or enhanced core density scenarios. The ratio of the red- and blue-shifted narrow components continue to drop if $v_{\rm boundary}$ is larger than about 1500 km s$^{-1}$. This phenomenon suggests the sample especially lacks objects with narrow component being red-shifted by $\lessapprox$ -1500 km s$^{-1}$, or has unusually enhanced number of objects with narrow components blue shifted by $\gtrapprox$ +1500 km s$^{-1}$. In the following, the possible reasons are discussed.

(i) $Residual~opacity~of~the~inner~ejecta$. The imbalance of the red- and blue-shifted narrow component can be possibly explained by the radiative transfer effect.  \citet{jerk15} found the opacity of their He star models is not negligible at around 200 days. For a He star with $M=4M_{\rm \odot}$, the escape probability of a photon ($\lambda$=6300$\rm \AA$) passing through 3500 km s$^{-1}$ material is $\sim$0.85. If the photon is emitted from rear side, it experiences twice the effective opacity. If the inner ejecta is optically thick, the emission from the rear side will be effectively scattered or absorbed, which possibly explains the lack of red-shifted narrow component.

However, this interpretation does not seem realistic for the reasons below: (1) The effect of radiative transfer decreases with the column density, which scales as $t^{-2}$, we therefore expect to see more red-shifted narrow component at later phases. However, no correlation can be discerned between the narrow component offset and the spectral phase ($\rho$ = 0.05, $p~\textless$ 0.64). Further, if the analysis is restricted to the objects observed later than 220 days, the imbalance is not eased ($N_{\rm{red}}$/$N_{\rm{blue}}$ $\sim$ 0.47), and the overall blue shift, which is about -12 $\rm {\AA}$, is still too large for the enhanced core density or the moving blob scenarios. In Figure \ref{fig:boundary_ratio}, we already find the $N_{\rm{red}}$/$N_{\rm{blue}}$ of NC objects is a decreasing function of $v_{\rm boundary}$, while no statistical difference can be discerned from the mean phases when $v_{\rm boundary}$ varies from 500 to 3000 km s$^{-1}$ (red solid line in Figure \ref{fig:boundary_ratio}).  (2) If the opaque ejecta is scattering-dominated, the peak of the emission will be blue-shifted. This effect is usually not very large (see \citealt{tauben09} for some simple models), and can possibly contribute to the small blue-shift of the broad base. However, the effect of scattering is not enough to explain the large overall blue-shift of the narrow component. (3) The fractional flux of the narrow component, $\alpha_{\rm w}$, can be a rough estimation of the fractional mass of the moving blob. The average value of $\alpha_{\rm w}$ is about 0.15. If the effect of self-absorption is included, to absorb the light emitted from such a massive blob, the ejecta will be unrealistically optical thick, resulting in a flux deficit at the red-shifted part of the broad base. The line profile will accordingly be heavily distorted, which contradicts the observation.

The relation between $N_{\rm{red}}$/$N_{\rm{blue}}$ of NC objects and $v_{\rm boundary}$ in Figure \ref{fig:boundary_ratio} can put important constrain on the asphericity development of SESNe ejecta. The limited range of $v_{\rm boundary}$, within which $N_{\rm{red}}$/$N_{\rm{blue}}$ is balanced, suggests that the objects with narrow components shifted by -1500 to 1500 km s$^{-1}$ can be explained by the moving blob or enhanced core density scenarios. However, the velocity of the blob can not be too large, otherwise more red-shifted narrow components with large velocity would be expected.

(ii) $Mis-classification$. Another solution is to use other profiles to fit the AS objects with the extremely blue-shifted narrow component. In the lower panels of Figure \ref{fig:relative_shift}, we take two AS objects, SNe 2000ew and 2008ax, which have narrow component offsets $\textless$ -2000 km s$^{-1}$, as examples. The line profile classification in this work is dependent on the initial guess of the fitting, as described in \S 2.4. For these two objects, the initial guess of case (4), i.e. broad base plus a blue shifted narrow component, indeed gives lower residual than the other cases, therefore is the \emph{numerically} best fit. However, the initial guess of case (1), i.e., blue- and red-shifted components with equal width and intensity, also gives reasonable good fit, as plotted by the green solid lines in the lower panels of Figure \ref{fig:relative_shift}.

Taking SNe 2000ew and 2008ax as examples, we are inclined to believe at least some of the AS objects are mis-classified, especially those with a trough located at $\sim$6300$\rm \AA$. The mis-classification explains the unusual enhancement of $N_{\rm blue}$. It turns out that, if a fraction of AS objects are re-classified to DP, the imbalance of the red- and blue-shifted narrow components can be eased. However, for a specific object, it is difficult to decide which profile is more appropriate, as both the DP and AS profiles give similarly good fits, while the geometry origins are totally different. The fitting procedure also has internal shortcoming; each component is assumed to be emitted by the Gaussian distributed emitter, while the intrinsic profile can be much more complicated. This also introduces uncertainty to the geometry interpretation of the line profile.

\section{Conclusions}
We have conducted a systematic study on the statistical properties of the SESNe nebular spectra. The sample includes 26 SNe IIb, 31 SNe Ib, 32 SNe Ic, 9 SNe Ic-BL and 5 SNe Ib/c. The investigation involves the morphology of the doublet [O {\sc i}] $\lambda\lambda$6300,6364, [O {\sc i}] width and the [O {\sc i}]/[Ca {\sc i}{\sc i}] ratio. The [O {\sc i}] $\lambda\lambda$6300,6364 is emitted from the oxygen-rich region, the amount of which is closely related to the properties of the core of the progenitor, especially its mass. Moreover, the [O {\sc i}] line is also one of the strongest emissions in the nebular spectrum of SESN, and is usually unblended, making it an ideal tracer of the geometry of the O-rich ejecta.

The measurement of the line width is based on the fractional flux of the line, and the result is in good agreement with those estimated in previous works. Although we have discussed the line profile of the [O {\sc i}], the measurement method of its width applied in this work does not assume any specific profile of the emission, allowing a more general discussion on the velocity scale of the ejecta.

To investigate the geometry of the oxygen-rich ejecta, a multi-Gaussian fitting is applied to the [O {\sc i}] $\lambda\lambda$6300,6363 of all the nebular spectra in the sample. The same classification scheme of \citet{tauben09} is applied, and according to the best-fit parameter, the line profiles are classified as: Gaussian (GS), narrow-core(NC, characterized by a Gaussian broad base plus a narrow component with center wavelength $\mid v_{\rm shift}\mid\textless1000$ km s$^{-1}$), double-peaked (DP, characterized by a horn-like profile, i.e., blue- and red-shifted components with similar widths and intensities) and asymmetry (AS, characterized by a Gaussian broad base plus a narrow component with center wavelength $\mid v_{\rm shift}\mid\textgreater1000$ km s$^{-1}$).

We then conduct a statistical analysis on the [O {\sc i}] profile, [O {\sc i}] width and the [O {\sc i}]/[Ca {\sc i}{\sc i}] ratio, along with the mutual relations between these quantities. For convenience, the observational findings are concluded as follows:
\begin{enumerate}

    \item Although the sample size in this work is about 2.5 times as large as that of \citet{tauben09}, the distribution of the line profile fractions are similar. The similarity between the results of the two samples suggests the sample size is sufficiently large to allow statistical study.

    \item For SNe IIb/Ib/Ic, the distributions of the line profiles are consistent with each other, which indicates the effects of the helium-rich layer and the small amount of the residual hydrogen envelope of SNe IIb are limited. On the other hand, there is a hint (at $1 \sigma$ level) that the distribution of the line profiles of SNe Ic-BL is different from canonical SESNe (SNe IIb/Ib/Ic).

    \item The distributions of the [O {\sc i}]/[Ca {\sc i}{\sc i}] ratio of SNe IIb and Ib are similar, but the average ratio for SNe IIb/Ib is significantly smaller than SNe Ic/Ic-BL. This result is consistent with the finding in \citet{fang19}. 
    
    \item The [O {\sc i}] width shows a similar sequence: it is larger for SNe Ic/Ic-BL than SNe IIb/Ib. The average velocity of SNe Ic-BL, inferred from the line width, is only slightly larger than the canonical SNe. It seems that the velocity of the innermost region is not strongly correlated with the velocity of the outermost ejecta. We leave the systematic investigation on the relation between the velocities measured from the early- and nebular-phase spectra to future works.

    \item A significant correlation between the [O {\sc i}]/[Ca {\sc i}{\sc i}] ratio and the [O {\sc i}] width is discerned, where objects with large [O {\sc i}]/[Ca {\sc i}{\sc i}] tend to have fast-expanding ejecta. The correlation is dependent on the SN sub types. For SNe IIb/Ib, the correlation is significant, but cannot be discerned for SNe Ic/Ic-BL. 
    
    \item The above correlation between the [O {\sc i}]/[Ca {\sc i}{\sc i}] ratio and the [O {\sc i}] width is found to be strong for objects showing a specific line profile. Among the line profile classes, NC objects have the tightest correlation, followed by DP/AS, then GS. 

    \item The dependence of the line profile on the [O {\sc i}]/[Ca {\sc i}{\sc i}] ratio is also observed. The average [O {\sc i}]/[Ca {\sc i}{\sc i}] of GS is the largest, followed by AS/NC, then DP objects. By dividing the sample into 5 groups with an equal number of members and calculate the fractions of the line profiles in each group, we find a steadily increasing tendency for the fraction of GS objects when [O {\sc i}]/[Ca {\sc i}{\sc i}] increases, while the fraction of DP objects goes to the opposite direction. Meanwhile, the fractions of NC/AS objects are not monotonic functions of [O {\sc i}]/[Ca {\sc i}{\sc i}]. 
\end{enumerate}

To interpret the observational results, it is crucial to connect the observables to the theoretical models. In this work, we use the [O {\sc i}]/[Ca {\sc i}{\sc i}] ratio as the measurement of progenitor CO core mass, as predicted by several nebular SESN models (\citealt{fransson89,jerk15}). The line fitting procedure is applied to the bipolar explosion models of \citet{maeda08} to qualitatively constrain the ejecta geometry.
The observational results can be interpreted as follows:

\begin{enumerate}
    \item For the canonical SNe, the material above the CO core (helium-rich layer and the residual hydrogen envelope) has limited effect on the ejecta geometry. 
    \item More than 50\% of the objects cannot be interpreted by the spherically symmetric ejecta. The deviation from spherical symmetry is  commonly seen for all types of SESNe. 
    \item The fraction of the DP objects is too low for the `strictly' bipolar explosion to be a majority. However, if we discard the condition of perfect axisymmetry and symmetry between the two hemispheres, and further combine the GS and NC profiles as `single-peak' profile and the DP and AS profiles as `non-single' profile, the bipolar explosion can account for the observed line profile fractions of the full sample. If this is the case, the deviation of the observed line profiles from the specific bipolar model sequence can be used to further constrain the nature of the explosion. We conclude that a large fraction of SESNe should have an non-axisymmetry configuration or imbalance in the two hemisphere to explain the distribution of the line profiles.
    \item The progenitors of SNe Ic/Ic-BL have on average a more massive CO core than SNe IIb/Ib. The helium-rich layer is most likely stripped by the mass-dependent stellar wind.
    \item The correlation between the CO core mass and expansion velocity of the ejecta, inferred from the line width, can not be explained by the constant kinetic energy for different progenitors. In a forthcoming work (Fang et al. in preparation), we will show that the correlation can be explained by assuming the kinetic energy is tightly correlated with the progenitor CO core mass. 
    \item Taking the DP profile as an indicator of the non-spherical ejecta, especially those with bipolar configurations, the relation between the [O {\sc i}]/[Ca {\sc i}{\sc i}] and DP fraction suggests the ejecta geometry is dependent on the progenitor CO core mass. However, the profile of
    [O {\sc i}] itself is not enough to reveal the geometry of the full ejecta. To firmly establish the relation between the progenitor CO core mass and the ejecta geometry, we thus need another probe of the ejecta with bipolar configuration, which should be independent from [O {\sc i}]. The investigation on this topic will be presented in a forthcoming work (Fang et al., in preparation).
\end{enumerate}

There remain uncertainties of the theoretical interpretation to the observational relations. (1) Our understanding on the important observable, [O {\sc i}]/[Ca {\sc i}{\sc i}], along with its relation with the physical properties (CO core mass, kinetic energy, calcium pollution, etc.), is highly dependent on the current He star model spectra. (2) The line fitting procedure and the classification scheme proposed by \citet{tauben09} are empirical. The geometrical interpretation of the line profile are complicated by the degeneracy of the fitting, as exemplified by SNe 2000ew and 2008ax; they are originally classified as AS, but DP profile also provides reasonably good fit. The inference from the line profile to the ejecta geometry is not straightforward. 

To better connect the observation to the properties of the SESNe progenitor, a sophisticated radiative transfer modeling of the ejecta involving different geometrical configurations, viewing angles and randomly-distributed moving blobs, is required.

\begin{acknowledgements}
The authors would like to thank Takashi Nagao, Luc Dessart, and Tomoya Takiwaki for the very helpful and enlightening discussions. The authors would like to thank the anonymous reviewer for his/her comments, which helped to improve the manuscript.
Q.F. acknowledges support by Japan Society for the Promotion of Science (JSPS) KAKENHI Grant (20J23342). K.M. acknowledges support by JSPS KAKENHI Grant (20H00174, 20H04737, 18H05223). M.T. acknowledges support by MEXT/JSPS KAKENHI Grant (17H06363, 19H00694, 20H00158, 20H00179). H.K. is funded by the Academy of Finland projects 324504 and 328898.

This research is based [in part] on data collected at Subaru Telescope, which is operated by the National Astronomical Observatory of Japan.
\end{acknowledgements}

\software{IRAF \citep{tody86,tody93}; LAcosmic \citep{dokkum01}; SciPy \citep{scipy}; NumPy \citep{numpy}; Astropy \citep{astropy13,astropy18}; Matplotlib \citep{matplotlib}, \texttt{R} \citep{teamr21}, \texttt{locfit} \citep{loader99,loader18}}

\appendix
\clearpage
\newpage

\setcounter{table}{0}
\renewcommand{\thetable}{A\arabic{table}}

\setcounter{figure}{0}
\renewcommand{\thefigure}{A\arabic{figure}}

\section{Lists of SNe in this work}

\begin{deluxetable}{cccccccc}
\tablecaption{SNe IIb in this work.}
\label{tab:listIIb}
\tablehead{
\colhead{SN name} & \colhead{Host} & \colhead{date}&\colhead{phase\tablenotemark{a}}&\colhead{redshift}&\colhead{$E(B-V)$\tablenotemark{b}}&\colhead{profile}&\colhead{references\tablenotemark{c}}
}
\startdata
1987K&NGC 4651&1988/02/24&211&0.0027&0.36(+)&NC&F88\\
1993J&NGC 3031&1993/11/07&203&-0.0001&0.19&AS&M00,J15\\
1996cb&NGC 3510&1997/07/01&176&0.0030&0.03&AS&Q99\\
2001ig&NGC 7424&2002/10/08&274&0.0066&0.10&AS&M07a,S09\\
2003bg&MCG-05-10-15&2003/11/29&254&0.0049&0.02&AS&H09\\
2006G&NGC 521&2006/06/30&169&0.0171&0.36(+)&AS&This work\\
2006T&NGC 3054&2006/11/26&284&0.0086&0.08&DP&M07b,M08,M14\\
2007ay&UGC 4310&2007/11/05&$\gtrapprox$ 190&0.0147&0.36(+)&GS&This work\\
2008aq&PGC 43458&2008/06/26&108&0.0075&0.36(+)&AS&M14\\
2008ax&NGC 4490&2008/11/24&245&0.0019&0.40&AS&T11,M14\\
2008bo&NGC 6643&2008/10/27&195&0.0053&0.08&AS&S19\\
2008ie&NGC 1070&2009/10/27&316&0.0136&0.36(+)&AS&This work\\
2009C&UGC 12433&2009/10/26&297&0.0226&0.36(+)&DP&This work\\
2009K&NGC 1620&2009/10/26&261&0.0113&0.36(+)&NC&This work\\
2009ka&Anon&2010/05/06&200&0.0175&0.36(+)&NC&This work\\
2010as&NGC 6000&2010/08/05&130&0.0078&0.44&DP&F14\\
2011dh&NGC 5194&2011/12/24&187&0.0020&0.07&NC&S13,E14,E15\\
2011ei&NGC 6925&2012/06/18&311&0.0089&0.24&GS&M13\\
2011fu&UGC 1626&2012/07/20&282&0.0185&0.10&AS&MG15\\
2011hs&IC 5267&2012/06/21&211&0.0057&0.17&NC&B14\\
2012P&NGC 5806&2012/08/08&197&0.0045&0.29&DP&F16\\
2012dy&ESO 145-G4&2012/12/23&nebular&0.0103&0.36(+)&AS&Y12\\
2013ak&ESO 430-G20&2013/09/13&179&0.0035&0.30&NC&Y12\\
2013bb&NGC 5504&2014/03/02&332&0.0190&0.30&GS&Y12,S19\\
2013df&NGC 4414&2014/02/04&223&0.0024&0.10&NC&MG14,M15\\
ASASSN-14az&PGC 110136&2014/11/25&189&0.0067&0.36(+)&¥AS&S19\\\hline
\enddata
\tablenotetext{a}{Phase relative to the light curve maximum or discover date.}
\tablenotetext{b}{Objects labeled by ($+$) indicate the case where its extinction can not be calculated from light curve reported by literature or Na I D absorption. The average $E(B - V)$ of SN Ib/c (0.36 mag) is adopted for this case.}
\tablenotetext{c}{F88: \citet{fili88}; Q99: \citet{qiu99}; M00: \citet{matheson00}; M07a: \citet{maund07}; M07b: \citet{modjazphd}; M08: \citet{maeda08} H09: \citet{hamuy09}; S09: \citet{silverman09}; T11: \citet{tauben11}; Y12: \citet{yaron12}; M13: \citet{milisa13}; S13: \citet{shivvers13}; B14: \citet{bufano14}; E14: \citet{ergon14}; F14: \citet{folatelli14}; M14: \citet{modjaz14}; MG14: \citet{morales14}; E15: \citet{ergon15}; J15: \citet{jerk15}; M15: \citet{maeda15}; MG15: \citet{morales15}; F16: \citet{fremling16}; S19: \citet{shivvers19}}.
\end{deluxetable}

\begin{deluxetable}{cccccccc}
\tablecaption{SNe Ib in this work.}
\label{tab:listIb}
\tablehead{
\colhead{SN name} & \colhead{Host} & \colhead{date}&\colhead{phase}&\colhead{redshift}&\colhead{$E(B-V)$}&\colhead{profile}&\colhead{references}
}
\startdata
1985F&NGC 4618&1985/04/01&280&0.0002&0.23&NC&F86\\
1990I&NGC 4650&1991/04/21&357&0.0097&0.12&NC&E04\\
1990U&NGC 7479&1991/01/06&189&0.0081&0.52&DP&G94,M01,T09,M14\\
1997X&NGC 4691&1997/05/13&103&0.0035&0.18&GS&G02,T09\\
1999dn&NGC 7714&2000/09/01&379&0.0090&0.10&GS&B11\\
2000ew&NGC 3810&2001/03/17&110&0.0033&0.36(+)&AS&T09\\
2002dz&MCG-01-01-52&2002/08/10&nebular&0.0184&0.36(+)&DP&S19\\
2004ao&UGC 10862&2004/11/14&250&0.0059&0.12&DP&E11,S19\\
2004dk&NGC 6118&2005/05/11&263&0.0052&0.34&AS&M08a,D11,M14,S17\\
2004gn&NGC 4527&2005/07/06&217&0.0061&0.36(+)&AS&M08b\\
2004gq&NGC 1832&2005/08/26&249&0.0059&0.25&AS&M08a,M08b,D11,M14\\
2004gv&NGC 856&2005/08/26&242&0.0200&0.25&GS&M08a,M08b,M14\\
2005bf&MCG+00-27-05&2005/12/11&213&0.0186&0.14&DP&F06,M14\\
2006F&NGC 935&2006/06/30&175&0.0139&0.54&NC&G06,M08b,D11\\
2006ep&NGC 214&2006/12/24&104&0.0152&0.36(+)&NC&This work\\
2006gi&NGC 3147&2007/02/10&145&0.0094&0.38&NC&T09,E11\\
2006ld&UGC 348&2007/07/17&258&0.0140&0.36(+)&AS&T09\\
2007C&NGC 4981&2007/06/20&155&0.0056&0.64&AS&T09,D11,M14\\
2007Y&NGC 1187&2007/09/22&200&0.0040&0.11&AS(NC)&S09\\
2007uy&NGC 2770&2008/06/06&141&0.0063&0.79&NC(AS)&R13,M14\\
2008D&NGC 2770&2008/06/07&140&0.0072&0.65&AS&M09\\
2008fd&ESO 466-G24&2009/07/23&330&0.0181&0.36(+)&GS&This work\\
2008im&UGC 2906&2009/08/18&232&0.0090&0.36(+)&GS&This work\\
2009jf&NGC 7479&2010/06/19&245&0.0068&0.12&NC&S11,V11,M14\\
2012au&NGC 4790&2012/12/19&284&0.0045&0.06&AS(NC)&M13\\
iPTF13bvn&NGC 5806&2014/02/21&234&0.0045&0.07&AS(NC)&F16\\
2014C&NGC 7331&2014/08/25&221&0.0029&0.75&DP&S19\\
2014ei&MCG-01-13-50&2015/03/27&142&0.0148&0.36(+)&GS&S19\\
2015Q&NGC 3888&2016/01/07&212&0.0078&0.36(+)&NC&S19\\
2015ah&UGC 12295&2016/01/07&152&0.0160&0.10&NC&S19\\
PS15bgt&NGC 6412&2015/12/17&147&0.0090&0.23&GS&S19\\
\enddata
\tablenotetext{c}{F86: \citet{fili86}; G94: \citet{gomez94}; M01: \citet{matheson01}; G02: \citet{gomez02}; E04: \citet{elmhamdi04}; F06: \citet{folatelli06}; G06: \citet{green06}; M08a: \citet{modjaz08}; M08b: \citet{maeda08}; M09: \citet{modjaz09}; S09: \citet{stritzinger09}; T09: \citet{tauben09}; B11: \citet{benetti11}; D11: \citet{drout11}; E11: \citet{elmhamdi11}; S11: \citet{sahu11}; V11: \citet{valenti11}; M13: \citet{milisa13b}; R13: \citet{roy13}; M14: \citet{modjaz14}; F16: \citet{fremling16}; S17: \citet{shivvers17}; S19: \citet{shivvers19}}
\end{deluxetable}

\begin{deluxetable}{cccccccc}
\tablecaption{SNe Ic in this work.}
\label{tab:listIc}
\tablehead{
\colhead{SN name} & \colhead{Host} & \colhead{date}&\colhead{phase}&\colhead{redshift}&\colhead{$E(B-V)$}&\colhead{profile}&\colhead{references}
}
\startdata
1987M&NGC 2715&1988/02/25&157&0.0043&0.45&GS&F90,J91\\
1990aa&MCG+05-03-16&1991/01/23&140&0.0170&0.36(+)&GS&M01\\
1991A&IC 2973&1991/04/07&96&0.0105&0.42&GS&M01\\
1991N&NGC 3310&1992/01/09&$\gtrapprox$ 286&0.0035&0.12&GS&F91,M01,M08\\
1994I&NGC 5194&1994/09/02&146&0.0015&0.45&AS&F95,R96,M14\\
1996aq&NGC 5584&1997/04/02&228&0.0055&0.36(+)&AS&N96,T09\\
1996D&NGC 1614&1996/09/10&214&0.0149&0.36(+)&GS&D96,T09\\
1997B&IC 438&1997/09/23&252&0.0095&0.36(+)&GS&T09\\
1997dq&NGC 3810&1998/05/30&210&0.0033&0.11&DP&N97,M01,T09,M14\\
2003gf&MCG-04-52-26&2003/11/29&158&0.0087&0.36(+)&AS&S19\\
2004aw&NGC 3997&2004/11/14&232&0.0159&0.37&NC&T06,M14\\
2004fe&NGC 132&2005/07/06&240&0.0180&0.32&DP&M08,D11,M14\\
2004gk&IC 3311&2005/07/10&223&-0.0005&0.47&AS&M08,M14,E11\\
2004gt&NGC 4038&2005/05/24&152&0.0046&0.10&DP&GY05,T09,M14\\
2005aj&UGC 2411&2005/08/25&188&0.0085&0.36(+)&AS&This work\\
2005bj&MCG+03-43-05&2005/08/25&136&0.0222&0.36(+)&NC&This work\\
2005kl&NGC 4369&2006/06/30&213&0.0034&0.29&DP&M08,M14,D11\\
2005kz&MCG+08-34-32&2006/06/30&215&0.0278&0.46&AS&M08,D11\\
2006ck&UGC 8238&2007/01/24&246&0.0245&0.39&NC(AS)&C06,M08,M14\\
2007gr&NGC 1058&2008/02/12&170&0.0020&0.09&NC&S19\\
2007rz&NGC 1590&2008/04/01&115&0.0135&0.36(+)&AS&M14\\
2008fo&A164012+3943&2009/04/05&240&0.0289&0.36(+)&AS(NC)&This work\\
2007hb&NGC 819&2008/01/11&140&0.0222&0.36(+)&GS&M14\\
2008hh&IC 112&2009/08/18&269&0.0196&0.26&GS&This work\\
2009jy&NGC 3208&2010/05/06&204&0.0103&0.36(+)&AS&This work\\
2010mb&A160023+3744&2011/03/04&248&0.1325&0.01&NC&BA14\\
2011bm&IC 3918&2012/01/22&263&0.0212&0.06&AS&V12\\
PTF12gzk&SDSS J221241.53+003042.7&2013/06/10&299&0.0139&0.14&AS&S19\\
2013ge&NGC 3287&2014/04/28&156&0.0045&0.07&NC&D16\\
2014L&NGC 4254&2014/06/29&142&0.0078&0.67&NC&Z18,S19\\
2014eh&NGC 6907&2015/06/16&210&0.0106&0.36(+)&GS&S19\\
iPTF15dtg&Anon&2016/10/31&327&0.0544&0.06&DP&T16,T19\\
\enddata
\tablenotetext{c}{F90: \citet{fili90}; F91: \citet{fili91}; J91: \citet{jeffery91}; F95: \citet{fili95}; D96: \citet{drissen96}; N96: \citet{nakano96}; R96: \citet{richmond96}; N97: \citet{nakano97}; M01: \citet{matheson01}; GY05: \citet{galyam05}; C06: \citet{cole06}; T06: \citet{tauben06}; M08: \citet{maeda08}; T09: \citet{tauben09}; D11: \citet{drout11}; E11: \citet{elmhamdi11}; V12: \citet{valenti12}; BA14: \citet{benami14}; M14: \citet{modjaz14}; D16: \citet{drout16};  T16: \citet{taddia16}; Z18: \citet{zhang18}; S19: \citet{shivvers19}; T19: \citet{taddia19}.}
\end{deluxetable}

\begin{deluxetable}{cccccccc}
\tablecaption{SNe Ic-BL in this work.}
\label{tab:listIcBL}
\tablehead{
\colhead{SN name} & \colhead{Host} & \colhead{date}&\colhead{phase}&\colhead{redshift}&\colhead{$E(B-V)$}&\colhead{profile}&\colhead{references}
}
\startdata
1997ef&UGC 4107&1998/09/21&282&0.0117&0.00&NC&I00,M00,M01,M14\\
1998bw&ESO 184-G82&1998/11/26&198&0.0096&0.06&NC&P01,C11,M16\\
2002ap&NGC 628&2002/08/09&183&0.0021&0.08&NC&F03,Y03,M16\\
2005nb&UGC 7230&2006/06/30&183&0.0235&0.36(+)&AS&M08,M14\\
2006aj&A032139+1652&2006/09/21&206&0.0330&0.15&NC&M06,M14,M16\\
2007D&UGC 2653&2007/09/18&252&0.0232&0.91&NC&This work,D16\\
2007I&A115913-0136&2007/07/15&182&0.0215&0.36(+)&AS&B07,T09,M14\\
PTF10qts&SDSS J164137.53+285820.3&2010/04/27&231&0.0912&0.02&GS&W14\\
2012ap&NGC 1729&2012/09/23&272&0.0121&0.45&AS&M15\\
\enddata
\tablenotetext{c}{I00: \citet{iwamoto00}; M00: \citet{mazzali00}; M01: \citet{matheson01}; P01: \citet{patat01}; F03: \citet{foley03}; Y03: \citet{yoshii03}; M06: \citet{modjaz06}; B07: \citet{blondin07}; M08: \citet{maeda08}; T09: \citet{tauben09}; C11: \citet{clocc11}; M14: \citet{modjaz14}; M16: \citet{modjaz16} W14: \citet{walker14}; M15: \citet{milisa15}; D16: \citet{drout16}.}
\end{deluxetable}

\begin{deluxetable}{cccccccc}
\tablecaption{SNe Ib/c in this work.}
\label{tab:listIbc}
\tablehead{
\colhead{SN name} & \colhead{Host} & \colhead{date}&\colhead{phase}&\colhead{redshift}&\colhead{$E(B-V)$}&\colhead{profile}&\colhead{references}
}
\startdata
1990W&NGC 6221&1991/02/21&186&0.0042&0.36(+)&NC&T09\\
1990aj&NGC 1640&1991/03/10&180&0.0053&0.36(+)&NC&M01\\
1995bb&A001617+1224&1996/01/21&nebular&0.0055&0.36(+)&GS&M14\\
2005N&NGC 5420&2005/01/22&nebular&0.0163&0.36(+)&AS&H08\\
2012fh&NGC 3344&2012/11/14&nebular&0.0017&0.36(+)&DP&S19\\
\enddata
\tablenotetext{c}{M01: \citet{matheson01}; H08: \citet{harutyunyan08}; T09: \citet{tauben09}; M14: \citet{modjaz14}; S19: \citet{shivvers19}.}
\end{deluxetable}

\clearpage
\newpage

\setcounter{table}{0}
\renewcommand{\thetable}{B\arabic{table}}

\setcounter{figure}{0}
\renewcommand{\thefigure}{B\arabic{figure}}

\section{Line fitting Results}
\begin{figure*}[!hbt]
\epsscale{1.}
\plotone{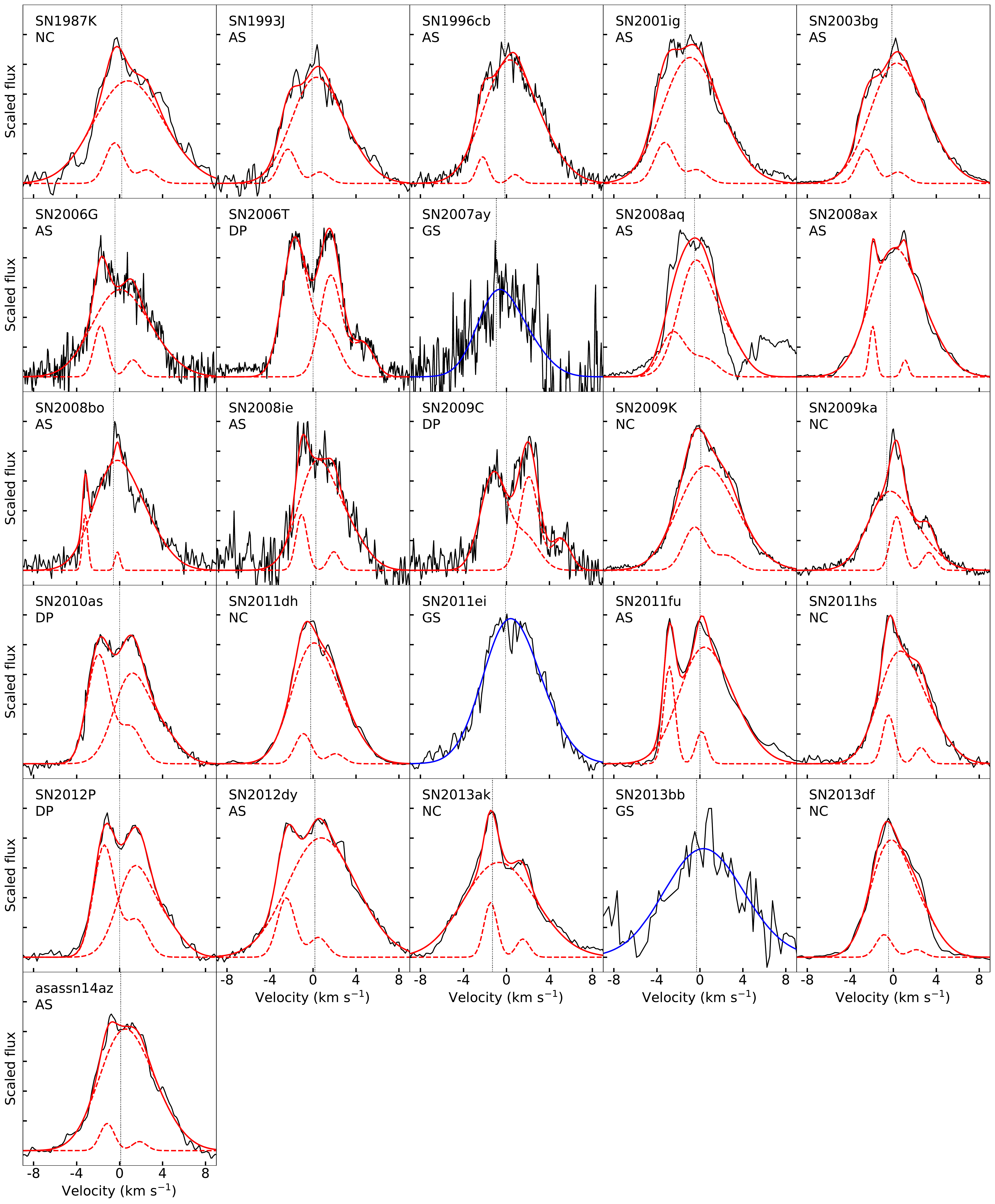}
\centering
\caption{Observed [O {\sc i}] of the SNe IIb in the sample fitted by multi-Gaussians. The spectra (black solid lines) are already subtracted by the background and the symmetric H$\alpha$/[N {\sc ii}]. The blue solid lines are the results of one-component fit. The red solid lines are the results of two-component fit, and the red dashed lines are the components. The vertical dotted lines are zero velocity (6300$\rm \AA$) for DP objects or the center wavelength of the Gaussian broad base for GS, NC or AS objects for references.}
\label{fig:IIb_fit}
\vspace{4mm}
\end{figure*}
\vspace{0.3cm}

\begin{figure*}[!hbt]
\epsscale{1}
\plotone{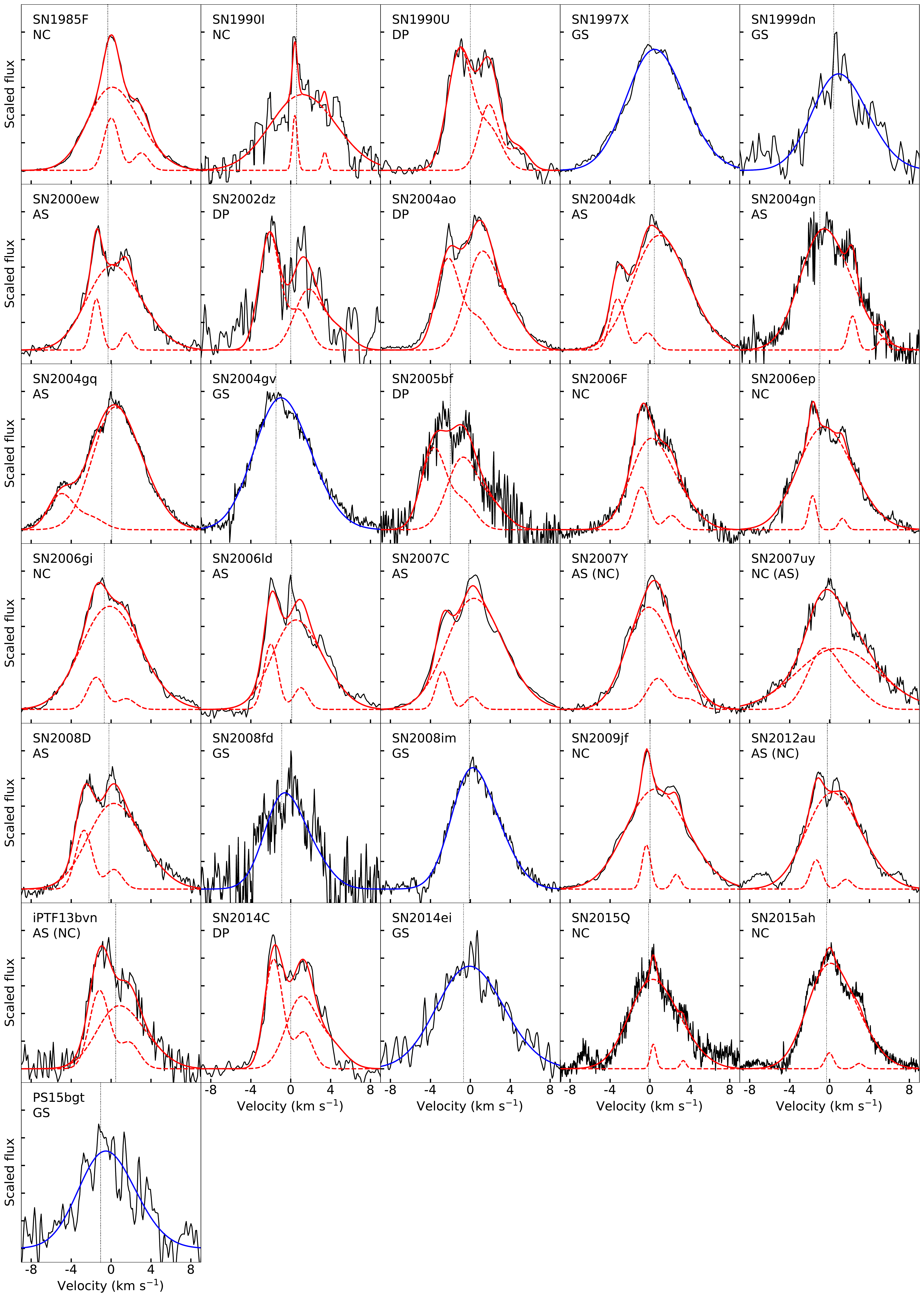}
\centering
\caption{Observed [O {\sc i}] of the SNe Ib in the sample fitted by multi-Gaussians. The spectra (black solid lines) are already subtracted by the background and the symmetric H$\alpha$/[N {\sc ii}]. The blue solid lines are the results of one-component fit. The red solid lines are the results of two-component fit, and the red dashed lines are the components. The vertical dotted lines are zero velocity (6300$\rm \AA$) for DP objects or the center wavelength of the Gaussian broad base for GS, NC or AS objects for references. SN 2005bf is an exception; the [Ca {\sc ii}] of this object is blue shifted by $\sim$2000 km s$^{-1}$, which is then taken as the "center" of SN 2005bf.}
\label{fig:Ib_fit}
\vspace{4mm}
\end{figure*}
\vspace{0.3cm}

\begin{figure*}[!hbt]
\epsscale{1}
\plotone{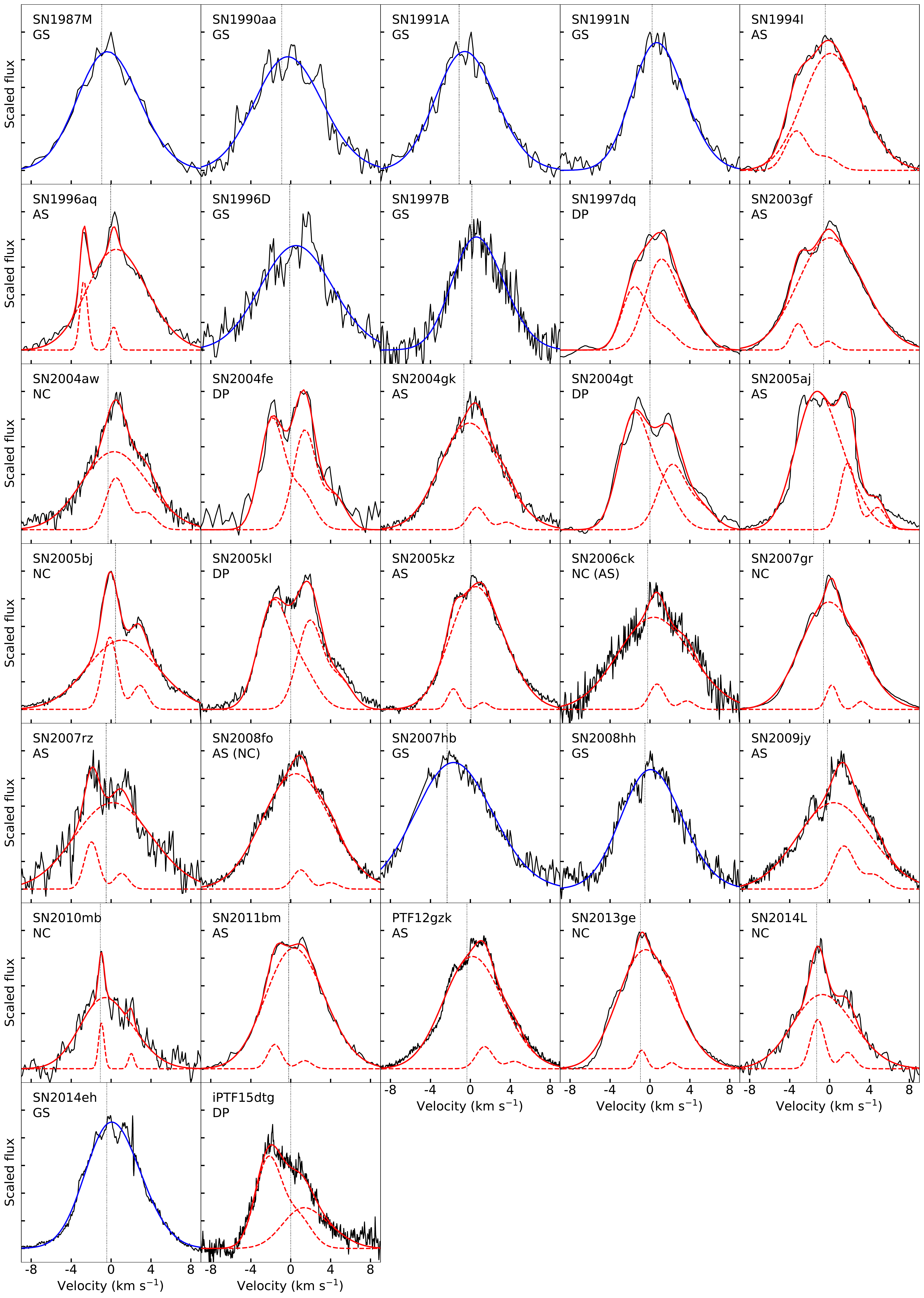}
\centering
\caption{Observed [O {\sc i}] of the SNe Ic in the sample fitted by multi-Gaussians. The spectra (black solid lines) are already subtracted by the background and the symmetric H$\alpha$/[N {\sc ii}]. The blue solid lines are the results of one-component fit. The red solid lines are the results of two-component fit, and the red dashed lines are the components. The vertical dotted lines are zero velocity (6300$\rm \AA$) for DP objects or the center wavelength of the Gaussian broad base for GS, NC or AS objects for references.}
\label{fig:Ic_fit}
\vspace{4mm}
\end{figure*}
\vspace{0.3cm}

\begin{figure*}[!hbt]
\epsscale{1}
\plotone{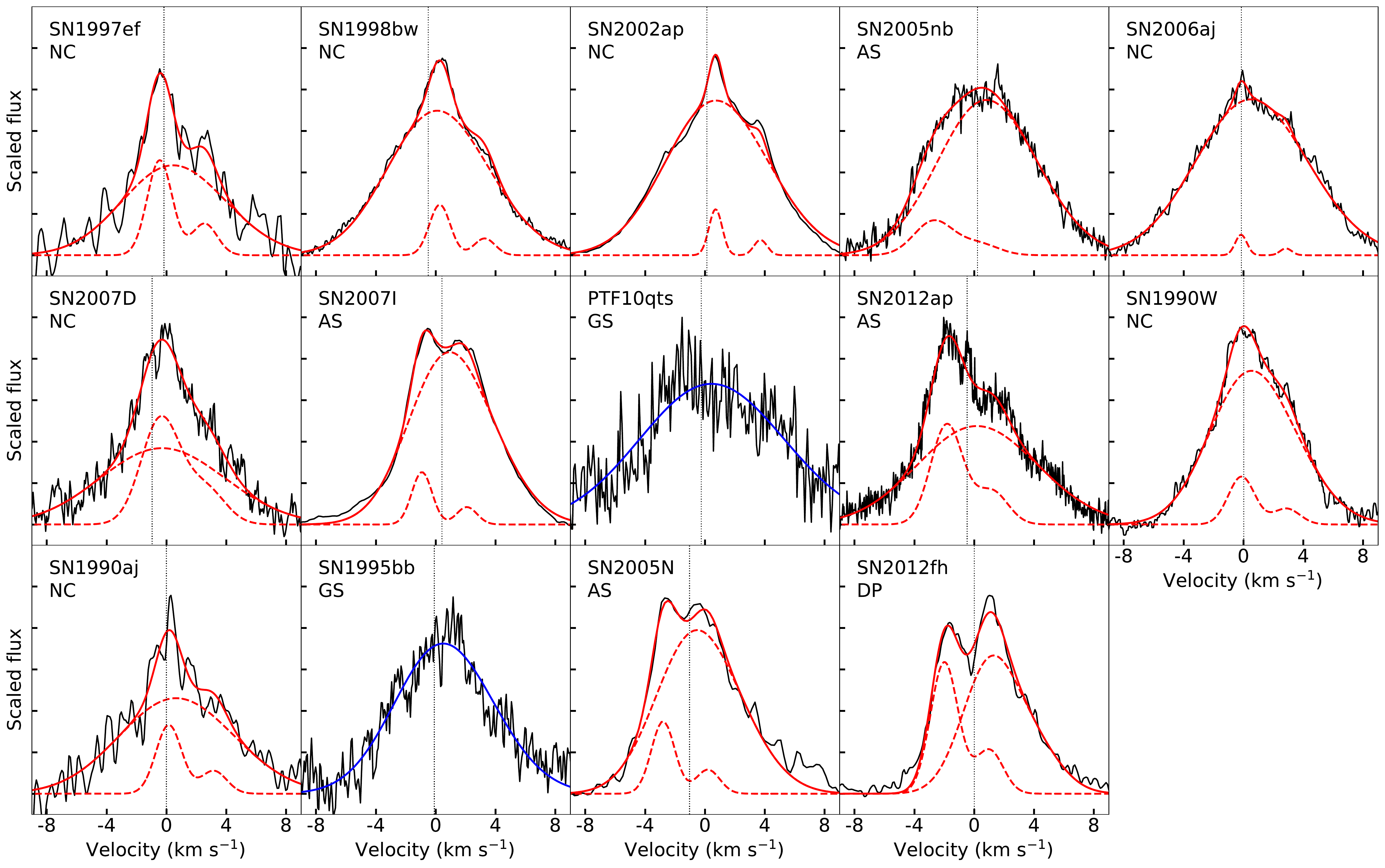}
\centering
\caption{Observed [O {\sc i}] of the SNe Ic-BL and SNe Ib/c in the sample fitted by multi-Gaussians. The spectra (black solid lines) are already subtracted by the background and the symmetric H$\alpha$/[N {\sc ii}]. The blue solid lines are the results of one-component fit. The red solid lines are the results of two-component fit, and the red dashed lines are the components. The vertical dotted lines are zero velocity (6300$\rm \AA$) for DP objects or the center wavelength of the Gaussian broad base for GS, NC or AS objects for references.}
\label{fig:BL_fit}
\vspace{4mm}
\end{figure*}
\vspace{0.3cm}

\clearpage
\newpage
\setcounter{table}{0}
\renewcommand{\thetable}{C\arabic{table}}

\setcounter{figure}{0}
\renewcommand{\thefigure}{C\arabic{figure}}

\section{Examples of multi-phase nebular spectra}
\begin{figure*}[!hbt]
\epsscale{1.1}
\plotone{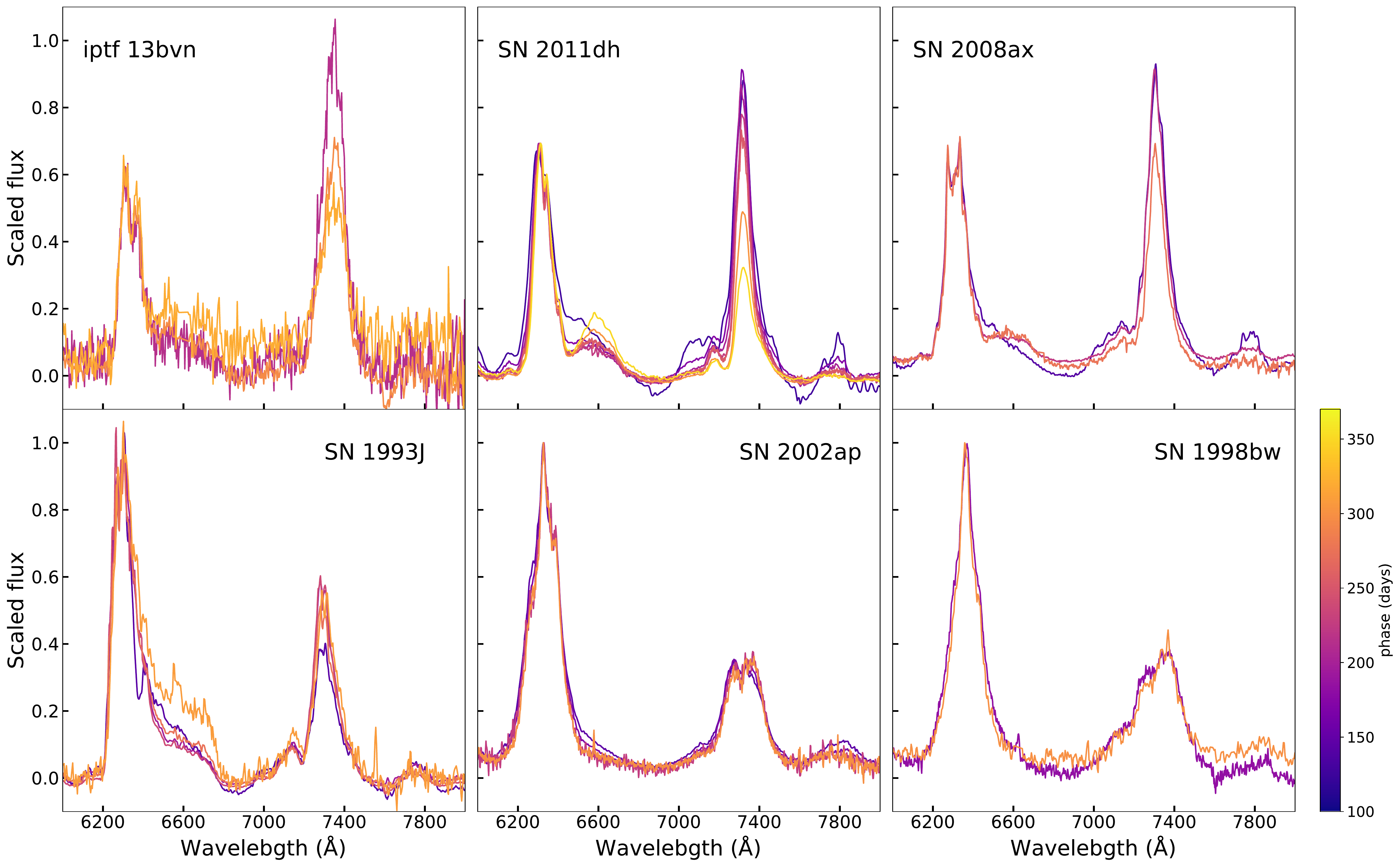}
\centering
\caption{The evolution of [O {\sc i}]/[Ca {\sc i}{\sc i}] of well-observed SNe. The spectra are scaled to the peak of the [O {\sc i}]. The colors of the lines indicate the phase of the spectra, with the late phase spectra plotted with the colors at the red-end. The SNe with spectra plotted in the upper panels have fast evolving [O {\sc i}]/[Ca {\sc i}{\sc i}] and relatively low [O {\sc i}]/[Ca {\sc i}{\sc i}] ($\textless$ 1) at $\sim$200 days. The SNe with spectra plotted in the lower panels are examples with relatively large [O {\sc i}]/[Ca {\sc i}{\sc i}] ($\textgreater$ 1.5), and their [O {\sc i}]/[Ca {\sc i}{\sc i}] hardly evolve from 100-350 days.}
\label{fig:multiphase_example}
\end{figure*}

{}
\end{CJK*}
\end{document}